# Simplicial Quantum Gravity

ACADEMISCH PROEFSCHRIFT

ter verkrijging van de graad van doctor
aan de Universiteit van Amsterdam
op gezag van de Rector Magnificus
prof. dr P.W.M. de Meijer
ten overstaan van een door het college van dekanen ingestelde
commissie in het openbaar te verdedigen in de Aula der Universiteit
op woensdag 13 september 1995 te 13.30 uur
door

## Bastiaan Valentijn de Bakker

geboren te Amsterdam

Promotor:              prof. dr J. Smit

Promotiecommissie:   prof. dr P.J. van Baal
                     prof. dr ir F.A. Bais
                     dr J.W. van Holten
                     prof. dr G. 't Hooft
                     prof. dr B. Nienhuis
                     prof. dr H.L. Verlinde

Faculteit der Wiskunde, Informatica,
            Natuur- en Sterrenkunde

Instituut voor Theoretische Fysica

«Что бы я хотел пожелать молодёжи моей Родины, посвятившей себя науке?

. . .

Второе — это скромность. Никогда не думайте, что бы уже всё знаете. И как бы высоко ни оценивали вас, всегда имейте мужество сказать себе: Я невежда.»

Иван Петрович Павлов
*Письмо к молодёжи*

# Table of contents

# Chapter One

# Introduction

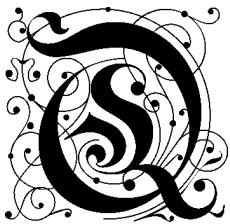 UANTUM gravity is one of the greatest unsolved problems in the physics of fundamental interactions. In this introduction I discuss some of the well known problems of quantum gravity and define simplicial quantum gravity, which is a relatively new approach to quantum gravity that tries to deal with some of these problems. I will also explain some general concepts of lattice field theory and Monte Carlo methods.

## 1.1   Quantum gravity

Quantum gravity is the name given to various theories that attempt to define a quantum theory of gravity that is consistent and gives the right answers in the classical limit. Although many people have been working on this problem, no satisfactory solution has yet been found. A few of the many attempts that have been made will be mentioned in section 1.2.

As is widely known, general relativity is a classical theory of gravity that is both very elegant and gives accurate predictions of all observed phenomena of gravity. Well known examples of such phenomena are Newtonian gravity (in the weak field limit of general relativity), the perihelion movement of Mercury and, a more recent discovery, the change in frequency of a binary pulsar due to the loss of energy to gravitational waves.

The action of general relativity without matter is called the Einstein-Hilbert action

$$S = \frac{1}{16\pi G} \int d^4x \sqrt{-g}(-2\Lambda + R),$$   (1.1)

where G is the gravitational constant, $\Lambda$ is the cosmological constant, g is the determinant of the metric $g_{\mu\nu}$ and R is the scalar curvature, which is a nasty function of the metric. This function is usually written in terms of the connection

$$\Gamma^{\sigma}{}_{\mu\nu} = \frac{1}{2} g^{\sigma\rho} \left( \partial_{\mu} g_{\nu\rho} + \partial_{\nu} g_{\mu\rho} - \partial_{\rho} g_{\mu\nu} \right), \tag{1.2}$$

as

$$R_{\mu\nu\rho}{}^{\sigma} = \partial_{\nu} \Gamma^{\sigma}{}_{\mu\rho} - \partial_{\mu} \Gamma^{\sigma}{}_{\nu\rho} + \Gamma^{\alpha}{}_{\mu\rho} \Gamma^{\sigma}{}_{\alpha\nu} - \Gamma^{\alpha}{}_{\nu\rho} \Gamma^{\sigma}{}_{\alpha\mu}, \tag{1.3}$$

$$R = g^{\mu\nu} R_{\mu\sigma\nu}{}^{\sigma}. \tag{1.4}$$

Because this function is so complicated, finding classical solutions to the action (1.1) is very difficult. General relativity allows wave-like solutions, which in a quantum theory would correspond to particles called gravitons. For more details on general relativity, see one of the many books on this subject, like [Misner *et al.* 1973].

The question arises whether quantum gravity is necessary at all. Considering that we have no experiments that detect quantum corrections to general relativity, one might try to describe gravity using only a classical theory. It is very hard, however, to conceive of a world where gravity is classical while all the matter is quantized. What would the metric couple to? If it would couple to the expectation value of the matter, strange situations would occur if after a measurement the matter configuration that we have measured is far away from its expectation value. Therefore it seems likely that the fundamental theory of gravitation is a quantum theory. Thus, the desire to find the fundamental laws of nature leads to a study of quantum gravity.

Another, somewhat more practical, reason to study quantum gravity is that general relativity predicts singularities, but breaks down at those same singularities. In particular the Big Bang could have produced anything, as far as classical general relativity is concerned. Quantum gravity might be able to produce some boundary conditions that can tell us why the universe looks like it does. Very roughly speaking, this is done by calculating the transition probabilities from nothing to the possible early universes. For the theory to have predictive power, the probabilities should be strongly peaked around one or more states of the universe.

As a related issue, quantum fluctuations in the very early universe may have produced fluctuations in the cosmic background radiation, such as were measured by the COBE satellite. Quantum gravity may predict the behaviour of these fluctuations. Because inflation (the rapid expansion of the early universe) would change the observed fluctuations, relating the pre-inflation quantum gravity predictions to the observations could provide information on the feasibility of inflation scenarios.

We will from the beginning restrict ourselves to Euclidean gravity, that is gravity where the metric is positive definite. In this approach, the action (1.1) becomes

$$S = \frac{1}{16\pi G} \int d^4x \sqrt{g}(2\Lambda - R). \qquad (1.5)$$

Strictly speaking Riemannian gravity would be a better name, because space certainly does not need to be flat, but Euclidean gravity is the commonly used name for this approach. We will discuss the consequences of this change in section 1.8.

## 1.2  Problems

The usual approach to quantum field theories is perturbation theory. One of the main problems of quantizing Einstein gravity (that is gravity with the normal Einstein action (1.5)) is that perturbation theory does not work, because the resulting perturbation expansion is not renormalizable. This can be quite easily seen from the fact that the coupling constant G has a negative mass dimension. This means that higher order loop corrections will generate an infinity of counterterms of ever higher dimensions.

One hope for standard perturbation theory might be that the theory is finite in every order, due to cancellations of divergences. In perturbation theory around flat space, this does indeed happen for the one loop diagrams ['t Hooft & Veltman 1974], but not for the two loop diagrams [Goroff & Sagnotti 1986, van de Ven 1992]. Also, the divergences no longer cancel at one loop if the background space is not flat or a scalar particle is added to the theory.

Not being renormalizable does not make the theory useless. Einstein gravity could be a low energy effective theory for some unknown underlying theory. This can be compared to the situation for pions. One can write down an effective theory for pions, but this theory is not renormalizable. In this case the underlying theory is QCD, which is renormalizable. At the mass of the $\rho$ particle the effective pion theory is not valid anymore and new physics arises.

In path integral quantization, one tries to calculate the Euclidean path integral over metrics, which formally looks like

$$Z = \int \mathcal{D}g_{\mu\nu} \exp(-S[g_{\mu\nu}]). \qquad (1.6)$$

The main problems in this case are that it is not clear how to define the measure $\mathcal{D}g_{\mu\nu}$ and that the action $S[g_{\mu\nu}]$ is not bounded from below, which (dependent on the definition of the measure) might lead to a diverging integral.

Many alternatives have been created. All of them have their attractions and problems. I will briefly mention some of them. This is not meant to imply anything about those I have left out.

In what is called $R^2$ gravity [Stelle 1977], one adds a term proportional to $R^2$ to the action. This makes the action bounded from below, making the path integral converge, assuming the measure can be defined. In perturbation theory, one can use the $R^2$ term to modify the propagator in such a way that the Feynman integrals become convergent. This theory, however, appears to be non-unitary [Stelle 1977, Johnston 1988].

The "Ashtekar variables" [Ashtekar 1986, Ashtekar 1987] result from a way to cast the gravity action in Hamiltonian form, which is then canonically quantized. This results in a large number of constraints, a solution of which is given by the loop representation, where states are represented as functionals on sets of closed loops. This is a non-perturbative approach.

Supergravity (see [Nieuwenhuizen 1981] for a review) adds to all fields a partner field with opposite statistics. To the graviton would correspond a spin 3/2 particle called the gravitino. The hope, which has now diminished somewhat, was that all the divergences would exactly cancel those of the partner fields.

A more radical approach are the famous strings and their supersymmetric counterparts, the superstrings (see [Green *et al.* 1987] for one of the many books on this subject). In string theory, the fundamental objects are not particles, but one-dimensional extended objects called strings. This allows the theory to be finite. The particles as we know them are excitations of the string. One of the important attractions of string theory is that gravity comes out of it in a natural way as the interaction of a spin 2 excitation on the string which is identified with the graviton.

Last but not least there is simplicial gravity, which is what this thesis is all about. It is a non-perturbative lattice formulation of quantum gravity. Several theories have been developed where spacetime itself is discrete in some way (e.g. ['t Hooft 1988, Garay 1995]), but we will take a more traditional point of view and consider the lattice only as a regularization, which eventually has to be removed.

Before explaining what simplicial gravity is in section 1.4, I will first review some basic information about lattice theories.

## 1.3  Lattice field theory

As an example of lattice field theory we will consider a system with a single scalar field $\phi(x)$ and the action

$$S = \int d^4x \left[ \frac{1}{2} \partial_\mu \phi(x) \partial_\mu \phi(x) + \frac{m_0^2}{2} \phi(x)^2 + \lambda_0 \phi(x)^4 \right], \qquad (1.7)$$

where $m_0$ is the bare mass and $\lambda_0$ the bare self-coupling. This system is commonly known as "$\lambda \phi^4$ theory". We can put this model on a four-dimensional hypercubic lattice with lattice spacing $a$. We can then discretize the action by replacing derivatives with differences and integrals with sums over points times the volume of a lattice hypercube. This gives us the discretized action

$$S = a^4 \sum_{x,\mu} \frac{1}{2} \left( \frac{\phi_{x+\mu a} - \phi_x}{a} \right)^2 + a^4 \sum_x \left( \frac{m_0^2}{2} \phi_x^2 + \lambda_0 \phi_x^4 \right). \qquad (1.8)$$

The bare parameters in the action have a priori little to do with the energy of the excited states of the system. The energy gap of the first excited state of the system is interpreted as the mass of a single particle. This mass is the renormalized mass $m_R$. The connected two-point function of the scalar field behaves for large distances like

$$\langle \phi(x)\phi(y) \rangle_c \propto \exp \left( \frac{|x - y|}{\xi a} \right). \qquad (1.9)$$

The correlation length $\xi$ (which is a dimensionless number of lattice spacings) is related to the renormalized mass $m_R$ as

$$\xi = \frac{1}{m_R a}. \qquad (1.10)$$

Similarly, we can define a renormalized coupling $\lambda_R$ from the connected four-point function. We see from (1.10) that in the continuum limit, where the lattice spacing $a$ goes to zero, the correlation length at fixed renormalized mass $m_R$ needs to go to infinity. This means that the system must become critical. If the system shows no critical behaviour, we cannot define a physical continuum limit. As the system becomes critical, the short distance (lattice) details become irrelevant and the behaviour of the system only depends on a few parameters. The large scale behaviour does not change if, e.g., we start with a different lattice or discretize the action in another way. This phenomenon is known as universality.

Figure 1.1. Qualitative phase diagram of lattice $\lambda\phi^4$ theory.

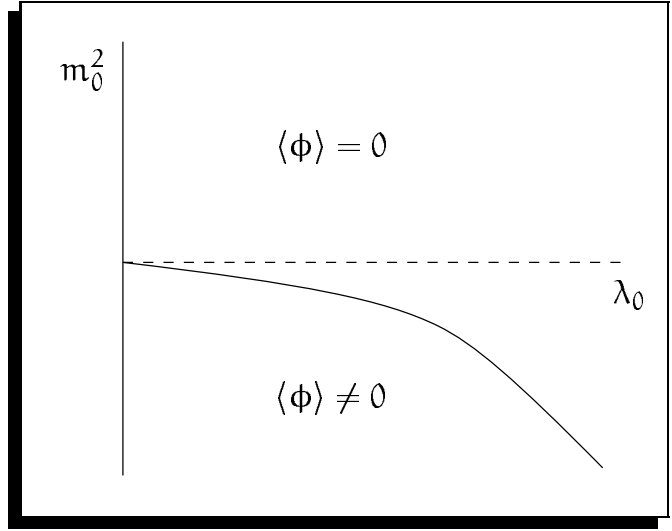

The model we are considering does indeed have a critical line, which is the curved line sketched in figure 1.1. Above this line the system is in a symmetric phase with $\langle\phi\rangle = 0$ and below the line in a spontaneously broken phase with $\langle\phi\rangle \neq 0$.

In this particular model, if we take the continuum limit the renormalized coupling vanishes, no matter where or how we approach the critical line. The theory becomes a free theory. This phenomenon is known as triviality. We can, however, still get a good interacting theory by not taking the continuum limit, but only approaching it closely. The coupling $\lambda_R$ only goes to zero as $(\ln a)^{-1}$. In renormalization group language it is a marginal operator. Other lattice effects, corresponding to irrelevant operators, go to zero as $a^2$. This means that by taking the lattice distance $a$ sufficiently small, we get proper continuum behaviour while still having the freedom to choose $\lambda_R$ within some bounds.

## 1.4   Simplicial gravity

Simplicial gravity is an attempt to define quantum gravity as a path integral over metrics by discretizing spacetime (henceforth often called space). Usually this is done in terms of simplices, but this is not necessary. A simplex in d dimensions is a solid object (a polytope) with d + 1 vertices which are pairwise connected by edges. It is in some sense the simplest d dimensional object. In two dimensions it is a triangle and in three dimensions a tetrahedron. Each d-simplex has d + 1 faces. See for an illustration figure C.5 on page 142.

We can glue a number of simplices together at the faces, to get a simplicial manifold, a space which is piecewise flat. This idea originated with Regge [Regge 1961] for classical general relativity. The idea was to approximate a smooth manifold, although in the quantum gravity version we do not think of a particular simplicial space as approximating any particular smooth space.

Simplicial quantum gravity attempts to define the path integral over metrics by a suitable integral or sum over simplicial manifolds. There are two main trends within this scheme, which are commonly known as Regge calculus and dynamical triangulation.

In Regge calculus one takes a simplicial complex and considers the integral over all edge lengths, keeping the connections between the simplices fixed. Similar to the continuum case, one of the main problems is what measure to use in this integral. A review of work in this direction can be found in [Williams & Tuckey 1992]. Recently there have been cast some doubts on this method [Holm & Janke 1994, Bock & Vink 1995], at least for the measures normally used.

A similar formulation [Ponzano & Regge 1968] specific for three dimensions labels the edges of a fixed simplicial complex with representations of the rotation group. The partition sum then consists of a sum over labelings of edges of a product of the 6j-symbols formed by the six edges of each simplex. This method has been extended to four dimensions using 12j-symbols in [Carfora *et al.* 1993].

In dynamical triangulation [Weingarten 1982, David 1985, Kazakov *et al.* 1985, Agishtein & Migdal 1992a, Ambjørn & Jurkiewicz 1992], the method used throughout this thesis, one takes equilateral simplices of the same size, and considers the sum over all possible ways to connect the simplices. One could conceive of combining the approaches and taking both the sum over connections and the integral over edge lengths. An approach that does this was put forward in [Shamir 1994].

In almost all cases, one only considers sums over simplicial complexes with the same topology. This is not because this is believed to be the right thing to do physically, but because such sums are easier to define and to simulate on the computer. In chapter seven I will briefly discuss the generalization to sums over topologies.

Dynamical triangulation has several nice features. It is well-defined, at least at the regularized level (see section 1.6). It is non-perturbative. It allows changes in the topology of spatial sections, i.e. spacetime does not have to be of the form $\mathcal{M}^3 \times \mathbb{R}$, and might be extendible to fluctuating spacetime topologies. It has the potential to give actual predictions through computer simulations.

Figure 1.2. A configuration with two edges between the same pair of vertices.

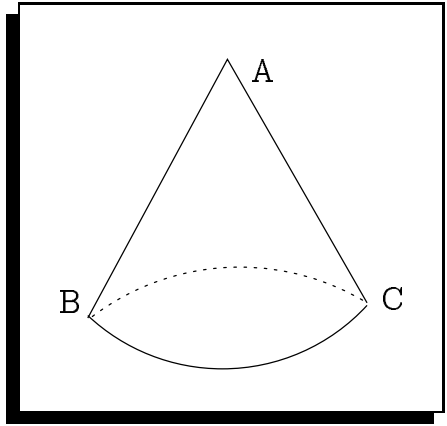

## 1.5   Discretization of gravity

The partition function of dynamical triangulation can be written as a sum over triangulations $\mathcal{T}$. These should not be considered as triangulations of smooth manifolds, but just as ways to glue equilateral simplices together such that the resulting space has the topology $S^4$.

$$Z = \sum_{\mathcal{T}} \exp(-S[\mathcal{T}]). \qquad (1.11)$$

We only allow those glueings where no two simplices of the same dimension have the same set of vertices. E.g. there cannot be two distinct edges between one pair of vertices. For an example of such a configuration in two dimensions, see figure 1.2. Two triangles have been glued along the sides AB and AC, creating a small cone. There are now two distinct edges between the vertices B and C. Although configurations violating this constraint are not simplicial complexes in the mathematical sense, there does not seem to be a physical argument in favour of or against allowing them. In the continuum limit it should not matter whether we include those configurations or not, but this has not been tested in four dimensions. The reason we do make this restriction is mainly for numerical convenience.

Figure 1.3. Two configurations with different orders of symmetry.

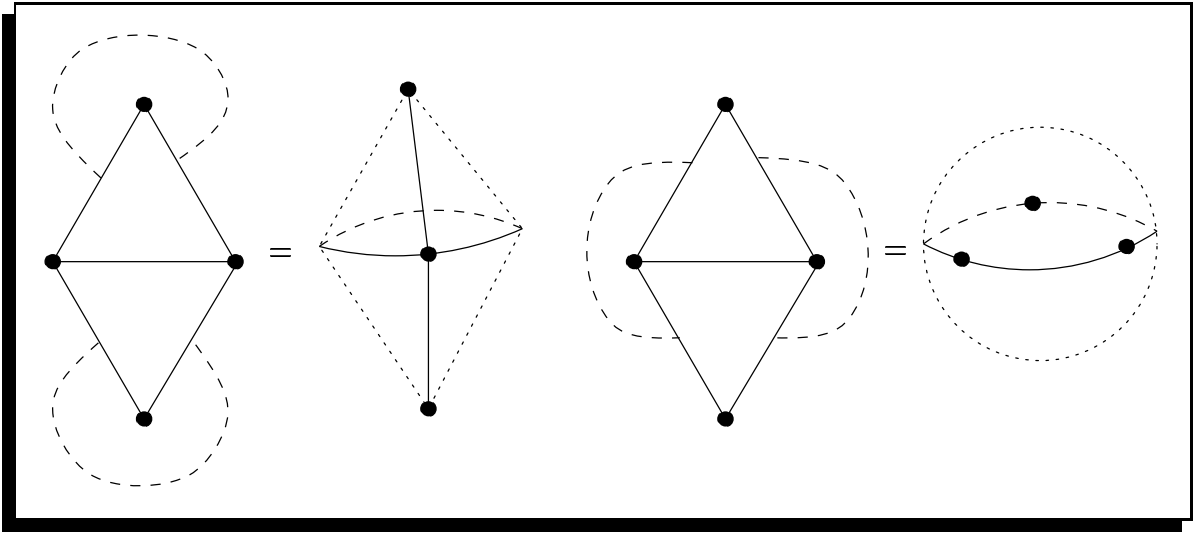

There is a natural measure on the space of triangulations: use each triangulation just once. There is some ambiguity in the word "once", i.e. in what triangulations are considered the same. In practice, symmetrical configurations are counted less often by a factor of the order of their symmetry. Let us again illustrate this in two dimensions. To simplify the example we temporarily drop the condition explained in the previous paragraph. There are two ways to make a two-sphere out of two triangles. I have drawn these in figure 1.3. The symmetry group of the left hand one has only two elements, the identity and turning it upside down. The right hand one can also be rotated over 120 or 240 degrees and its symmetry group (which is $D_3$) has six elements. In the simulations the first one will be counted three times as often as the second one. Considering that for a large number of simplices only a very small fraction of the triangulations is symmetrical this is not expected to make a difference. Anyway, the current simulation practice is identical to the situation in the matrix models explained in section 1.7.

The next question is what the action $S[\mathcal{T}]$ should be. To find the answer we need to find the discrete version of the Einstein-Hilbert action for the continuum (1.5). The term with the cosmological constant $\Lambda$ is easy, as it is proportional to the total volume. In the discrete system the volume is simply the number of simplices times the volume of a single simplex. We can therefore substitute

$$\int \sqrt{g} d^4 x \implies N_4 V_4, \tag{1.12}$$

where $N_i$ is the number of $i$ dimensional simplices and $V_i$ is the $i$-volume of one such simplex. If the edge length equals $\ell$, the volumes are

$$V_i = \frac{\ell^i \sqrt{(i+1)}}{i! \sqrt{2^i}}. \tag{1.13}$$

We now need to know what the scalar curvature is in the discrete system. We cannot assign a finite curvature to any point of the complex, because the simplices are flat inside and the curvature is concentrated in a $\delta$-function like way. But we can determine the integrated curvature. Curvature is defined from the rotation of a vector which is parallel transported around a small closed loop. For a small closed loop not to be flat, it needs to go around a $(d-2)$-dimensional object, in our case a triangle. The rotation of a vector will be proportional to the deficit angle around this triangle, that is the angle that is missing from $2\pi$. Also, the curvature integrated over a region around such a triangle will be proportional to the volume of a triangle $V_2$. The result is that this integrated curvature around a single triangle is given by

$$\int_\triangle R\sqrt{g}d^4x = 2V_2(2\pi - \theta n_\triangle), \tag{1.14}$$

where $n_\triangle$ is the number of simplices around a triangle and $\theta$ is the angle between two faces of a four simplex, which is

$$\theta = \arccos\left(\frac{1}{4}\right) \approx 1.318. \tag{1.15}$$

I have given only a plausibility argument and not a real derivation as this would be quite long. See for details [Cheeger *et al.* 1984, Friedberg & Lee 1984]. Summing the relation (1.14) over all triangles in the simplicial complex tells us what to substitute for the total curvature in the discrete system.

$$\int R\sqrt{g}d^4x \Longmapsto 4V_2(\pi N_2 - 5\theta N_4), \tag{1.16}$$

where we used the fact that every four-simplex contains 10 triangles and therefore $\sum n_\triangle = 10N_4$. Performing the substitutions (1.12) and (1.16) in the action (1.5) gives us the discrete action

$$S[\mathcal{T}] = \frac{1}{16\pi G}\left(2V_4\Lambda N_4 - 4\pi V_2 N_2 + 20\theta V_2 N_4\right), \tag{1.17}$$

which is usually written as

$$S[\mathcal{T}] = \kappa_4 N_4 - \kappa_2 N_2, \tag{1.18}$$

where

$$\kappa_2 = \frac{V_2}{4G},$$ (1.19)

and

$$\kappa_4 = \frac{V_4\Lambda + 10\theta V_2}{8\pi G}.$$ (1.20)

This action is known as the Regge-Einstein action, specialized to equilateral simplices.

Due to conditions on the glueing of the simplices there are several relations between the $N_i$, making them dependent. First of all, each simplex has five faces which are glued pairwise, giving the relation

$$5N_4 = 2N_3.$$ (1.21)

Second, we want the triangulations to have a fixed topology and therefore a fixed Euler number $\chi$, which can be expressed in the $N_i$ as

$$N_0 - N_1 + N_2 - N_3 + N_4 = \chi.$$ (1.22)

For the hypersphere $S^4$ that we usually consider we have $\chi = 2$.

The third condition is known as the manifold condition. We cannot just glue together simplices at the faces and expect the result to be a simplicial manifold. We have to impose the restriction that the neighbourhood of each point is homeomorphic to a (simplicial) ball. To see how this could be violated, look at figure 1.4(a) on the following page for an illustration in three dimensions. Two tetrahedrons are glued together. Four outer sides come together in the upper vertex. Now take these four faces and glue together each pair of opposite ones, in both cases such that the upper vertex is glued to itself. A neighbourhood of this vertex will not be a ball, but something with a two-dimensional torus as its boundary. This is illustrated in figure 1.4(b), which depicts such a boundary. It is the intersection of figure 1.4(a) with a horizontal plane just below the upper vertex.

Consider now in four dimensions the set of points which have a fixed small (compared to the lattice spacing) distance from a particular vertex of the triangulation. Because the intersection of this set with each four-simplex surrounding this vertex is equivalent to a three-simplex, this set of points will be a three-dimensional

Figure 1.4. Simplices glued together but not creating a simplicial manifold.

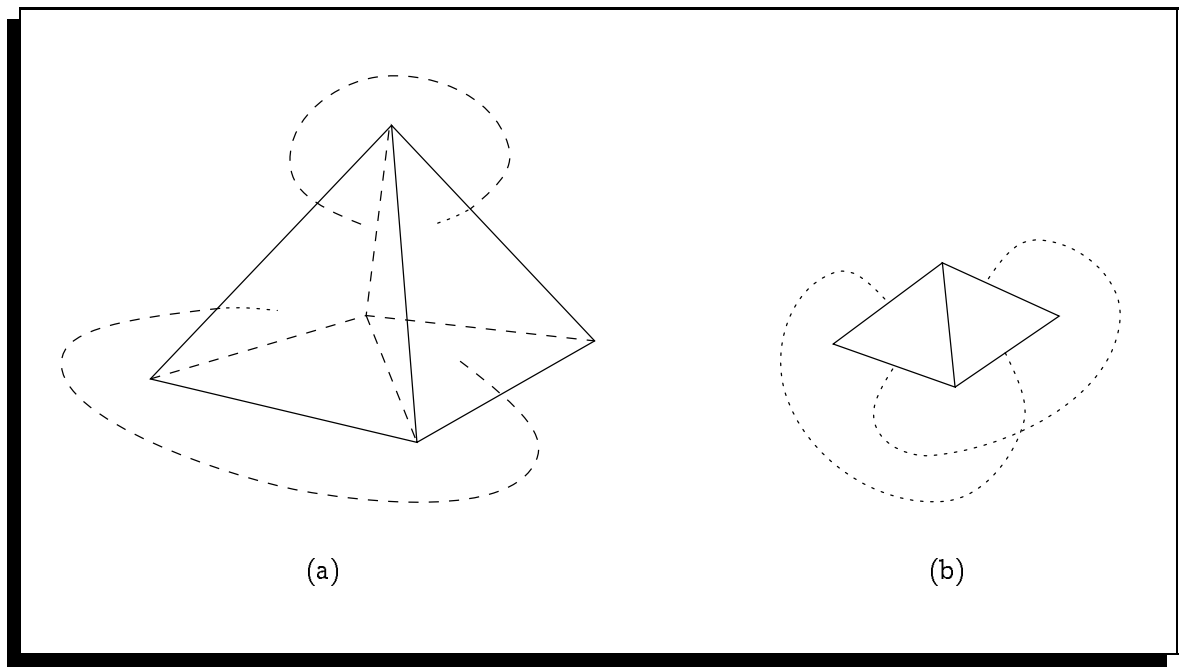

(a)                       (b)

simplicial complex. For this three-dimensional complex with numbers of simplices $M_i$ we demand

$$M_0 - M_1 + M_2 - M_3 = 0, \tag{1.23}$$

because the Euler number of an odd-dimensional manifold is zero. We can now sum this relation over all vertices of the four-dimensional complex, taking into account that a $d$ simplex of the three-manifold is part of a $d + 1$ simplex of the four-manifold and that each such $d + 1$ simplex joins $d + 2$ vertices. This results in

$$2N_1 - 3N_2 + 4N_3 - 5N_4 = 0. \tag{1.24}$$

Note that this equation is a necessary, but not a sufficient condition for the simplicial complex to be a simplicial manifold. As I will explain on page 99 in chapter seven, a manifold which satisfies (1.24) does satisfy (1.23) at each vertex. But because the Euler number of any three dimensional manifold is zero, the neighbourhood of a vertex that satisfies (1.23) does not have to be a simplicial ball.

The equations (1.21), (1.22) and (1.24) are three independent relations between five numbers $N_i$, $i = 0 \ldots 4$. Therefore, only two of the $N_i$ are independent, which we can choose to be $N_2$ and $N_4$. This shows that the action (1.18) is the most general action linear in the $N_i$. We can also impose a manifold condition on the two-spheres that we can draw around an edge of the complex, but the relation we get from this is dependent on the other three.

These manifold conditions can easily be extended to other dimensions. In d dimensions they become

$$N_i = \sum_{j=i}^{d} (-1)^{d-j} \binom{j+1}{i+1} N_j, \tag{1.25}$$

which are known as the Dehn-Sommerville relations. For fixed topology there is also the relation that fixes the Euler number $\chi$

$$\chi = \sum_{i=0}^{d} (-1)^i N_i. \tag{1.26}$$

For odd number of dimensions d, the Euler number $\chi$ is always zero. In that case, this relation is not independent from (1.25), but follows from them by taking the sum over $i$ with an additional factor $(-1)^i$, after which the right hand side of the sum can be shown to vanish.

## 1.6   Is this well-defined?

The question is now whether the theory is well-defined. This question has two parts. The first part is whether the partition function is well-defined in the discrete system. The second part is whether the theory can give any continuum predictions.

The first part is the easiest. The question is whether the partition sum

$$Z = \sum_{\mathcal{T}} \exp(\kappa_2 N_2 - \kappa_4 N_4) \tag{1.27}$$

converges for some values of the coupling constants. It does if the number of triangulations is bounded by an exponential function of $N_4$. This has been investigated both analytically and in numerical simulations. The result is that it does indeed converge [Ambjørn & Jurkiewicz 1994, Brügmann & Marinari 1995, Bartocci *et al.* 1994]. I will discuss this extensively in chapter two.

The hard part of the question is whether the theory can give continuum predictions, in other words, whether the theory shows scaling. This means that for small lattice spacings the behaviour of the system becomes independent of the lattice spacing. Some encouraging results are presented in chapter three.

One might think that a theory which is perturbatively non-renormalizable will not have a sensible continuum limit. This does not need to be true. The canonical counterexample is the four-dimensional $O(4)$ non-linear sigma model. It is a theory with a four component scalar field $\phi^a$ that satisfies the condition $\phi^a\phi^a = 1$. In the continuum the action is

$$S = f^2 \int d^4x \partial_\mu \phi^a \partial_\mu \phi^a \qquad (1.28)$$

This theory also has a coupling constant with a dimension, just like gravity. For the same reasons as in gravity it is not renormalizable in perturbation theory. As a non-perturbative lattice model, however, it has a well-defined continuum limit [Lüsher & Weisz 1988, Lüsher & Weisz 1989, Heller 1994].

One might also think that because the continuum action is not bounded from below, the model will not make sense, because the configurations with the largest curvature will dominate the partition function. Those configurations, however, may have very low entropy. In continuum language: the measure may behave in such a way as to make the path integral converge, despite the action being unbounded from below. This is indeed what happens in the simulations. Except for large values of $\kappa_2$, the typical configurations are not those with very high curvature. Similar results are reported in Regge calculus [Berg 1986]. The effects of the measure on the semiclassical fluctuations have also been investigated in the continuum [Mottola 1995].

Although the simulations seem to indicate that the model is alright, this does not imply that it actually describes gravity. In lattice field theory there is no guarantee that the model that comes out in the continuum limit is the same model as the one that we started discretizing. We will discuss this further in chapter nine.

## 1.7 Two dimensions

An important indication that this simplex stuff might have something to do with gravity comes from work in two dimensions. In two dimensions many results have been found analytically, both in the continuum theory [Polyakov 1981, Knizhnik *et al.* 1988] and in dynamical triangulation. The latter through what are called matrix models.

In its simplest form, a matrix model is a model of a matrix field in zero dimensions. In other words, a single matrix $M_{ab}$ of dimension $N_m$. This matrix is

Figure 1.5. Feynman diagram of a matrix model.

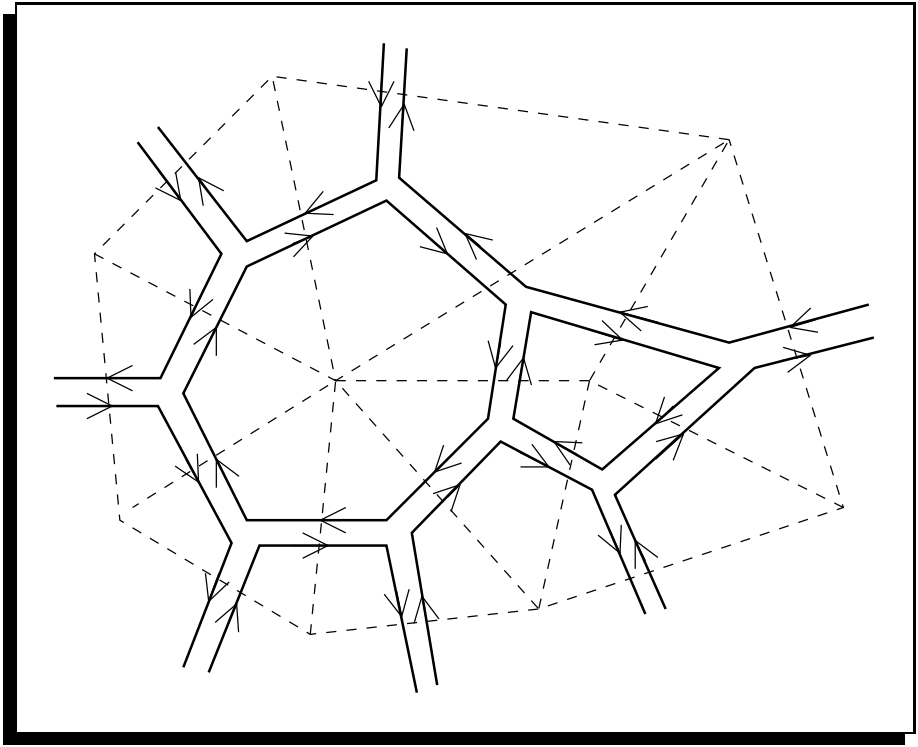

usually taken to be hermitian. The partition function of this model is [Brézin *et al.* 1978]

$$Z = \int dM_{ab} \exp(-S(M_{ab})), \qquad (1.29)$$

with the action

$$S(M_{ab}) = \frac{1}{2} M_{ab} M_{ab}^* + \frac{g}{\sqrt{N_m}} M_{ab} M_{bc} M_{ca}, \qquad (1.30)$$

where $g$ is a coupling constant. Due to the cubic term in the action, this partition function is not well defined for real $g$. We can however still define it by its perturbation expansion. If we expand this partition function in powers of $g$ we generate Feynman diagrams like the one shown in figure 1.5 with solid lines. The propagators carry two matrix indices and are therefore drawn as double lines. The two point function is

$$\langle M_{ab} M_{cd} \rangle = \delta_{ad} \delta_{bc}, \qquad (1.31)$$

which translates into the connection of the double lines from two vertices in such a way that an outgoing arrow from one vertex matches up with an ingoing one from the other vertex and vice versa.

The dual of such a Feynman diagram, shown in the figure using dashed lines, is precisely a triangulation of a two-dimensional orientable surface. The contribution of a diagram is

$$\left(\frac{g}{\sqrt{N_m}}\right)^{N_2} N_m^{N_0} = g^{N_2} N_m^\chi, \tag{1.32}$$

where $N_0$ and $N_2$ are the numbers of vertices and triangles on the dual lattice, corresponding to loops and vertices of the Feynman digram. The parameter $\chi$ is the Euler number of the two dimensional surface. Taking the limit $N_m \to \infty$ keeps only the configurations with highest $\chi$, which are those that have the topology of the two-sphere $S^2$. The partition function (1.29) becomes equal to the sum over all triangulated surfaces, weighted by the Boltzmann weight with the Regge-Einstein action. This sum includes disconnected diagrams, corresponding to disconnected surfaces. Taking the logarithm of (1.29) results in the usual way in a sum over only connected diagrams. This sum is exactly the partition function of two-dimensional dynamical triangulation with cosmological constant $\kappa_2 = \ln(g)$.

All the results which exist in both the continuum and the matrix model approach coincide. This is not only true for pure gravity, but also for gravity coupled to conformal matter. In particular we have what is called KPZ-scaling [Knizhnik *et al.* 1988]. In dynamical triangulation this means that the partition function at fixed number of triangles $N_2$ behaves like

$$Z \sim N_2^{\gamma-3} \exp(\kappa_2^c N_2), \tag{1.33}$$

where the string susceptibility $\gamma$ depends on the central charge $c$ of the matter and the Euler number $\chi$ of the surface as

$$\gamma = \frac{c - 25 - \sqrt{(c-1)(c-25)}}{24}\chi + 2. \tag{1.34}$$

The constant $\kappa_2^c$ in (1.33) is not universal and depends for instance on whether we use triangles or squares in the model.

It has been suggested that the model works better in two than in four dimensions, because in two dimensions one can build flat space and local deviations from it out of triangles. In four dimensions one cannot fill flat space with equilateral simplices. This argument does not hold, because the two-dimensional model works equally well using pentagons, heptagons or other regular polygons [Kazakov 1989]. Also, one could use hypercubes instead of simplices in four dimensions [Weingarten 1982], although it has not been verified that this indeed gives the same results. Obviously, there can be other reasons why the model in two dimensions is different from the one in four dimensions and the fact that the model works in the former case is no guarantee for the latter one. It is often suggested that the topological nature of the two-dimensional action is such a difference. I do not think this is a very compelling argument. First, the two-dimensional model also works if we add matter, making the action no longer topological. Second, the difficult part is not the action, but the measure, and that comes out correctly in two dimensions.

## 1.8 Euclidean approach

As was mentioned in section 1.1, what we are discussing is Euclidean quantum gravity, as opposed to Lorentzian quantum gravity. This means that we consider metrics with signature $(1, 1, 1, 1)$ instead of $(-1, 1, 1, 1)$. The main reason to do so is that we do not know how to simulate Lorentzian gravity. Although it might be possible to define a discretization by glueing pieces of Minkowski space together, we cannot simulate the partition sum $\sum \exp(iS)$, because we cannot do Monte Carlo simulations (see section 1.9) with complex probabilities.

In lattice gauge theory the Minkowskian and Euclidean formulations are related by a Wick rotation where time is replaced by imaginary time. In quantum gravity the situation is far from clear. We cannot simply analytically continue Lorentzian metrics to Riemannian ones, but perhaps it is possible to analytically continue the Green functions in some way.

It has even been suggested [Hawking 1978b] that reality is described by a Euclidean path integral and that that is the formulation that others should be proven equivalent to instead of vice versa. What seems less implausible is that both signatures of the metric can exist and that one is chosen by some dynamical process. A mechanism by which the Lorentzian signature, as well as the dimension of spacetime, comes out was described in [Greensite 1993]. If this is the case, this mechanism is probably not described by the dynamical triangulation method used here.

We mentioned in section 1.6 that the unboundedness from below of the Euclidean action could be compensated by the measure. There is however still a problem with the effective action describing semiclassical fluctuations around the average spacetime we measure in the simulations. We will discuss this problem in chapter nine.

## 1.9   Monte Carlo simulations

Almost no analytic results are known for four dimensional dynamical triangulation. Most of our knowledge stems from computer simulations, using Monte Carlo methods. In these simulations we measure the expectation value of observables A,

$$\langle A \rangle = \frac{\sum_{\mathcal{T}} A \exp(-S[\mathcal{T}])}{\sum_{\mathcal{T}} \exp(-S[\mathcal{T}])}, \tag{1.35}$$

by generating a suitably weighted (i.e. with their relative probabilities proportional to $\exp(-S[\mathcal{T}])$) set of triangulations. In Monte Carlo jargon, these are usually referred to as configurations. We then calculate (or "measure") the value of the observable A on each triangulation. The average is an approximation for $\langle A \rangle$ that becomes better as we take more configurations.

The sets of configurations are generated by starting with an arbitrary configuration and generating new ones by performing local changes in the configuration. These changes are called moves. If the probabilities which are used to choose a particular move to perform satisfy a (sufficient, but not necessary) condition called detailed balance, the probability distribution of the configurations will converge to the desired distribution, where the probabilities are proportional to $\exp(-S[\mathcal{T}])$. This process of convergence is called thermalization.

When the configuration has thermalized, we can start to perform the measurements. Two configurations that are only one move apart, however, will usually show almost the same value of the observable. In other words, the measurements are highly correlated. To get a good estimation of the expectation value of the observable and the statistical error we make we need to be able to produce uncorrelated configurations. The number of moves that is needed between two configurations to make them uncorrelated is called the autocorrelation time. To be more precise, the correlation between two observables behaves like

$$\langle O(t)O(t') \rangle - \langle O \rangle^2 \propto \exp(-|t - t'|/\tau), \tag{1.36}$$

where $\tau$ is the autocorrelation time. This time can depend very much on the observable that is used.

One of the major practical problems in these simulations is a phenomenon called critical slowing down. This means that the autocorrelation time defined above becomes very large when the system becomes critical. As was explained in section 1.3, this is just where we want to study the system. Critical slowing down can be explained intuitively as follows. If the Monte Carlo moves are local, any change in the system will propagate through the system like a random walk. Before the system can become uncorrelated from its previous state, this change will need to propagate for a distance of the order of the correlation length $\xi$. Because the distance travelled by a random walk goes like the square root of its length, the autocorrelation time of the system will behave like $\xi^2$, which, by definition, becomes very large for a critical system.

Ways to get around this problem depend very much on the system under consideration. Most of these ways consist of defining Monte Carlo moves which make non-local changes in the system. For the particular case of dynamical triangulation, I will briefly discuss this on page 109 in chapter eight.

## 1.10  Outline

The reader who is interested in the generation of the triangulations on the computer can find the details in chapter eight. Most of the other chapters deal with the results of the various measurements done on these triangulations and do not require knowledge of these methods.

Chapter two describes the region of parameter space where the model is defined and the two phases that occur in the model. It discusses the distinguishing features of those phases and the nature of the transition between them.

Chapter three describes the spaces generated in the model by defining a curvature and an effective (global) dimension at scales large compared to the lattice spacing. Both features indicate that at the transition the space resembles the hypersphere $S^4$. This chapter also presents evidence for scaling, i.e. the idea that as we go to small lattice spacings the features of the model become independent of the lattice spacing.

In chapter four we measure two-point correlation functions of the scalar curvature and something we call the local volume. Peculiar effects occur because both the local observables and the distance between them depend on the geometry. It turns out that at the transition and in the elongated phase the correlations have a long range, i.e. they fall off like a power of the distance.

Chapter five goes into the measurement of gravitational attraction and binding of particles. It presents a possible way of defining the renormalized gravitational constant $G_R$ and thereby measuring the Planck length $\ell_P = \sqrt{G_R}$ in lattice units.

An interesting feature of the dynamical triangulation model is that the results are non-computable, in the recursion theoretical sense of the word. This could create problems for the numerical simulations. Chapter six tries to quantify the effects of this feature, which seem to be negligible.

A model that does not fix the topology of the four dimensional space is briefly considered in chapter seven. It is equivalent to a tensor model generalization of the matrix model described in section 1.7.

Finally, chapter nine gives a general discussion of the current status of dynamical triangulation as a model for four-dimensional quantum gravity.

# Chapter Two

# Phases

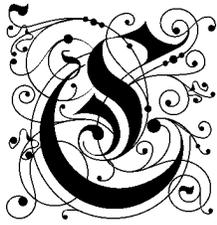 CRITICAL behaviour of a lattice system is necessary to define a good continuum limit. Such behaviour is generally found at a second order phase transition. This chapter describes the phase structure of the dynamical triangulation model of four dimensional quantum gravity. Numerical simulations show that the model has two phases with very different properties. We will usually just call these phases, but in chapter three we will discuss the possibility that the system has an entire critical line and that therefore these phases are not phases in the usual sense.

## 2.1  Partition function

As has been explained in chapter one, the partition function of four dimensional dynamical triangulation without matter is

$$Z(\kappa_4, \kappa_2) = \sum_{\mathcal{T}} \exp(\kappa_2 N_2 - \kappa_4 N_4). \tag{2.1}$$

The sum is over all the ways $\mathcal{T}$ to glue together four-simplices such that the resulting simplicial complex is a simplicial manifold with the topology of the four-sphere $S^4$ and $N_i$ is the number of i-simplices in the resulting complex. To facilitate the discussion below, we define a canonical partition function $Z(N_4, \kappa_2)$ at fixed volume $N_4$.

$$Z(N_4, \kappa_2) = \sum_{\mathcal{T}(N_4)} \exp(\kappa_2 N_2), \tag{2.2}$$

which we can use to rewrite (2.1) as

$$Z(\kappa_4, \kappa_2) = \sum_{N_4} Z(N_4, \kappa_2) \exp(-\kappa_4 N_4). \tag{2.3}$$

The question is now whether the sum in (2.1) converges for some values of the coupling constants. We see in (2.3) that this will be the case if the canonical partition function has an exponential bound

$$Z(N_4, \kappa_2) \leqslant \exp(\kappa_4^c(\kappa_2) N_4), \tag{2.4}$$

for some critical $\kappa_4^c(\kappa_2)$. This will make the grand canonical partition function converge for $\kappa_4 > \kappa_4^c(\kappa_2)$ and perhaps if $\kappa_4 = \kappa_4^c(\kappa_2)$. Because the $N_4$ simplices have only $5N_4$ faces to connect in pairs and each connection can be made in at most 12 ways, $Z(N_4, \kappa_2)$ certainly has the factorial bound[*] $(5N_4 - 1)!! \times 12^{5N_4/2}$.

The existence of an exponential bound for the canonical partition function (2.2) does not depend on $\kappa_2$. This can be shown as follows. Because each simplex has 10 triangles and each triangle is shared by at least 3 simplices, the number of triangles is bounded as $N_2 \leqslant 10N_4/3$. Therefore, if $\kappa_2 \geqslant 0$

$$Z(N_4, \kappa_2 \geqslant 0) = \sum_{\mathcal{T}(N_4)} \exp(\kappa_2 N_2)$$
$$\leqslant \sum_{\mathcal{T}(N_4)} \exp(10\kappa_2 N_4/3) = Z(N_4, \kappa_2 = 0) \exp(10\kappa_2 N_4/3), \quad (2.5)$$

and if $\kappa_2 \leqslant 0$

$$Z(N_4, \kappa_2 \leqslant 0) = \sum_{\mathcal{T}(N_4)} \exp(\kappa_2 N_2) \leqslant \sum_{\mathcal{T}(N_4)} = Z(N_4, \kappa_2 = 0). \tag{2.6}$$

Or, in words, if the exponential bound exists for $\kappa_2 = 0$ (or any other specific value), it exists for all $\kappa_2$. To investigate the question further we define an $N_4$-dependent critical $\kappa_4$ as

$$\kappa_4^c(N_4, \kappa_2) = \frac{\partial \ln Z(N_4, \kappa_2)}{\partial N_4}. \tag{2.7}$$

If this object converges as $N_4 \to \infty$, its limit will be $\kappa_4^c(\kappa_2)$ and the exponential bound exists. If, on the other hand, it diverges, there is no exponential bound.

---

[*] $n!!$ stands for $n(n-2)(n-4)\ldots$.

To measure the value of this $\kappa_4^c(N_4, \kappa_2)$, we simulate on the computer the modified system

$$Z'(\kappa_4, \kappa_2, \gamma, V) = \sum_{\mathcal{T}} \exp(\kappa_2 N_2 - \kappa_4 N_4 - \gamma(N_4 - V)^2), \qquad (2.8)$$

$$= \sum_{N_4} Z(N_4, \kappa_2) \exp(-\kappa_4 N_4 - \gamma(N_4 - V)^2). \qquad (2.9)$$

The extra term $\gamma(N_4 - V)^2$ in the action certainly makes the sum converge (for positive $\gamma$), because the number of triangulations can never rise faster than factorially in $N_4$, which is slower than $\exp(\gamma N_4^2)$. A constant term in the action is irrelevant, which allows us to get rid of $\kappa_4$ by a suitable shift in $V$ or vice versa, but we will not do so.

We can now make a saddle point expansion of (2.9). Let us define $\overline{N_4}$ as the value of $N_4$ that gives the most important contribution to the sum (2.9). Using (2.7) we see that

$$\overline{N_4} = V + \frac{\kappa_4^c(\overline{N_4}, \kappa_2) - \kappa_4}{2\gamma}. \qquad (2.10)$$

The saddle point expansion of (2.9) around $\overline{N_4}$ shows that in this system $\langle N_4 \rangle = \overline{N_4} + O((\overline{N_4}\gamma)^{-1})$. Consequently, we can measure $\kappa_4^c(N_4, \kappa_2)$ using

$$\kappa_4^c(N_4, \kappa_2) \equiv \frac{\partial \ln Z(N_4, \kappa_2)}{\partial N_4}$$

$$= \kappa_4 + 2\gamma(\langle N_4 \rangle - V) + O\left(\frac{1}{\langle N_4 \rangle}\right). \qquad (2.11)$$

In figure 2.1 on the following page we have plotted the measured value of $\kappa_4^c(N_4, \kappa_2)$ as a function of the volume $\langle N_4 \rangle$ at $\kappa_2 = 0$, where all configurations of a particular volume contribute equally to the partition function. Within the errors, the data at $\kappa_2 = 2$ show just a horizontal line. This means that at large $\kappa_2$, we measure no dependence of $\kappa_4^c(N_4, \kappa_2)$ on $N_4$.

I fitted the data at $\kappa_2 = 0$ to a logarithm and a converging power. The logarithm corresponds to a diverging $\kappa_4^c(N_4, \kappa_2)$ and no exponential bound, while the power corresponds to the existence of the exponential bound. The results of the logarithmic fit were

$$\kappa_4^c(\langle N_4 \rangle, \kappa_2) = a + b \ln\langle N_4 \rangle, \qquad (2.12)$$

$$a = 0.816(10), \qquad (2.13)$$

Figure 2.1. The value of $\kappa_4^c$ as a function of the volume at $\kappa_2 = 0$.

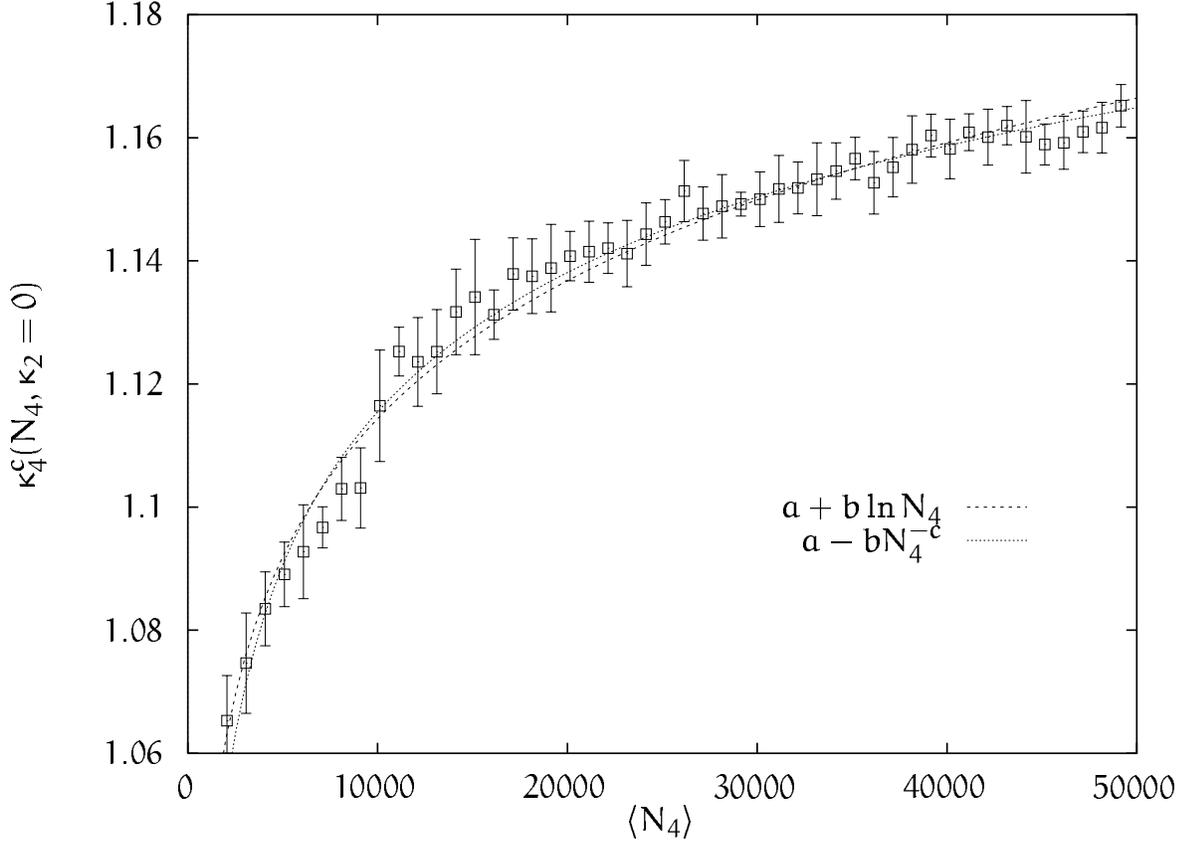

$$b = 0.0324(10), \qquad (2.14)$$

$$\chi^2 = 24 \text{ at } 46 \text{ d.o.f.}, \qquad (2.15)$$

and those of the power fit were

$$\kappa_4^c(\langle N_4 \rangle, \kappa_2) = a - b\langle N_4 \rangle^{-c}, \qquad (2.16)$$

$$a = 1.36(12), \qquad (2.17)$$

$$b = 0.88(14), \qquad (2.18)$$

$$c = 0.14(7), \qquad (2.19)$$

$$\chi^2 = 20 \text{ at } 45 \text{ d.o.f.} \qquad (2.20)$$

These results are not conclusive. The $\chi^2$ of the power law is somewhat lower, but considering that it has more parameters this is not surprising. Data that has been found by other groups [Ambjørn & Jurkiewicz 1994, Brügmann & Marinari 1995], with $N_4$ up to 128000, now quite strongly favors a $\kappa_4^c$ converging according to (2.16), which means that there is an exponential bound on the number of configurations and that the grand canonical partition function is well defined. A handwaving argument for an exponent of $c = 1/4$ was presented in [Ambjørn & Jurkiewicz 1994]. From the latest data one now finds an exponent of 0.36(4) [Brügmann & Marinari 1995].

Figure 2.2. The value of $\kappa_4^c$ as a function of $\kappa_2$ for various volumes.

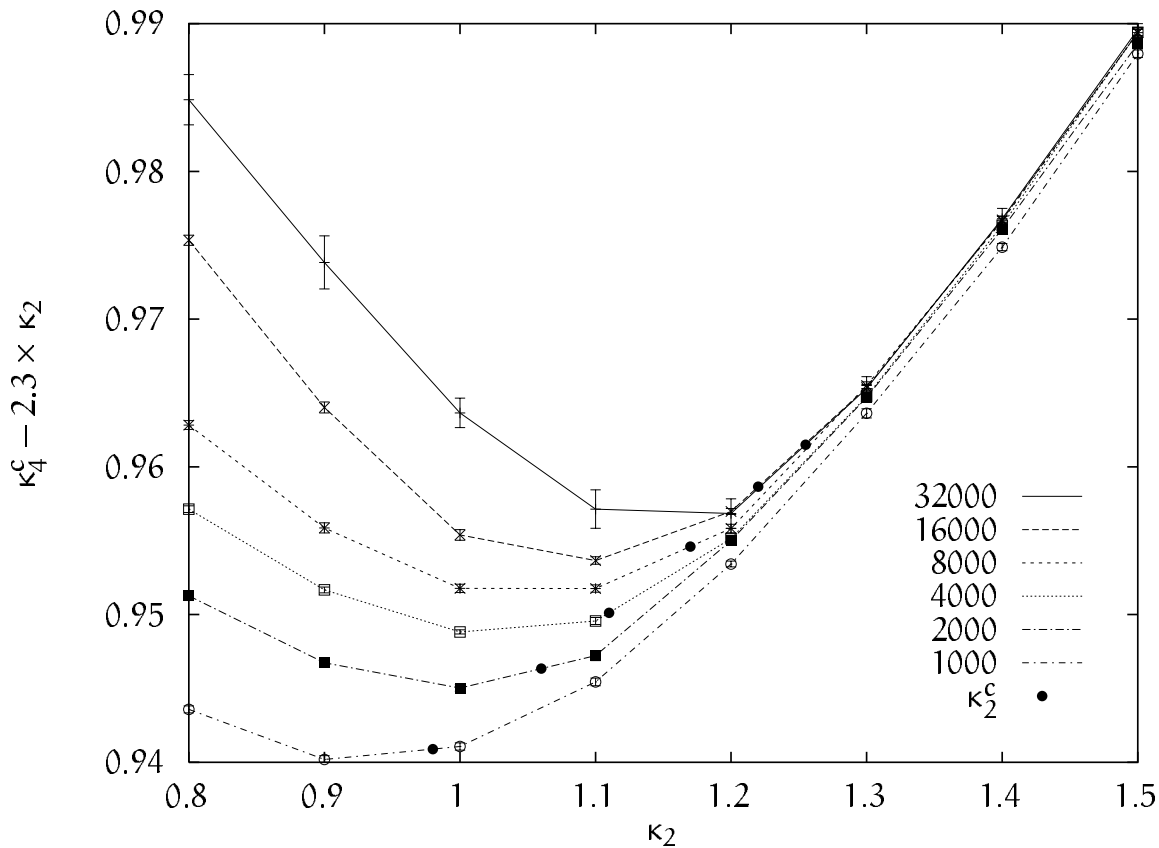

Recently there has also appeared an analytical proof that the number of configurations has an exponential bound for all dimensions and (fixed) topologies [Bartocci *et al.* 1994]. This proof, however, applies to "metric ball coverings". While some arguments can be made that this also provides a bound for our triangulations, this is not yet completely clear.

We will now take a closer look at the $\kappa_2$ dependence of $\kappa_4^c$. I have plotted its value at various fixed volumes in figure 2.2. To increase the vertical resolution, a linear function $2.3\kappa_2$ has been subtracted from the measured values. The black dots are the critical values of $\kappa_2$ discussed below in section 2.3. As we have already mentioned above, at large values of $\kappa_2$ there is virtually no dependence of $\kappa_4^c$ on the volume.

There is a close relation between the slopes in the figure and the average curvature. We see from equation (1.16) that up to an additive and a multiplicative constant the curvature per unit volume is simply $N_2/N_4$. We have plotted this quantity in figure 2.3. Its relation with the slopes of figure 2.2 on the page before is

$$\frac{\partial \langle N_2 \rangle}{\partial N_4} = \frac{\partial^2 \ln Z(N_4, \kappa_2)}{\partial N_4 \partial \kappa_2} = \frac{\partial \kappa_4^c(N_4, \kappa_2)}{\partial \kappa_2}. \tag{2.21}$$

Figure 2.3. The average curvature $\langle N_2/N_4 \rangle$ as a function of $\kappa_2$ for various volumes.

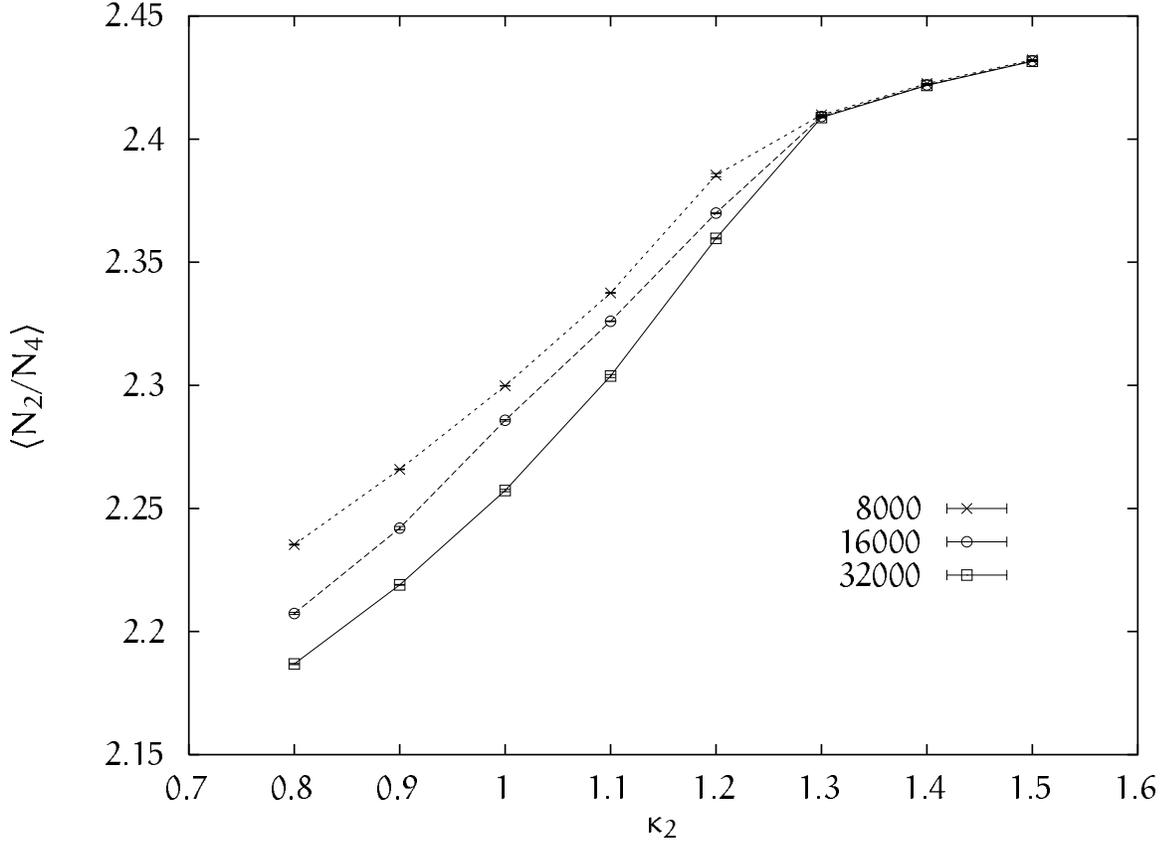

Using this relation, we can see that those slopes are bounded by deriving bounds on the number of triangles. It follows from the relations between the $N_i$ which were derived in chapter one that

$$N_0 = \frac{N_2}{2} - N_4 + \chi. \tag{2.22}$$

Therefore, as $N_0 \geqslant 6$ and $\chi = 2$, we have $N_2 \geqslant 2N_4 + 8$. On the other hand, as we have shown above, $N_2 \leqslant 10N_4/3$. In the simulations it turns out that in fact $N_2 \leqslant 5N_4/2 + O(1)$, but I do not know how to prove this from geometrical arguments. These bounds fit with the slopes of $\kappa_4^c(\kappa_2)$. In [Brügmann & Marinari

1995] a larger range of $\kappa_2$ values was used than in our simulations and the authors concluded that

$$\kappa_4^c(\kappa_2 \geqslant 4) = 2.497\kappa_2 + \text{constant}, \qquad (2.23)$$

$$\kappa_4^c(\kappa_2 \leqslant -4) = 2.002\kappa_2 + \text{constant}. \qquad (2.24)$$

Although the grand canonical partition function would not be well-defined if this exponential bound did not exist, this need not invalidate the model, as discussed in [de Bakker & Smit 1994]. One can also define the model using the canonical partition function and then take the number of simplices to infinity. Alternatively, one could try to make a suitable expansion in the inverse gravitational constant. In the case of two dimensions and varying topology the latter approach is known as the double scaling limit [Brézin & Kazakov 1990, Douglas & Shenker 1990, Gross & Migdal 1990].

## 2.2 Canonical partition function

Because the grand canonical ensemble (2.1) is very difficult to simulate, simulations are always done using the canonical ensemble (2.2). We cannot, however, directly simulate the canonical ensemble on the computer. The reason is that no set of usable moves is known that keeps the volume $N_4$ constant and is ergodic in the canonical ensemble. It seems probable that no such set of moves exists (see chapter six for more discussion on this topic).

To get around the problem explained above we can rewrite the canonical partition function (2.2) as

$$Z(N_4, \kappa_2) = \exp(\Delta S(N_4)) \sum_{\mathcal{T}} \exp(\kappa_2 N_2 - \Delta S(N_4[\mathcal{T}])) \, \delta_{N_4[\mathcal{T}], N_4}, \qquad (2.25)$$

where $N_4$ is fixed and $N_4[\mathcal{T}]$ means the number of simplices in the triangulation $\mathcal{T}$. This allows us to calculate observables in the canonical ensemble using grand canonical moves. The precise form of $\Delta S(N_4)$ is not important, as long as it makes the sum without the $\delta$-function factor convergent. In practice we use

$$\Delta S(N_4) = \kappa_4 N_4 + \gamma(N_4 - V)^2, \qquad (2.26)$$

where $\kappa_4$, $\gamma$ and $V$ are free parameters. We should now ideally only use the configurations with $N_4[\mathcal{T}] = N_4$ in the measurements. In practice, we don't do this, but use all the configurations. We can control the volume fluctuations by changing the parameter $\gamma$ according to

$$\langle N_4^2 \rangle - \langle N_4 \rangle^2 = \left( 2\gamma - \frac{\partial \kappa_4^c}{\partial N_4}(\overline{N_4}) \right)^{-1}. \tag{2.27}$$

This follows from the saddle point approximation of (2.8). Due to the dependence of $\overline{N_4}$ on $\gamma$ the right hand side always stays positive. In our simulations we used $\gamma = 5 \cdot 10^{-4}$. This makes $\partial \kappa_4^c / \partial N_4(\overline{N_4}) \ll 2\gamma$. Thus, the volume fluctuations were approximately 30.

It is often argued that using the canonical ensemble for large volumes approximates taking the limit $\kappa_4 \to \kappa_4^c$, because the volume diverges for $\kappa_4 > \kappa_4^c$. Unfortunately, this is not at all clear. Let us write the canonical partition function (2.2) as

$$Z(N_4, \kappa_2) = f(N_4, \kappa_2) \exp(\kappa_4^c(\kappa_2)N_4). \tag{2.28}$$

At $\kappa_4 = \kappa_4^c$, the grand canonical partition function (2.3) behaves like

$$Z(\kappa_4^c, \kappa_2) = \sum_{N_4} f(N_4, \kappa_2), \tag{2.29}$$

and therefore

$$\langle N_4 \rangle(\kappa_2) = \frac{\displaystyle\sum_{N_4} N_4\, f(N_4, \kappa_2)}{\displaystyle\sum_{N_4} f(N_4, \kappa_2)}. \tag{2.30}$$

Depending on $f(N_4, \kappa_2)$, there are two scenarios. First, $\langle N_4 \rangle$ might diverge and the argument is alright. Second, $\langle N_4 \rangle$ might converge to a finite value. In this case, there is no way to tune $\kappa_4$ such that $\langle N_4 \rangle$ becomes large. Recent data [Ambjørn & Jurkiewicz 1995b] show that $f(N_4, \kappa_2)$ behaves approximately like

$$f(N_4, \kappa_2) \propto \begin{cases} N_4^{\gamma(\kappa_2)-3} & \text{if } \kappa_2 > \kappa_2^c, \\ \exp(-cN_4^\alpha) & \text{if } \kappa_2 < \kappa_2^c. \end{cases} \tag{2.31}$$

with values of $\gamma(\kappa_2) \leqslant 1/2$. This would imply the second scenario. This might change if we add matter and/or change the topology of the system. In two dimensions this happens according to (1.33) and (1.34).

Figure 2.4. The curvature susceptibility as a function of $\kappa_2$ for 8000 and 16000 simplices. R is defined as $N_2/N_4$.

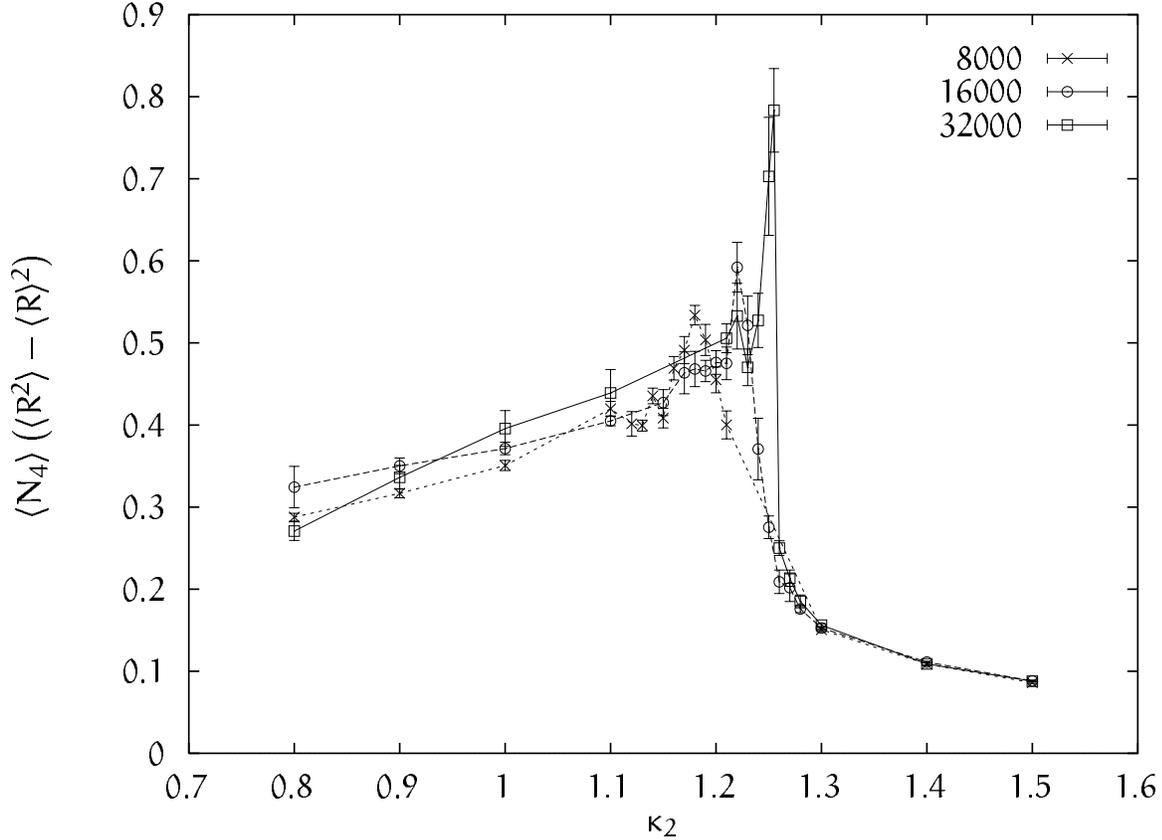

## 2.3 Phase transition

We have seen that we need to tune $\kappa_4$ or, equivalently, V to keep the system around the volume we would like. This leaves us with only one parameter to play around with, which is $\kappa_2$. Remember that this constant is inversely proportional to the bare Newton constant G by equation (1.19).

A strong indication that there is a phase transition comes from the curvature susceptibility, which turns out to have a diverging peak. This curvature susceptibility is defined as

$$\chi_R \equiv \frac{1}{N_4} \frac{\partial \langle N_2 \rangle}{\partial \kappa_2} = \frac{1}{N_4} \frac{\partial^2 \ln Z(N_4, \kappa_2)}{\partial \kappa_2^2} = \frac{1}{N_4} \left[ \langle N_2^2 \rangle - \langle N_2 \rangle^2 \right]. \qquad (2.32)$$

But because we let the volume fluctuate, the actual fluctuations in $N_2$ are dominated by the volume fluctuations. We therefore study the fluctuations in the curvature per unit volume

$$\chi_R = \langle N_4 \rangle \left[ \left\langle \left( \frac{N_2}{N_4} \right)^2 \right\rangle - \left\langle \frac{N_2}{N_4} \right\rangle^2 \right],$$  (2.33)

This quantity is plotted in figure 2.4 on the preceding page. Although the picture is somewhat messy, the peak in this susceptibility clearly increases with the volume and also moves somewhat to the right. A more detailed analysis using multihistogramming methods [Catterall *et al.* 1994a] has shown that the peak increases as

$$\chi_R^c \propto \langle N_4 \rangle^\Delta,$$  (2.34)

with a critical exponent $\Delta = 0.259(7)$. This indicates a second or higher order phase transition. A first order transition would have $\Delta = 1$.

The value of the critical coupling $\kappa_2^c$ changes with the volume. These values have been drawn as black dots in figure 2.2 on page 31. As with $\chi_R^c(N_4)$, one can investigate the behaviour of $\kappa_2^c(N_4)$. Using a different method to determine $\kappa_2^c(N_4)$, it was found in [Ambjørn & Jurkiewicz 1995b] that it moves like

$$|\kappa_2^c(N_4) - \kappa_2^c| \propto N_4^{-\delta},$$  (2.35)

with an exponent $\delta = 0.47(3)$. Again, a first order transition would have $\delta = 1$.

There are several features that qualitatively distinguish the system on both sides of the phase transition. The most notable is that the phase at low $\kappa_2$ is highly connected; the average distance between two simplices is small. Therefore this phase is called the crumpled phase. The other phase has long thin branches and is called the elongated phase. In the elongated phase the system resembles a branched polymer.

Figure 2.5 on the facing page shows the average distance between two simplices as a function of $\kappa_2$. The distance between two simplices is defined to be the minimum number of steps we need to take from simplex to directly connected simplex to get from the first simplex to the second. To distinguish this definition from others, this distance is often called the geodesic distance. The figure shows a sharp crossover between the regions of small and large distance. On the left side of the transition, at $\kappa_2 < \kappa_2^c$, the average distance increases as the logarithm of the volume, while in the elongated phase it increases as the square root of the volume. The latter behaviour supports the statement that the system behaves like a branched polymer in that phase, because the internal fractal dimension of a branched polymer is two [David 1992]. It is remarkable that, due to the change of $\kappa_2^c$ with the volume, for some values of $\kappa_2$ and volume ranges the average distance can actually decrease by adding more simplices.

Figure 2.5. The average distance between two simplices as a function of $\kappa_2$ for various volumes.

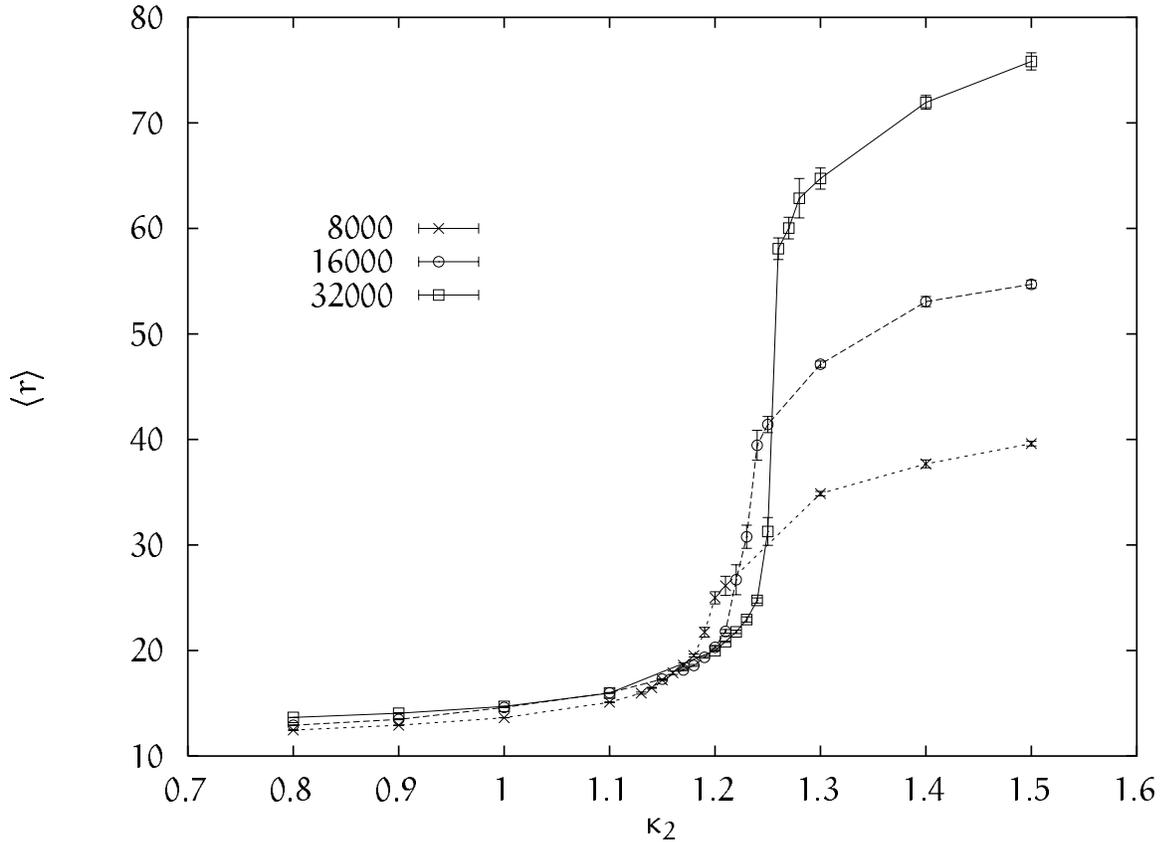

Not only the distance itself, but also its fluctuations show a very marked difference between the phases. By fluctuations in the average distance we mean that we take the average of the distances of one configuration and then measure its fluctuations across configurations. As we can see in figure 2.6 on the next page, in the elongated phase the average distance fluctuates wildly. The squared fluctuations $(\langle r^2 \rangle - \langle r \rangle^2)$ are approximately equal to $\langle r \rangle$. On the other hand, in the crumpled phase there is hardly any change. Note the logarithmic vertical scale in the picture. Unlike the fluctuations in the curvature, there is no peak and the fluctuations remain high for $\kappa_2 > \kappa_2^c$.

A nice illustration of the nature of the elongated phase can be made by calculating the massless scalar propagator on such a configuration by solving the Laplace equation

$$\Box \phi_x = \delta_{x,0}, \tag{2.36}$$

Figure 2.6. Fluctuations in the average distance as a function of $\kappa_2$.

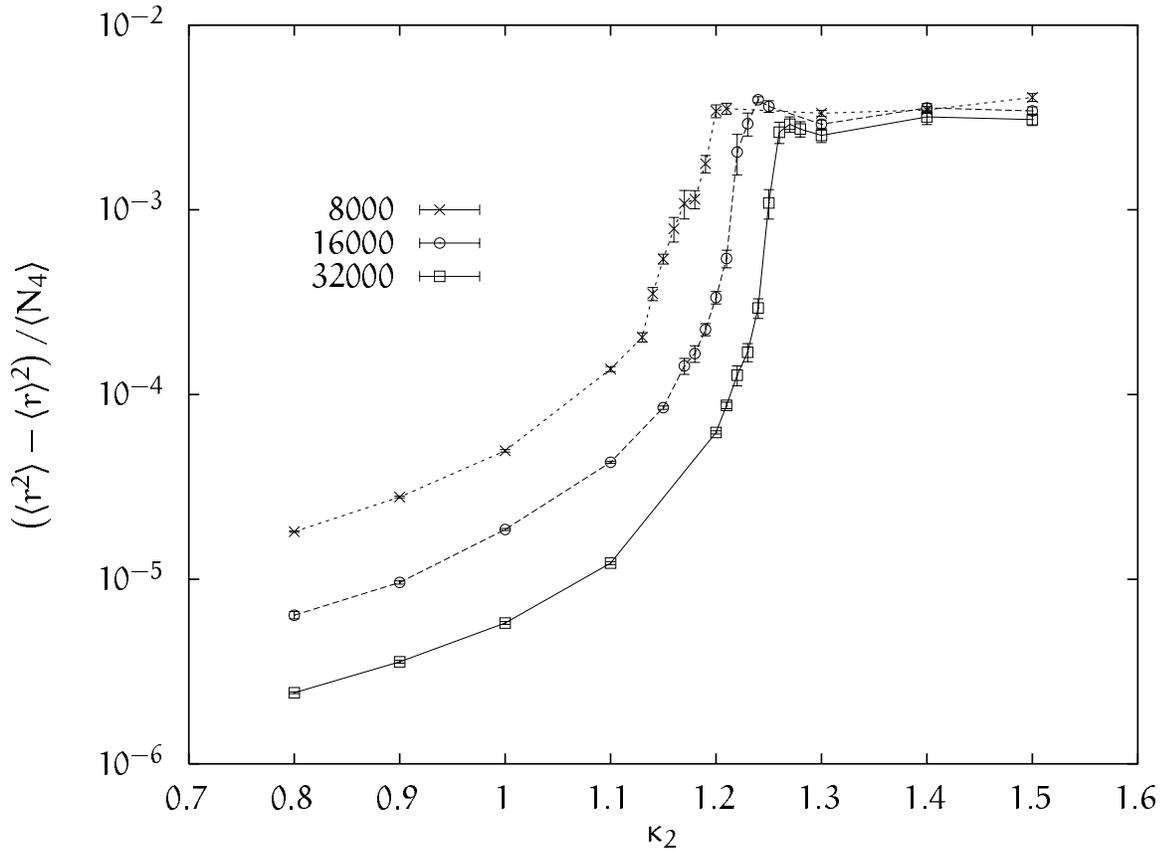

for some origin $0$. We can then plot for each simplex the value of the propagator against the geodesic distance. This has been done in figure 2.7 on the facing page for 16000 simplices and $\kappa_2 = 2.2$. If there are long tubes sticking out of the space, the geodesic distance along these tubes will increase, but the propagator will be constant. Such effects can be clearly seen in the figure.

## 2.4 Clusters

This section is about an idea that is still rather speculative and its arguments are mostly intuitive. Considering that there is a phase transition in the model one might wonder whether there is a symmetry that is spontaneously broken in one of the phases, creating long range order.

It will be suggested in this section that this symmetry is the reversal of the orientation. Popularly speaking, the symmetry of turning the configuration inside out. Imagine a part of a two dimensional triangulated surface consisting of a vertex with less than six triangles around it, i.e. less than in flat space. Keeping

Figure 2.7. The massless propagator against the geodesic distance in the elongated
phase.

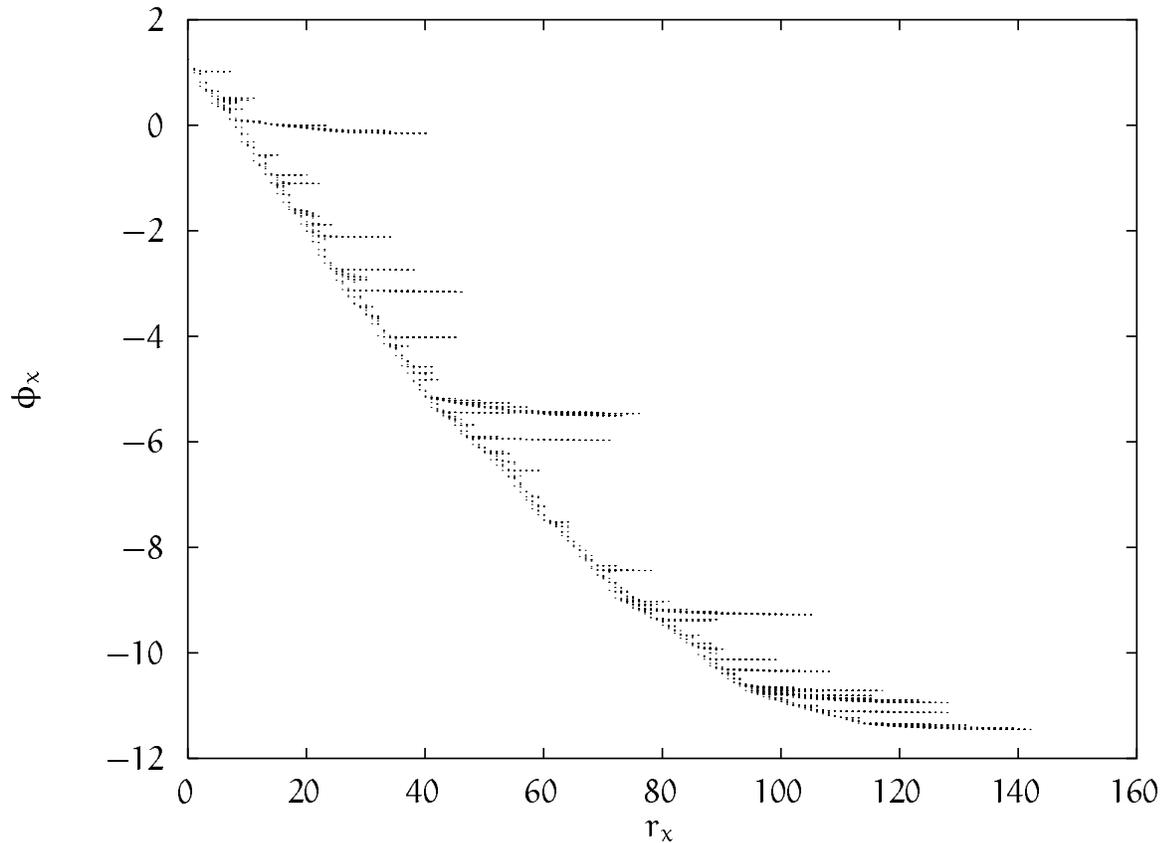

the triangles rigid, but allowing them to turn around the edges, it is not possible
to flip the vertex through the circle of edges of the configuration to the other
side. With more than six triangles around the vertex, this is possible. If two
vertices with less then six triangles each are connected by an edge, both sets of
triangles must point in the same direction, because the triangles next to this edge
are contained in both sets. This is illustrated in figure 2.8 on the next page.

This means that if most of the vertices with less than six triangles are connected
by edges in a large cluster, all these vertices will point in the same general direction,
outside or inside. The configuration will not be flexible enough to continuously
deform it to turn it inside out. If however, those vertices exist only in isolated
patches, such a deformation seems possible. As the space is not embedded, imagine
it to be able to freely pass through itself. Let us compare this with the familiar
Ising model. In the symmetric phase, we can flip the whole configuration (i.e.
turn all the spins upside down) by locally changing spins without going through
a large energy barrier. On the other hand, in the broken phase (the phase with spontaneous magnetization) this is not possible. Similarly, not being able to turn the dynamical triangulation configuration inside out is an indication of symmetry breaking.

Figure 2.8. Two neighbouring vertices with less than six triangles must point in the same direction.

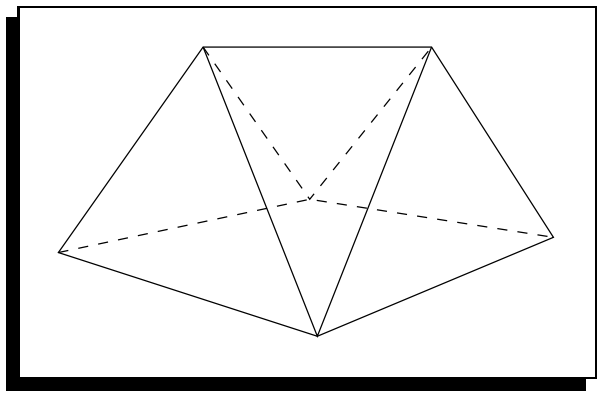

In four dimensions we can now similarly look for clusters of vertices with less simplices around them than in flat space. To do this we first have to calculate how many simplices there would be around a vertex to completely fill the spatial angle around it.

We see from equation (1.16) that for a space with zero average curvature we have

$$\frac{N_2}{N_4} = \frac{5\theta}{\pi} \approx 2.098, \tag{2.37}$$

and therefore, using (2.22),

$$\frac{N_4}{N_0} = \frac{2\pi}{5\theta - 2\pi} \approx 20.44, \tag{2.38}$$

where the $\chi$ term in (2.22) vanishes because we are considering flat space. Considering that each simplex has five vertices, the number of simplices around each vertex at zero curvature will be 5 times (2.38) or approximately 102.2. We get this same number, but not as easily, by calculating the four dimensional spatial angle $\Omega$ of a simplex at one vertex and dividing the total spatial angle $2\pi^2$ by $\Omega$.

In figure 2.9 on the facing page I have plotted the average cluster size as a function of $\kappa_2$ for volumes of 8000, 16000 and 32000 simplices. The average cluster

Figure 2.9. Average cluster size as a function of $\kappa_2$ for 8000, 16000 and 32000 simplices.

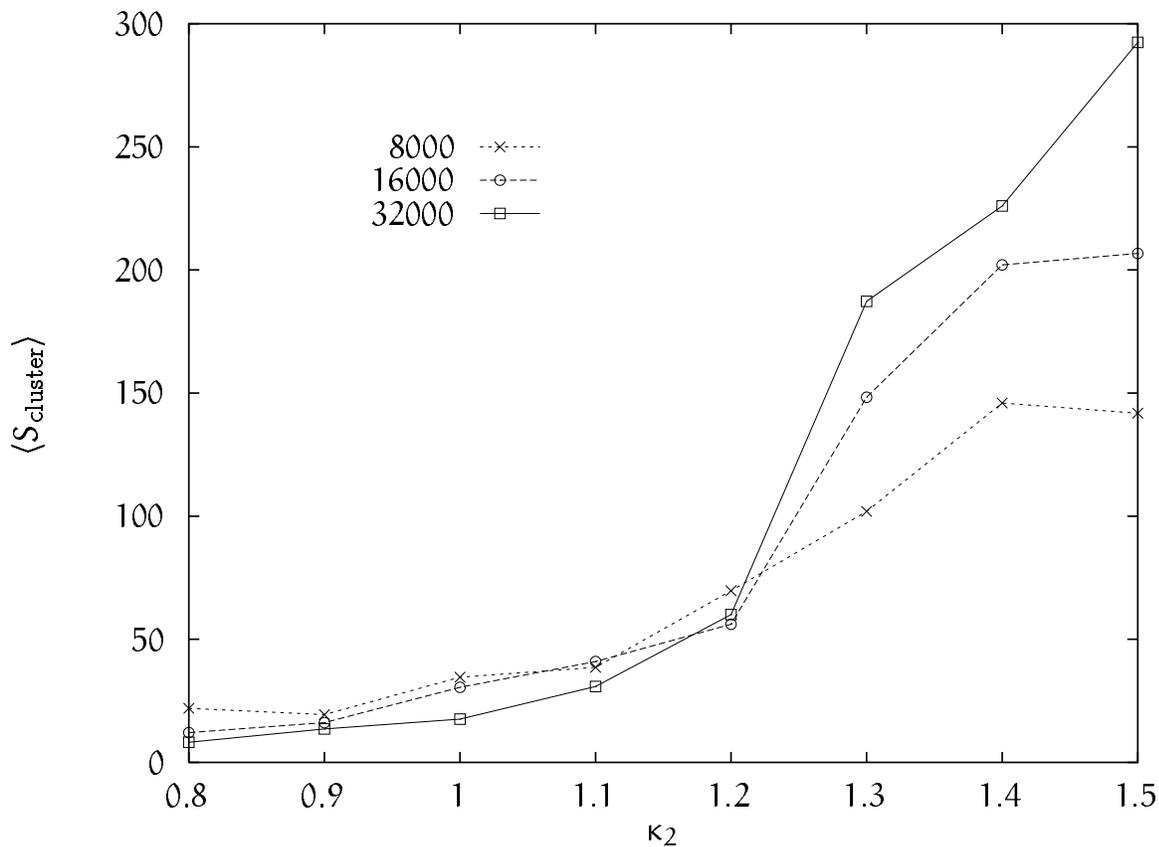

size is defined as

$$\langle S_{\text{cluster}} \rangle = \frac{\sum_x{}' S_{\text{cluster}}(x)}{\sum_x{}' 1}, \tag{2.39}$$

where the sum is taken only over those vertices $x$ that can be in clusters (all vertices of order $\leqslant 102$) and $S_{\text{cluster}}(x)$ is the size (i.e. the number of vertices) of the cluster containing the vertex $x$. We see that in the crumpled phase the average size is much smaller and decreases with the volume, while in the elongated phase it is much larger and increases with the volume. Extrapolating this to large volumes we see that the change in $\langle S_{\text{cluster}} \rangle$ between the phases will become very sharp. This is even more remarkable if we consider that in the crumpled phase the connectedness is much higher, making any kind of percolation process much easier.

One should keep in mind that there are various models where the percolation transition does not coincide with the phase transition. For instance, although the two-dimensional Ising model on a square lattice becomes percolating at the phase transition this is no longer true on a triangular lattice. It can therefore be dangerous to draw conclusions about phase transitions from percolation arguments. Nevertheless, considering the geometrical nature of dynamical triangulation, it seems likely that in this case the two are related.

A completely different symmetry breaking mechanism was described in [Shamir 1994] in the model mentioned in section 1.4 where both the connectivity and the edge lengths of the simplicial complex were varied. In this model there is a $SL(4)$ symmetry which could spontaneously break to $O(4)$. The Goldstone bosons of this symmetry breaking would then be the gravitons.

# Chapter Three

# Curvature and scaling

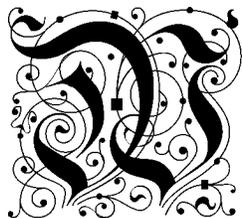 E STUDY the average number of simplices $N'(r)$ at geodesic distance $r$ in the dynamical triangulation model of Euclidean quantum gravity in four dimensions. We use $N'(r)$ to explore definitions of curvature and of effective global dimension. An effective curvature $R_V$ goes from negative values for low $\kappa_2$ (the inverse bare Newton constant) to slightly positive values around the transition $\kappa_2^c$. Far above the transition $R_V$ is hard to compute. This $R_V$ depends on the distance scale involved and we therefore investigate a similar explicitly $r$ dependent 'running' curvature $R_{eff}(r)$. This increases from values of order $R_V$ at intermediate distances to very high values at short distances.

A global dimension $d$ goes from high values in the region with low $\kappa_2$ to $d = 2$ at high $\kappa_2$. At the transition $d$ is consistent with 4. We present evidence for scaling of $N'(r)$ and introduce a scaling dimension $d_s$ which turns out to be approximately 4 in both weak and strong coupling regions. We discuss possible implications of the results, the emergence of classical euclidean spacetime and a possible 'triviality' of the theory.

## 3.1  Introduction

The dynamical triangulation model is a very interesting candidate for a non-perturbative formulation of four-dimensional euclidean quantum gravity [Agishtein & Migdal 1992a, Agishtein & Migdal 1992b, Ambjørn & Jurkiewicz 1992]. The configurations in the model are obtained by glueing together equilateral four-dimensional simplices in all possible ways such that a simplicial manifold is obtained. A formulation using hypercubes was pioneered in [Weingarten 1982]. The simplicial model with spherical topology is defined as a sum over triangulations $\mathcal{T}$ with the topology of the hypersphere $S^4$ where all the edges have the same length $\ell$. The partition function of this model is

$$Z(N, \kappa_2) = \sum_{\mathcal{T}, N_4 = N} \exp(\kappa_2 N_2). \tag{3.1}$$

Here $N_i$ is the number of simplices of dimension $i$ in the triangulation $\mathcal{T}$. The weight $\exp(\kappa_2 N_2)$ is part of the Regge form of the Einstein-Hilbert action $-\int \sqrt{g}R/16\pi G_0$,

$$-S = \frac{1}{16\pi G_0} \sum_{\triangle} V_2 \, 2\delta_{\triangle} = \kappa_2(N_2 - \rho N_4), \tag{3.2}$$

$$\kappa_2 = \frac{V_2}{4G_0}, \tag{3.3}$$

$$\rho = \frac{10 \arccos(1/4)}{2\pi} \approx 2.098, \tag{3.4}$$

Here $V_2$ is the volume of 2-simplices (triangles $\triangle$) and $\delta_{\triangle}$ is the deficit angle around $\triangle$. The volume of an i-simplex is

$$V_i = \ell^i \sqrt{(i+1)/2^i}/i!. \tag{3.5}$$

Because the $N_i$ ($i = 0 \ldots 4$) satisfy three constraints only two of them are independent. We have chosen $N_2$ and $N_4$ as the independent variables. For comparison with other work we remark that if $N_0$ is chosen instead of $N_2$ then the corresponding coupling constant $\kappa_0$ is related to $\kappa_2$ by $\kappa_0 = 2\kappa_2$. This follows from the relations between the $N_i$, which imply that

$$N_0 - \frac{1}{2}N_2 + N_4 = \chi, \tag{3.6}$$

where $\chi$ is the Euler number of the manifold, which is 2 for $S^4$.

Average values corresponding to (3.1) can be estimated by Monte Carlo methods, which require varying $N_4$ [Agishtein & Migdal 1992a, Agishtein & Migdal 1992b, Ambjørn & Jurkiewicz 1992]. One way to implement the condition $N_4 = N$ is to base the simulation on the modified partition function [Agishtein & Migdal 1992a]

$$Z'(N, \kappa_2) = \sum_{\mathcal{T}} \exp(\kappa_2 N_2 - \kappa_4 N_4 - \gamma(N_4 - N)^2), \tag{3.7}$$

where $\kappa_4$ is chosen such that $\langle N_4 \rangle \approx N$ and the parameter $\gamma$ controls the volume fluctuations. The precise values of these parameters are irrelevant if the desired $N_4$ are picked from the ensemble described by (3.7). This is not done in practice but the results are insensitive to reasonable variations in $\gamma$. We have chosen the parameter $\gamma$ to be $5 \cdot 10^{-4}$, giving $\langle N_4^2 \rangle - \langle N_4 \rangle^2 \approx (2\gamma)^{-1} = 1000$, i.e. the fluctuations in $N_4$ are approximately 30.

Numerical simulations [Agishtein & Migdal 1992a, Agishtein & Migdal 1992b, Ambjørn & Jurkiewicz 1992, Ambjørn *et al.* 1993c, Brügmann 1993, Catterall *et al.* 1994a, Varsted 1994] have shown that the system (3.7) can be in two phases.[*] For $\kappa_2 > \kappa_2^c(N_4)$ (weak bare coupling $G_0$) the system is in an elongated phase with high $\langle \overline{R} \rangle$, where $\langle \overline{R} \rangle$ is the average Regge curvature

$$\overline{R} = \frac{4\pi V_2}{V_4}(\frac{N_2}{N_4} - \rho) \longleftrightarrow \frac{\int \sqrt{g}R}{\int \sqrt{g}}. \tag{3.8}$$

In this phase the system has relatively large baby universes [Ambjørn *et al.* 1993b] and resembles a branched polymer. For $\kappa_2 < \kappa_2^c(N_4)$ (strong coupling) the system is in a crumpled phase with low $\langle \overline{R} \rangle$. This phase is highly connected, i.e. the average number of simplices around a point is very large. The transition between the phases appears to be continuous.

As an example and for later reference we show in figure 2.4 on page 35 the susceptibility

$$\left[ \left\langle \left( \frac{N_2}{N_4} \right)^2 \right\rangle - \left\langle \frac{N_2}{N_4} \right\rangle^2 \right] N = \frac{V_4}{(4\pi V_2)^2} \left[ \langle \overline{R}^2 \rangle - \langle \overline{R} \rangle^2 \right] V, \tag{3.9}$$

where $V$ is the total volume $N_4 V_4$. The three curves are for $N = 8000$, $16000$ and $32000$ simplices.

The data for $N = 8000$ are consistent with those published in reference [Catterall *et al.* 1994a], where results are given for

$$\chi(N_0) = \frac{1}{N} \left( \langle N_0^2 \rangle - \langle N_0 \rangle^2 \right). \tag{3.10}$$

For fixed $N_4$ this is $1/4$ of our curvature susceptibility (3.9), as can be seen from (3.6).

The behavior of $Z(\kappa_2, N)$ as a function of $N$ for large $N$ has been the subject of recent investigations [Ambjørn & Jurkiewicz 1994, de Bakker & Smit 1994,

---

[*]In section 3.6 on page 64 we discuss the possibility that these are not phases in the sense of conventional statistical mechanics.

Brügmann & Marinari 1995, Catterall *et al.* 1994b]. In [de Bakker & Smit 1994] we discussed the possibility that $\kappa_2^c$ might move to infinity as $N \to \infty$ and argued that this need not invalidate the model. So far a finite limit is favoured by the data [Ambjørn & Jurkiewicz 1994, Brügmann & Marinari 1995], however.

It is of course desirable to get a good understanding of the properties of the euclidean spacetimes described by the probability distribution $\exp(-S)$. A very interesting aspect is the proliferation of baby universes [Ambjørn *et al.* 1993b]. Here we study more classical aspects like curvature and dimension, extending previous work in this direction [Agishtein & Migdal 1992a, Agishtein & Migdal 1992b, Ambjørn & Jurkiewicz 1992, Ambjørn *et al.* 1993c, Brügmann 1993, Catterall *et al.* 1994a, Varsted 1994]. The basic observable for this purpose is the average number of simplices at a given geodesic distance from the arbitrary origin, $N'(r)$. We want to see if this quantity can be characterized, approximately, by classical properties like curvature and dimension, and if for suitable bare couplings $\kappa_2$ there is a regime of distances where the volume-distance relation $N'(r)$ can be given a classical interpretation. It is of course crucial for such a continuum interpretation of $N'(r)$ that it scales in an appropriate way.

In section 3.2 we investigate the properties of an effective curvature for a distance scale that is large compared to the basic unit but small compared to global distances. Effective dimensions for global distances are the subject of section 3.3 and scaling is investigated in section 3.4. We summarize our results in section 3.5 and discuss the possible implications in section 3.6.

## 3.2 Curvature

A straightforward measure of the average local curvature is the bare curvature at the triangles, i.e. $\langle \overline{R} \rangle = 96\pi\sqrt{3/5}\,\ell^{-2}\langle N_2/N_4 - \rho \rangle$ (this follows from (3.8) and (3.5)). It has been well established that at the phase transition $\langle N_2/N_4 - \rho \rangle \approx 2.38 - 2.10 = 0.28$, practically independent of the volume [Agishtein & Migdal 1992a, Agishtein & Migdal 1992b, Ambjørn & Jurkiewicz 1992, Ambjørn *et al.* 1993c]. This means that $\langle \overline{R} \rangle \approx 65\ell^{-2}$ has to be divergent in the continuum limit $\ell \to 0$. The curvature at scales large compared to the lattice distance $\ell$, however, is not necessarily related to the average curvature at the triangles. One could imagine e.g. a spacetime with highly curved baby universes which is flat at large scales.

The scalar curvature at a point is related to the volume of a small hypersphere around that point. Expanding the volume in terms of the radius of the hypersphere

Table 3.1. Number of configurations used in the various calculations.

| $\kappa_2$ | 8000 | 16000 |
|------|------|-------|
| 0.80 | 29 | 16 |
| 0.90 | 34 | 19 |
| 1.00 | 38 | 25 |
| 1.10 | 37 | 38 |
| 1.12 | 13 | — |
| 1.13 | 24 | — |
| 1.14 | 28 | — |
| 1.15 | 21 | 13 |
| 1.16 | 36 | — |
| 1.17 | 44 | 16 |
| 1.18 | 90 | 8 |
| 1.19 | 26 | 22 |

| $\kappa_2$ | 8000 | 16000 |
|------|------|-------|
| 1.20 | 56 | 45 |
| 1.21 | 22 | 56 |
| 1.22 | — | 51 |
| 1.23 | — | 65 |
| 1.24 | — | 58 |
| 1.25 | — | 45 |
| 1.26 | — | 38 |
| 1.27 | — | 31 |
| 1.28 | — | 43 |
| 1.30 | 43 | 41 |
| 1.40 | 40 | 24 |
| 1.50 | 47 | 21 |

results in the relation

$$V(r) = C_n r^n (1 - \frac{R_V r^2}{6(n+2)} + O(r^4)), \tag{3.11}$$

$$C_n = \frac{\pi^{n/2}}{\Gamma(n/2+1)}, \tag{3.12}$$

for an $n$-dimensional manifold. We have written $R_V$ here to distinguish it from the small scale curvature at the triangles. Differentiating with respect to $r$ results in the volume of a shell at distance $r$ with width $dr$,

$$V'(r) = C_n n r^{n-1} (1 - \frac{R_V r^2}{6n} + O(r^4)). \tag{3.13}$$

We explore this definition of curvature as follows. We take the dimension $n = 4$, assuming that there is no need for a fractional dimension differing from 4 at small scales. For $r$ we take the geodesic distance between the simplices, that is the lowest number of hops from four-simplex to neighbour needed to get from one four-simplex to the other. Setting the distance between the centers of neighbouring simplices to 1 corresponds to taking a fixed edge length in the simplicial complex of $\sqrt{10}$, i.e. we will use lattice units with $\ell = \sqrt{10}$. For $V(r)$ and $V'(r)$ we take

$$V(r) = V_{eff} N(r), \tag{3.14}$$

$$V'(r) = V_{eff} N'(r), \tag{3.15}$$

$$N'(r) = N(r) - N(r-1), \tag{3.16}$$

where $N(r)$ is the average number of four-simplices within distance $r$ from the (arbitrary) origin, $N'(r)$ is the number of simplices at distance $r$ and we have allowed for an effective volume $V_{\text{eff}}$ per simplex which is different from $V_4$. Since $R_V$ is to be a long distance (in lattice units) observable we shall call it the effective curvature.

The effective curvature was determined by fitting the function $N'(r)$ to

$$N'(r) = ar^3 + br^5. \tag{3.17}$$

It then follows from (3.13) and (3.15) that $R_V$ is determined by

$$R_V = -24\frac{b}{a}, \tag{3.18}$$

and $V_{\text{eff}}$ by

$$V_{\text{eff}} = \frac{4C_4}{a}. \tag{3.19}$$

The constant $C_4$ in equation (3.11) is $\pi^2/2$. In flat space we would have $V_{\text{eff}} = V_4$, giving

$$a = \frac{4C_4}{V_4} = \frac{48\sqrt{5}\pi^2}{125} \approx 8.47, \tag{3.20}$$

where we used $V_4 = 25\sqrt{5}/24$ for $\ell = \sqrt{10}$. Such a space cannot be formed from equilateral simplices because we cannot fit an integer number of simplices in an angle of $2\pi$ around a triangle.

Figure 3.1 on the facing page shows $N'(r)$ for 16000 simplices. Three different values of $\kappa_2$ are shown, 0.8 (in the crumpled phase), 1.22 (close to the transition) and 1.5 (in the elongated phase). These curve can also be interpreted as the probability distribution of the geodesic length between two simplices. Such distributions were previously presented in [Ambjørn & Jurkiewicz 1992, Ambjørn *et al.* 1993c].

We determined $N'(r)$ by first averaging this value per configuration successively using each simplex as the origin. We then used a jackknife method, leaving out one configuration each time, to determine the error in $R_V$.

Our results in this paper are based on $N = 8000$ and 16000 simplices. Configurations were recorded every 10000 sweeps, where a sweep is defined as $N$ accepted moves. The time before the first configuration was recorded was also 10000 sweeps. We estimated the autocorrelation time in the average distance between two simplices to be roughly 2000 sweeps for $N = 16000$ and $\kappa_2 = 1.22$. For other values of $\kappa_2$ the autocorrelation time was lower. The number of configurations at the values of $\kappa_2$ used are shown in table 3.1 on page 47.

Figure 3.1. The number of simplices $N'(r)$ at distance $r$ from the origin at $\kappa_2 = 0.80$, 1.22 and 1.50, for $N = 16000$.

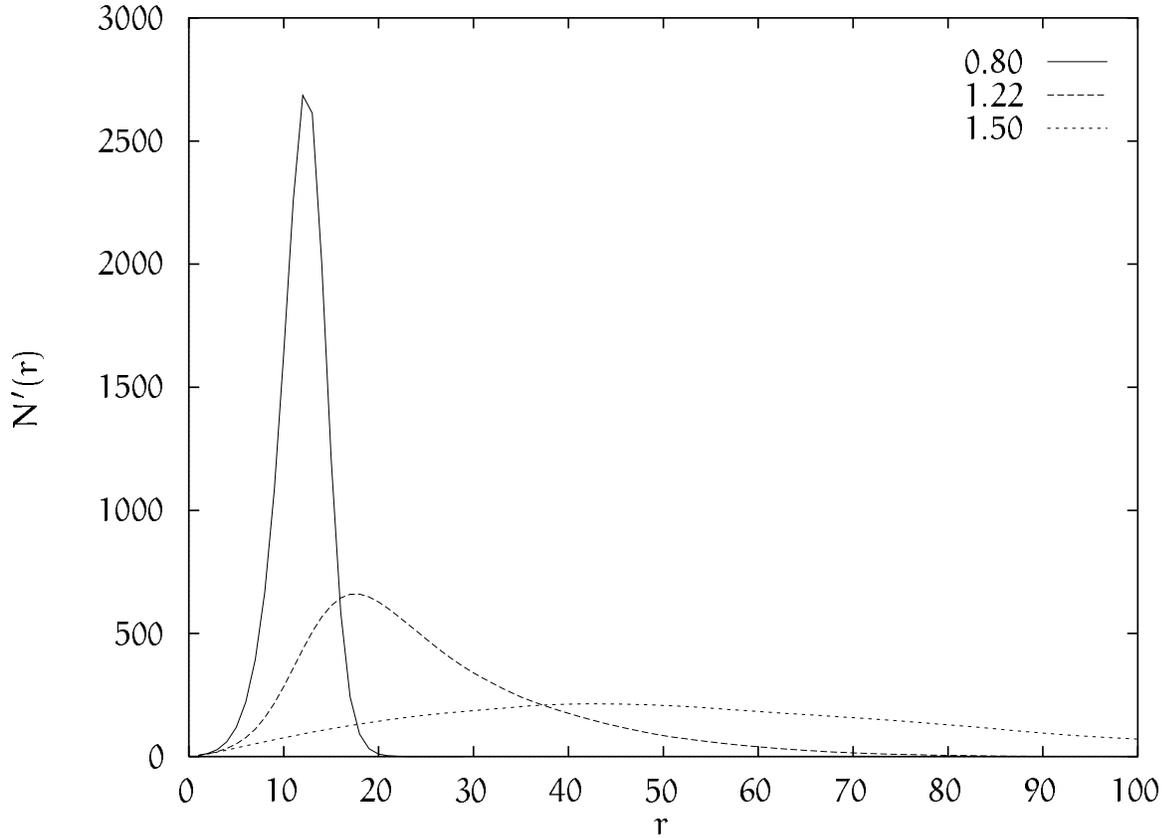

Figures 3.2–3.4 show effective curvature fits (continuous lines) in the crumpled phase ($\kappa_2 = 0.80$), near the transition ($\kappa_2 = 1.22$) and in the elongated phase ($\kappa_2 = 1.50$), together with global dimension fits (see the next section) at longer distances, for $N = 16000$. The curves are extended beyond the fitted data range, $r = 1$–11, to indicate their region of validity. A least squares fit was used, which suppresses the lattice artefact region where $N'(r)$ is small because it is sensitive to absolute errors rather than relative errors.

For $\kappa_2 \lesssim \kappa_2^c$ the fits were good even beyond the range of $r$ used to determine the fit (except obviously when this range already included all the points up to the maximum of $N'(r)$). This can be seen in figure 3.3 on page 51, where the fit is good up to $r = 15$. The fit with $ar^3 + br^5$ does not appear sensible anymore for $\kappa_2$ values somewhat larger than the critical value $\kappa_2^c$, because then $N'(r)$ goes roughly linear with $r$ down to small distances. This can be seen quite clearly in figure 3.4 on page 52 (so the $R_V$ fit in this figure should be ignored).

Figure 3.2. Effective curvature fit and global dimension fit in the crumpled phase at $\kappa_2 = 0.80$.

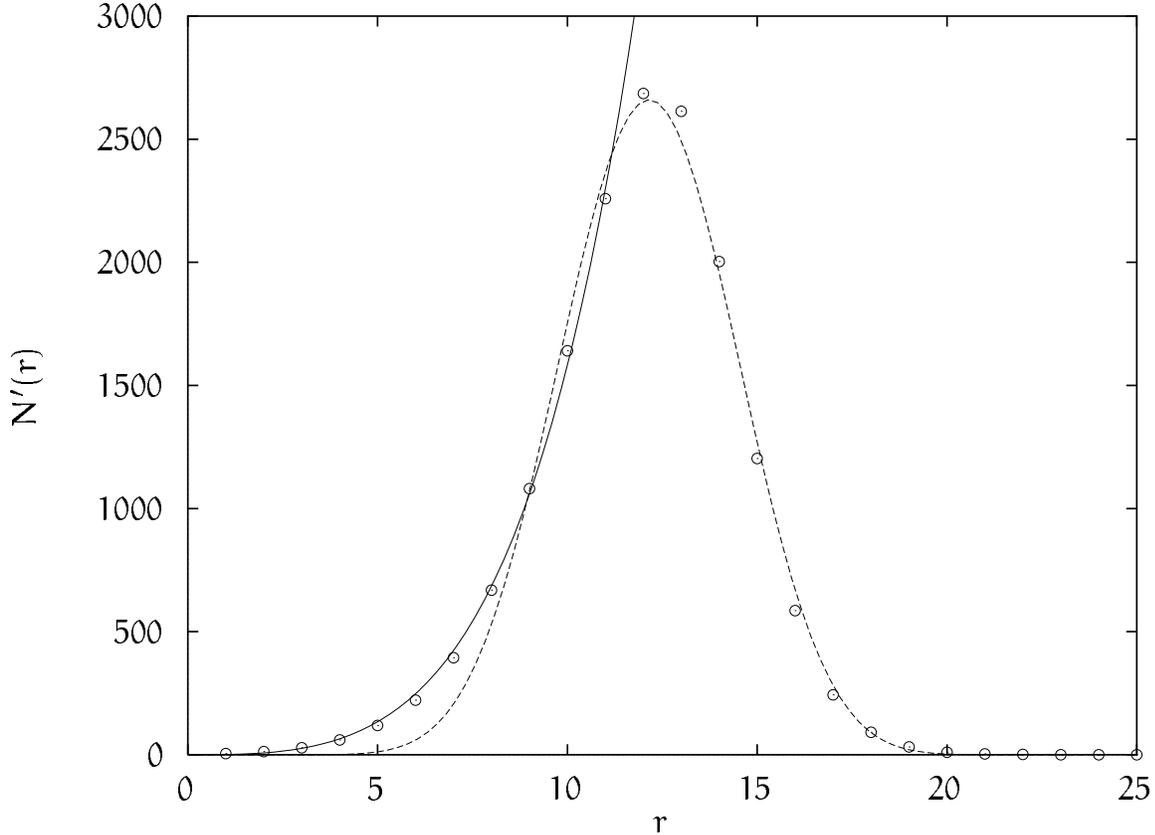

Figure 3.5 on page 53 shows the resulting effective curvature $R_V$ as a function of $\kappa_2$ for 8000 and 16000 simplices. For $N = 8000$ the fitting range was $r = 1-9$. We see that $R_V$ starts negative and then rises with $\kappa_2$, going through zero. In contrast, the Regge curvature at the triangles $\langle \overline{R} \rangle$ is positive for all $\kappa_2$ values in the figure.

The value of $a$ in (3.18) varied from about 0.9 at $\kappa_2 = 0.8$ to 0.4 near the transition. These numbers are much smaller than the flat value of 8.47 in equation (3.20), indicating an effective volume per simplex $V_{\text{eff}} \approx 20 - 50$, much larger than $V_4 \approx 2.3$. This is at least partly due to the way we measure distances. The distances are measured using paths which can only go along the dual lattice and will therefore be larger than the shortest paths through the simplicial complex.

Figure 3.3. Effective curvature fit and global dimension fit near the transition, $\kappa_2 = 1.22$.

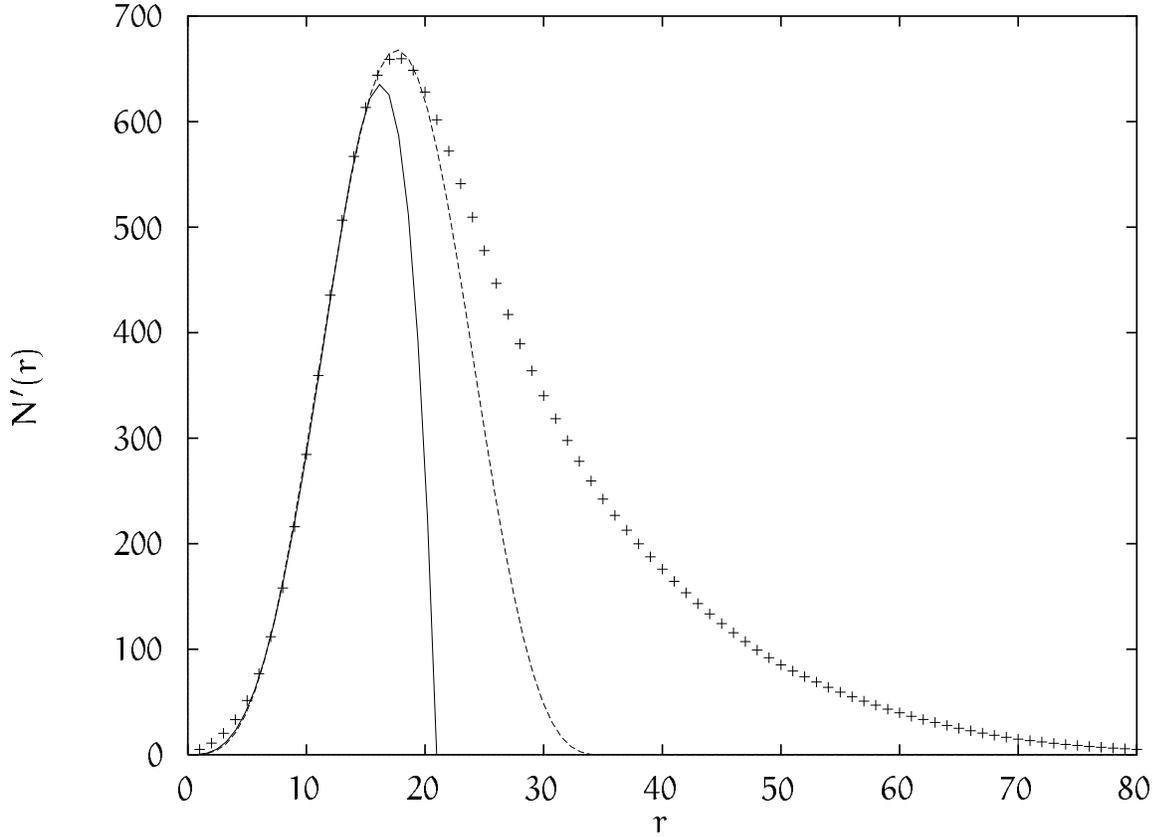

There is a strong systematic dependence of $R_V$ on the range of $r$ used in the fit. For example, for the $N = 16000$ data in figure 3.5 on page 53 we used a least squares fit in the range $r = 1$–$11$. If the range is changed to $r = 1$–$9$ the data for $R_V$ have to be multiplied with a factor of about 1.5. We can enhance this effect by reducing the fitting range to only two $r$ values and thereby obtain a 'running curvature' $R_{\text{eff}}(r)$ at distance $r$. We write

$$N'(r) = a(r)r^3 + b(r)r^5, \tag{3.21}$$

$$N'(r+1) = a(r)(r+1)^3 + b(r)(r+1)^5, \tag{3.22}$$

and define

$$R_{\text{eff}}(r + \tfrac{1}{2}) = -24b(r)/a(r), \tag{3.23}$$

which gives

$$R_{\text{eff}}(r + \tfrac{1}{2}) = 24 \frac{(r+1)^3 - r^3 N'(r+1)/N'(r)}{(r+1)^5 - r^5 N'(r+1)/N'(r)}. \tag{3.24}$$

Figure 3.4. Effective curvature fit and global dimension fit in the elongated phase at $\kappa_2 = 1.50$ (the effective curvature fit is not appropriate here).

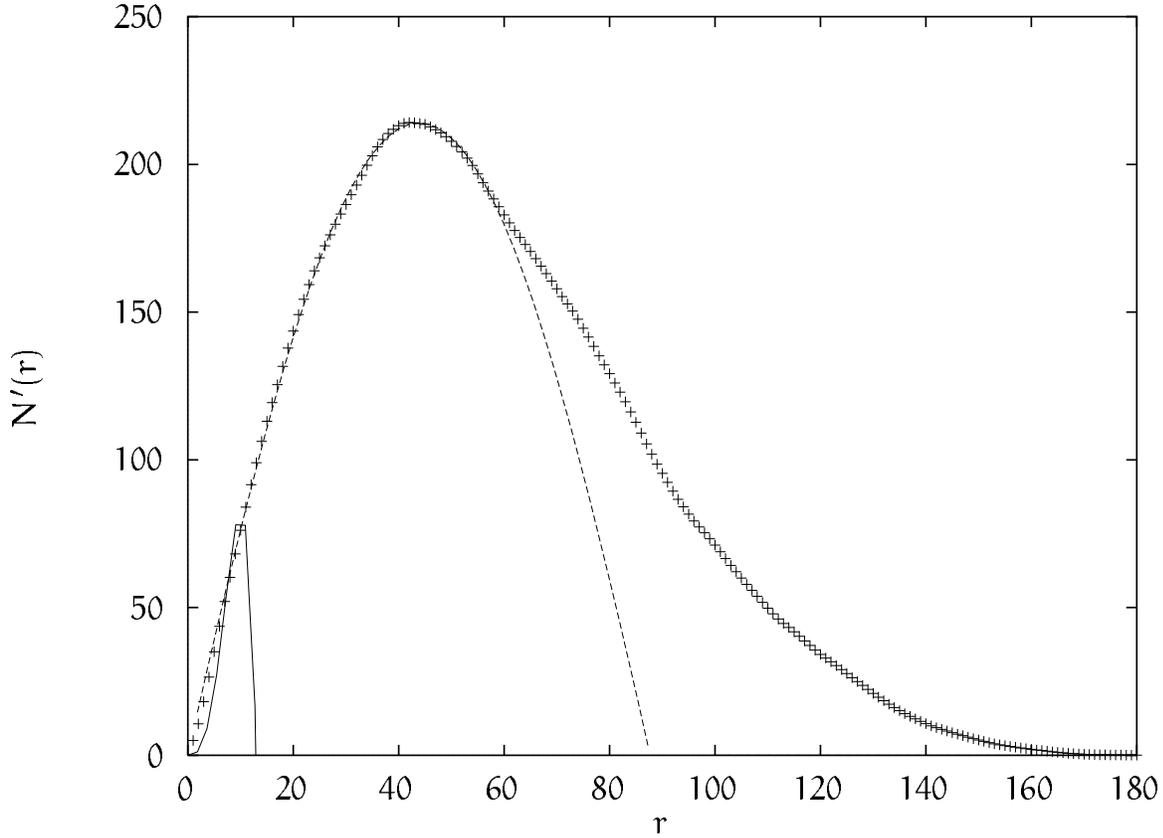

Figure 3.6 on page 54 shows the behavior of $R_{\text{eff}}(r)$ for various $\kappa_2$. It drops rapidly from large values $\approx 4.5$ (which is near $\langle \overline{R} \rangle \approx 6.5$) at $r = 0$ to small values at $r \approx 8$. For $\kappa_2 \lesssim \kappa_2^c$ the curves have a minimum and around this minimum the values of $R_{\text{eff}}$ average approximately to the $R_V$'s displayed earlier.

The idea of $R_{\text{eff}}(r)$ and $R_V$ is to measure curvature from the correlation function $N'(r)$ by comparing it with the classical volume-distance relation for distances $r$ going to zero, as long as there is reasonable indication for classical behavior at these distances. Clearly, we cannot let $r$ go to zero all the way because of the huge increase of $R_{\text{eff}}$. This seems to indicate a 'Planckian regime' where classical behavior breaks down.

## 3.3  Dimension

One of the interesting observables in the model is the dimension at large scales. A common way to define a fractal dimension is by studying the behaviour of the

Figure 3.5. The effective curvature $R_V$ as a function of $\kappa_2$ for 8000 and 16000 simplices.

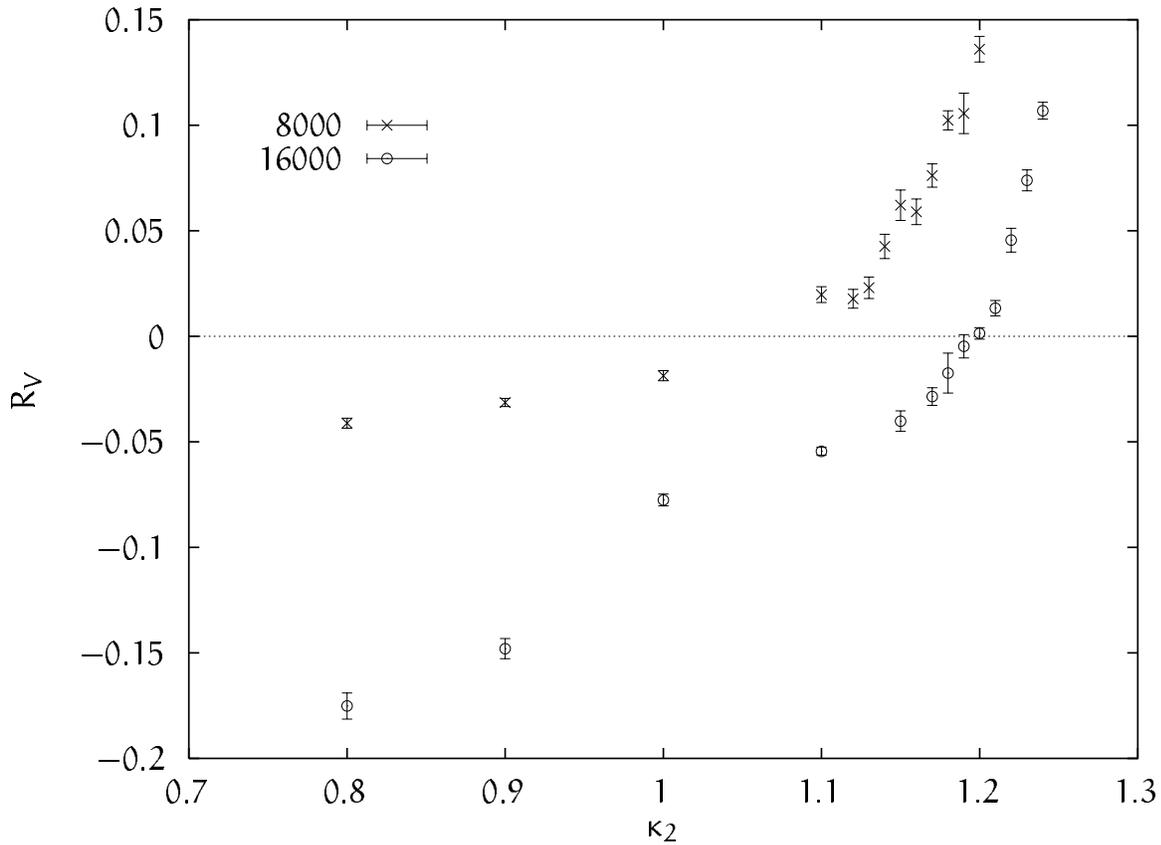

volume within a distance $r$ and identifying the dimension $d$ if the volume behaves like

$$V(r) = \text{const.} \times r^d. \tag{3.25}$$

Such a measurement has been done in [Agishtein & Migdal 1992a, Agishtein & Migdal 1992b], using the geodesic distance and a distance defined in terms of the massive scalar propagator [David 1992]. Arguments against the necessity of such use of the massive propagator were raised in [Filk 1992]. Although we feel that the issue is not yet settled, we shall use here the geodesic distance as in the previous section.

If the volume does go like a power of $r$, the quantity

$$d = \frac{d \ln V}{d \ln r} \longleftrightarrow \frac{\ln N(r) - \ln N(r-1)}{\ln(r) - \ln(r-1)}, \tag{3.26}$$

would be a constant. Figure 3.7 on page 55 shows this quantity for some values of $\kappa_2$ for a system with 16000 simplices. We see a sharp rise at distances $r = 1, 2$ independent of $\kappa_2$ which is presumably a lattice artefact. The curves then continue to rise until a maximum where we may read off an effective dimension, which is clearly different in the crumpled phase ($\kappa_2 = 0.80$–$1.20$) and in the elongated phase ($\kappa_2 = 1.30$–$1.50$). Instead of a local maximum one would of course like a plateau of values where $d \ln V / d \ln r$ is constant and may be identified with the dimension d. Only for $\kappa_2$ beyond the transition a range of $r$ exists where $d \ln V / d \ln r$ looks like a plateau. In this range, $d \approx 2$.

Figure 3.6. The running curvature $R_{eff}(r)$ for various $\kappa_2$ and 16000 simplices.

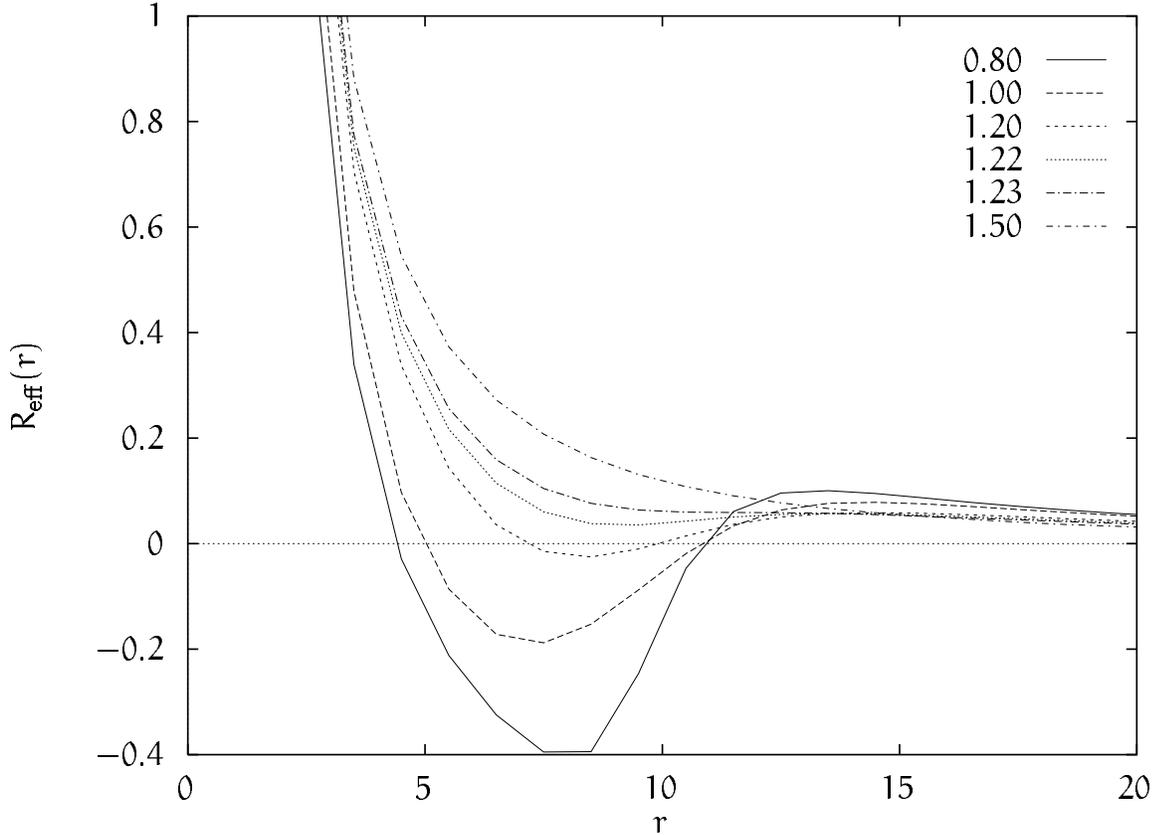

Similar studies have been carried out in two-dimensional dynamical triangulation where it was found that plateaus only appear to develop for very large numbers of triangles [Agishtein & Migdal 1991, Ambjørn et al. 1995, Kawamoto et al. 1992]. Our four-dimensional systems are presumably much too small for $d \ln V / d \ln r$ to shed light on a fractal dimension at large scales, if it exists. We feel, however, that the approximate plateau in the elongated phase with $d = 2$ should be taken seriously.

As we are studying a system with the topology of the sphere $S^4$, it seems reasonable to look whether it behaves like a d-dimensional sphere $S^d$. For such a

Figure 3.7. $\ln[(N(r)/N(r-1))]/\ln[r/(r-1)] \longleftrightarrow d\ln V/d\ln r$ as a function of $r$ for some values of $\kappa_2$.

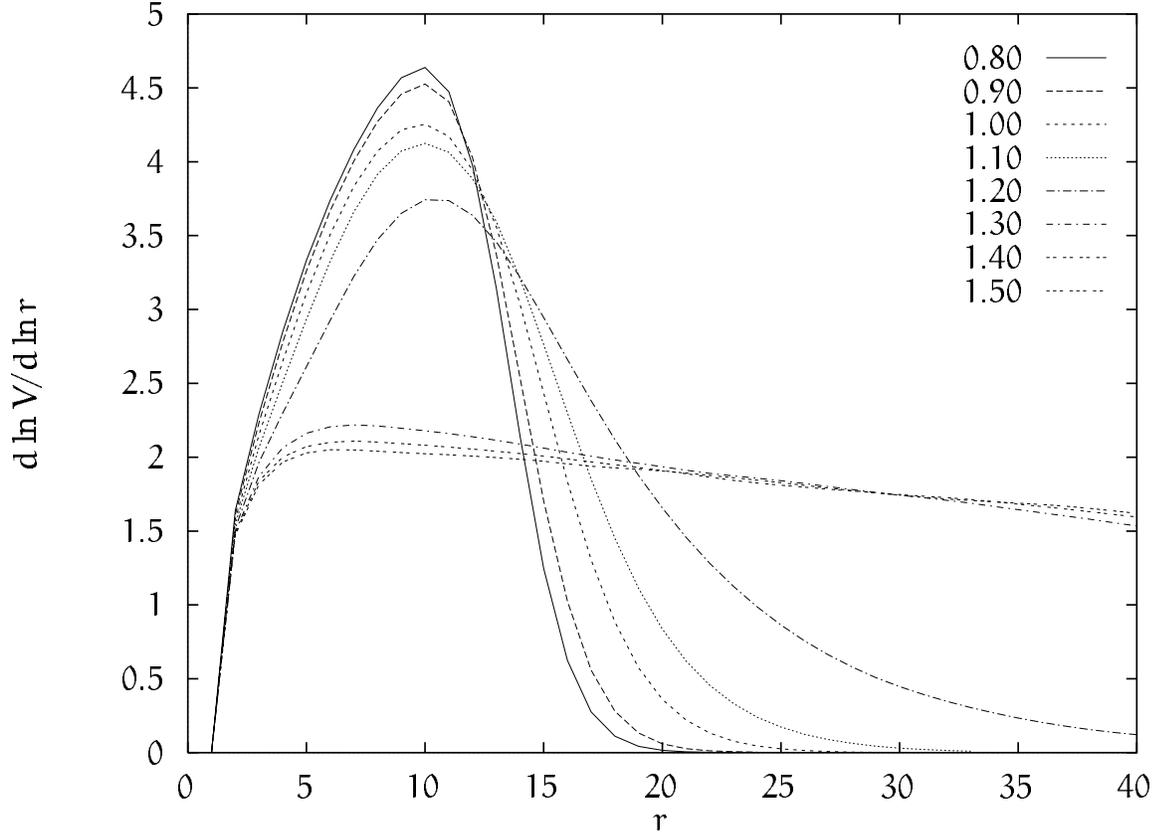

hypersphere with radius $r_0$, the volume behaves like

$$V'(r) = dC_d \, r_0^{d-1} (\sin \frac{r}{r_0})^{d-1}. \tag{3.27}$$

This prompts us to explore (3.27) as a definition of the dimension $d$, we shall call it the global dimension. For small $r/r_0$ this reduces to (3.25). On the other hand for $d = n$ the form (3.27) is compatible with the effective curvature form (3.13), with

$$R_V = \frac{n(n-1)}{r_0^2}. \tag{3.28}$$

To determine the dimension $d$ we fit the data for $N'(r)$ to a function of the form (3.27),

$$N'(r) = c \, (\sin \frac{r}{r_0})^{d-1}. \tag{3.29}$$

The free parameters are $r_0$, d and the multiplicative constant c. It is a priori not clear which distances we need to use to make the fit. At distances well below the maximum of $N'(r)$ (cf. figure 3.1 on page 49) the effective curvature fit appears to give a reasonable description of the data, but it will be interesting to try (3.27) also for these distances. Small distances are of course affected by the discretization. This is most pronounced at low $\kappa_2$ where the range of r-values is relatively small. On the other hand, for small $\kappa_2$ the fits turn out to be good up to the largest distances, indicating a close resemblance to a hypersphere, while at larger $\kappa_2$ the values of $N'(r)$ are asymmetric around the peak (cf. figure 3.1 on page 49) and fits turn out to be good only up to values of r not much larger than this peak. The system behaves like a hypersphere up to the distance where $N'(r)$ has its maximum, which would correspond to halfway the maximum distance for a real hypersphere. Above that distance it starts to deviate, except for small $\kappa_2$ where $N'(r)$ is more symmetric around the peak and the likeness remains.

For $\kappa_2 = 0.8$ (figure 3.2 on page 50) the global dimension fit (3.29) was performed to the data at r = 7–21, for $\kappa_2 = 1.22$ (figure 3.3 on page 51) to r = 4–20 and for $\kappa_2 = 1.50$ (figure 3.4 on page 52) to r = 3–60.

The two descriptions, effective curvature at lower distances and effective dimension at intermediate and larger distances, appear compatible. Notice that at $\kappa_2 = 0.8$ in the crumpled phase the local effective curvature $R_V$ is negative while the global structure resembles closely a (positive curvature) sphere with radius $r_0 = 7.6$ ($r_0 = 2r_m/\pi$, with $r_m$ the value where $N'(r)$ is maximal). At $\kappa_2 = 1.22$ near the transition the effective curvature and effective dimension descriptions appear to coincide. At $\kappa_2 = 1.50$, deep in the elongated phase, the effective curvature fit does not make sense anymore, its r-region of validity has apparently shrunk to order 1 or less. The effective dimension fit on the other hand is still good in this phase and the power behavior (3.25) with d $\approx$ 2 is extended by (3.29) to intermediate distances including the maximum of $N'(r)$.

Figure 3.8 on the next page shows the global dimension as a function of $\kappa_2$ for the total volumes of 8000 and 16000 simplices. For small values of $\kappa_2$ it is high and increases with larger volumes. For values of $\kappa_2$ beyond the transition it quickly goes to two, confirming the statement made earlier that in this region $N'(r)$ is approximately linear with r down to small r.

A most interesting value of the dimension is the one at the phase transition. To determine the value of $\kappa_2^c$ where the transition takes place we look at the curvature susceptibility of the system, figure 2.4 on page 35. For 8000 simplices the peak in the susceptibility is between $\kappa_2 = 1.17$ and 1.18 where the dimensions we measured are 4.2(1) and 3.8(1). For 16000 simplices the peak is between 1.22 and 1.23 where the dimensions are 4.2(1) and 3.6(1). Therefore the dimension at the transition is consistent with 4. As can be seen from these numbers the largest uncertainty in the dimension is due to the uncertainty in $\kappa_2^c$. The effective dimensions have some uncertainty due to the ambiguity of the range of $r$ used for the fit. Near the transition this generates an extra error of approximately 0.1.

Figure 3.8. The global dimension as a function of $\kappa_2$ for 8000 and 16000 simplices.

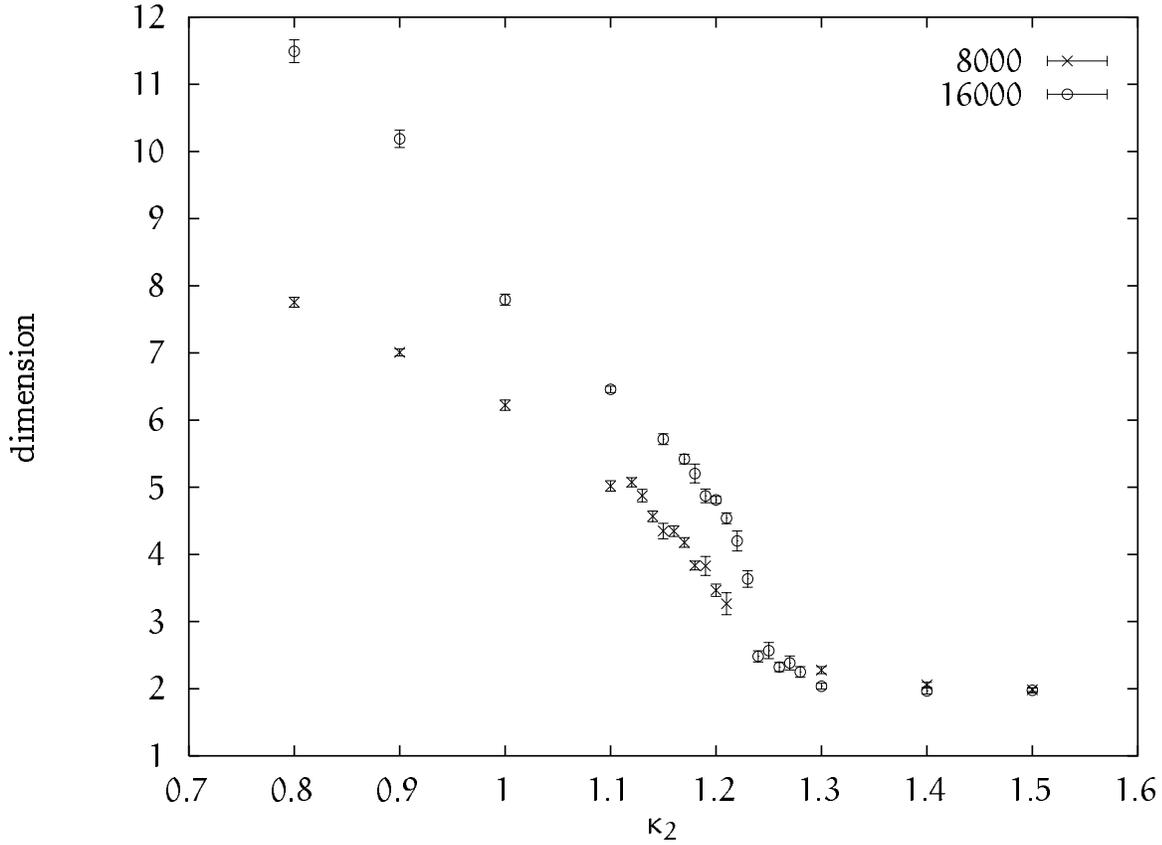

## 3.4 Scaling

To get a glimpse of continuum behavior it is essential to find scaling behavior in the system. We found a behavior like a $d$ dimensional hypersphere for $r$ values up to the value $r_m$ where $N'(r)$ is maximal. This suggests scaling in the form

$$N'(r) = r_m^{d-1} f(\frac{r}{r_m}, d), \qquad (3.30)$$

i.e. $N'(r)$ depends on $\kappa_2$ and $N$ through $d = d(\kappa_2, N)$ and $r_m = r_m(\kappa_2, N)$. The occurrence of $d$ in this formula is unattractive, however, since it is obtained by comparing $N'(r)$ to $\sin^{d-1}(r/r_0)$ at intermediate scales, which is a somewhat imprecise concept. We would like a model independent test of scaling.

Figure 3.9. The scaling function $\rho$ for $\kappa_2 = 0.8$ at $N = 8000$ and $\kappa_2 = 1.0$ at $N = 16000$.

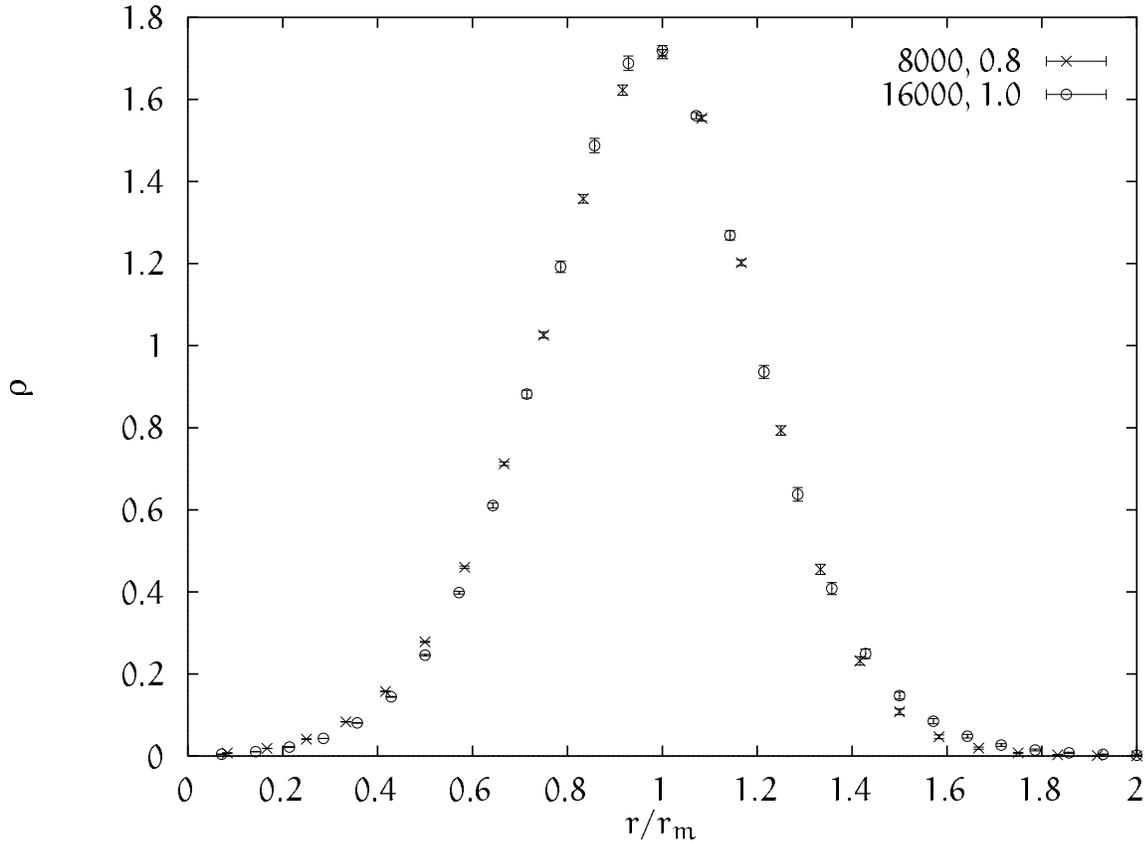

Consider the probability for two simplices to have a geodesic distance $r$,

$$p(r) = \frac{N'(r)}{N}, \tag{3.31}$$

$$1 = \sum_{r=1}^{\infty} p(r) \approx \int dr\, p(r), \tag{3.32}$$

which depends parametrically on $\kappa_2$ and $N$. It seems natural to assume scaling for this function in the form

$$p(r) = \frac{1}{r_m} \rho(\frac{r}{r_m}, \tau), \tag{3.33}$$

$$\int dx\, \rho(x, \tau) = 1, \tag{3.34}$$

where $x = r/r_m$ and $\tau$ is a parameter playing the role of $d$ in (3.30) which labels the different functions $\rho$ obtained this way. For instance, $\tau$ could be the value $\rho_m = \rho(1, \tau)$ at the maximum of $\rho$. This may give give problems if $\rho_m$ does not change appreciably with $\kappa_2$ (similar to $d$ in the elongated phase). Other possibilities are $\tau = \overline{r}p(r_m)$ or $\tau = \overline{r^k}/\overline{r}^k$ for some $k$ with $\overline{r^k} = \sum_r p(r)r^k$. In practice we may also simply take $\tau = \kappa_2 - \kappa_2^c(N)$ at some standard choice of $N$ and compare the probability functions with the $p(r)$ at this $N$.

Figure 3.10. The scaling function $\rho$ for $\kappa_2 = 1.18$ at $N = 8000$, $\kappa_2 = 1.22$ at $N = 16000$ and $\kappa_2 = 1.25$ at $N = 32000$.

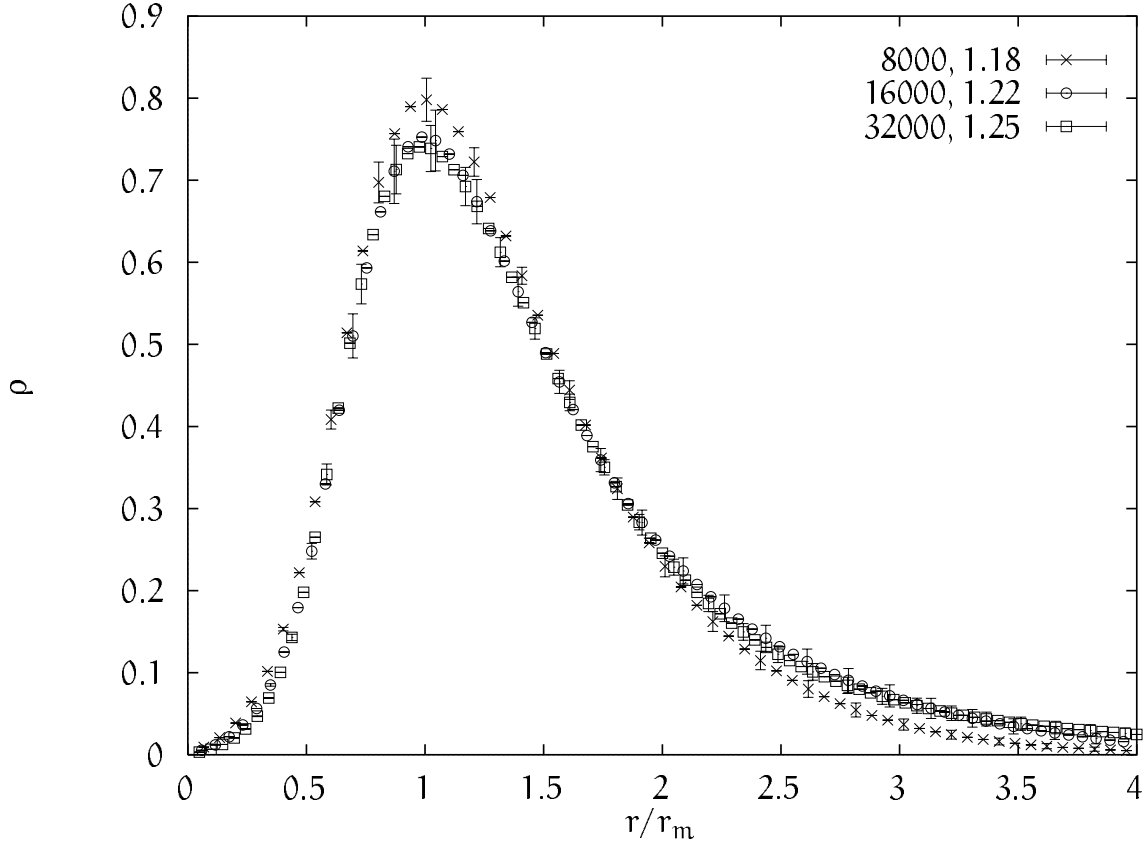

Matched pairs of $\rho(x, \tau)$ for $N = 8000$ and $16000$ and one curve at $N = 32000$ are shown in figures 3.9–3.11 on pages 58–60, respectively far in the crumpled phase, near the transition and in the elongated phase. For clarity we have left out part of the errors. Scaling appears to hold even for $\kappa_2$ values we considered far away from the transition.

The values of $\kappa_2(N)$ of the matched pairs in figures 3.9–3.11 are increasing with $N$ in the crumpled phase and decreasing with $N$ in the elongated phase. This suggests convergence from both sides to $\kappa_2^c(N)$ as $N$ increases. For current system sizes $\kappa_2^c(N)$ is still very much dependent on $N$, a power extrapolation estimate in [Catterall *et al.* 1994a] gives $\kappa_2^c(\infty) \approx 1.45$, while one in [Ambjørn &

Figure 3.11. The scaling function ρ for $\kappa_2 = 1.5$ at $N = 8000$ and $\kappa_2 = 1.3$ at $N = 16000$.

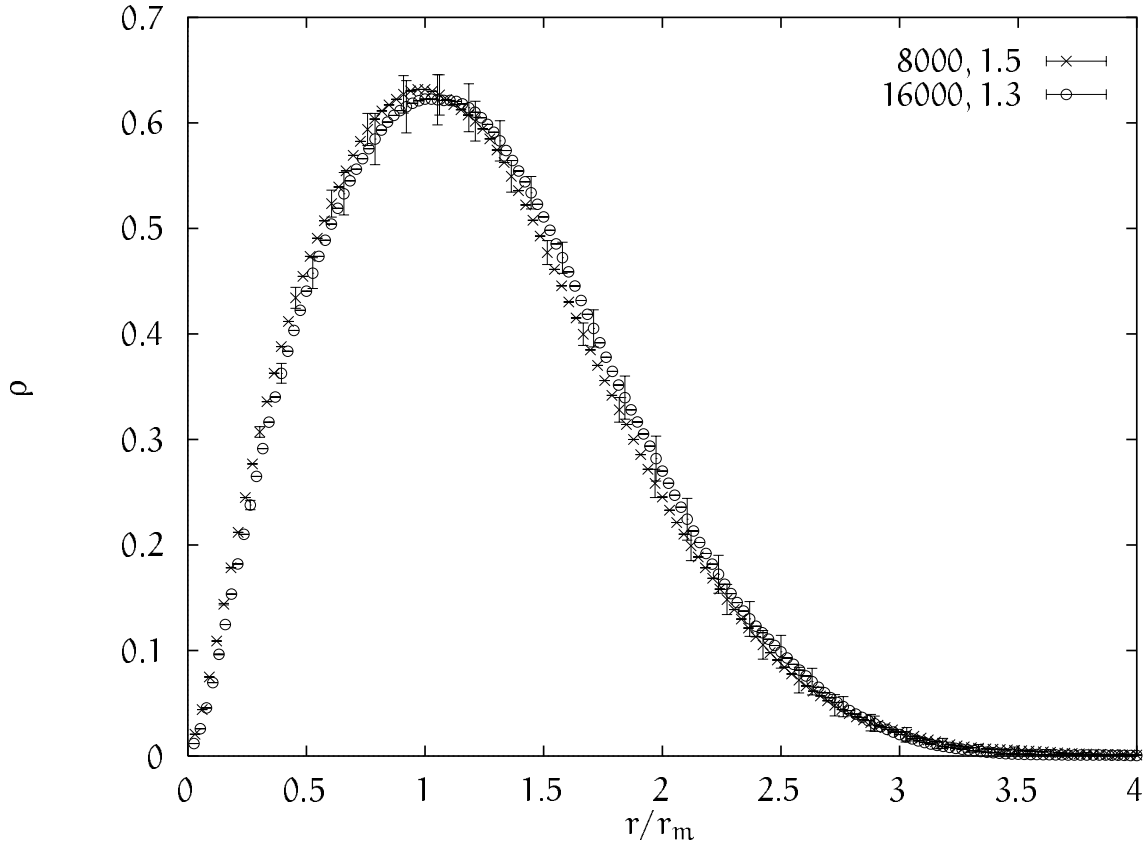

Jurkiewicz 1995b] gives 1.336(6), although the latter one uses a different criterion to determine the place of the transition.

We can use the matched pairs of ρ to define a scaling dimension $d_s$ by

$$N = \alpha\, r_m^{d_s}, \qquad (3.35)$$

where α and $d_s$ depend only on τ. Using non-lattice units, replacing the integer $r_m$ by $r_m/(\ell/\sqrt{10})$ where $r_m$ is now momentarily dimensionful, we can interpret (3.35) as

$$N \propto \left(\frac{r_m}{\ell}\right)^{d_s}, \qquad (3.36)$$

which shows that the scaling dimension characterizes the dimensionality of the system at small scales $\ell \to 0$ with $r_m$ fixed. Returning to lattice units, taking for $r_m$ the integer value of r where $N'(r)$ is maximal and using an assumed error of 0.5, these scaling dimensions would be 4.5(3), 3.8(2) and 4.0(1) respectively for figure 3.9 on page 58, figure 3.10 on page 59 and figure 3.11 on the preceding page. The scaling dimension between 16000 and 32000 simplices in figure 3.10 was 4.0(2). The largest errors in these figures probably arise due to the uncertainty in the values of $\kappa_2$ we need to take to get matching curves, i.e. to get the same value of $\tau$. As we do not have data for a continuous range of $\kappa_2$ values, we have to make do with what seems to match best from the values we do have. Nevertheless, the values of $d_s$ far away from the transition are strikingly close to 4 when compared to the values of the global dimension d, which are 7.8 and 2.0 for figures 3.9 on page 58 and 3.11 on the preceding page.

The scaling form (3.30) is in general incompatible with (3.33), except for $d_s =$ d. The evidence for $d_s = 4$ points instead to a scaling behavior of the form

$$N'(r) = r_m^3 f(\frac{r}{r_m}, \tau),  \tag{3.37}$$

with $f(x, \tau) = \alpha(\tau)\rho(x, \tau)$. This further suggests scaling of the volume $V'(r)$ at distance r with an effective volume $V_{eff}$ per simplex (cf. (3.15)) depending only on $\tau$.

A precise definition of the scaling form $\rho(x, \tau)$ may be given by

$$\rho(x, \tau) = r_m p(r_m x), \ \kappa_2 = \kappa_2(N) \ \text{such that} \ r_m p(r_m) = \tau, \ N \to \infty,  \tag{3.38}$$

where we used $\tau = \rho_m$ for illustration. Intuitively one would expect the convergence to the scaling limit (3.38) to be non-uniform, with the large x region converging first, and there may be physical aspects to such non-uniformity.

The scaling analogue of running curvature (3.24) is given by

$$\widetilde{R}(x) \equiv r_m^2 R_{eff}(xr_m) = \frac{24}{x^2} \frac{3 - d\ln\rho/d\ln x}{5 - d\ln\rho/d\ln x}.  \tag{3.39}$$

Figure 3.12 on the following page shows this function for the matched pair of figure 3.10 on page 59 in the transition region. We have also included the curve for $\kappa_2 = 1.23$ at N = 16000. The curves do not match in the region around x = 0.5 and $R_{eff}(r)$ is apparently a sensitive quantity for scaling tests. Still, figure 3.12 on the following page suggests reasonable matching for a value of $\kappa_2(16000)$ somewhere between 1.22 and 1.23 and even the steep rise as far as shown appears to be scaling approximately, with $\widetilde{R}(16000)$ somewhat below $\widetilde{R}(8000)$. We find similar scaling behavior for the matched pair in the elongated phase. The number of our $\kappa_2$ values in the crumpled phase is too limited to be able to draw a conclusion there.

The steep rise appears to move to the left for increasing N (a scaling violation). A most interesting question is whether the onset of the rise (e.g. the x value where $\widetilde{R} = 50$) continues to move towards $x = 0$ as $N \to \infty$. Such behavior is needed for a classical region to open up from $x$ around 1 towards the origin $x = 0$. In other words, the 'Planckian regime' would have to shrink in units of the size $r_m$ of the 'universe'. Then $r_m^2 R_V$ could be defined as a limiting value of $\widetilde{R}$, $x \to 0$.

Figure 3.12. Scaling form $r_m^2 R_{eff}(xr_m)$ versus $x = r/r_m$ near the the transition, for $N = 8000$, $\kappa_2 = 1.18$ (middle) and $N = 16000$, $\kappa_2 = 1.22$, $1.23$ (lower and upper). The hypothetical limiting form corresponding to $S^4$ is also shown.

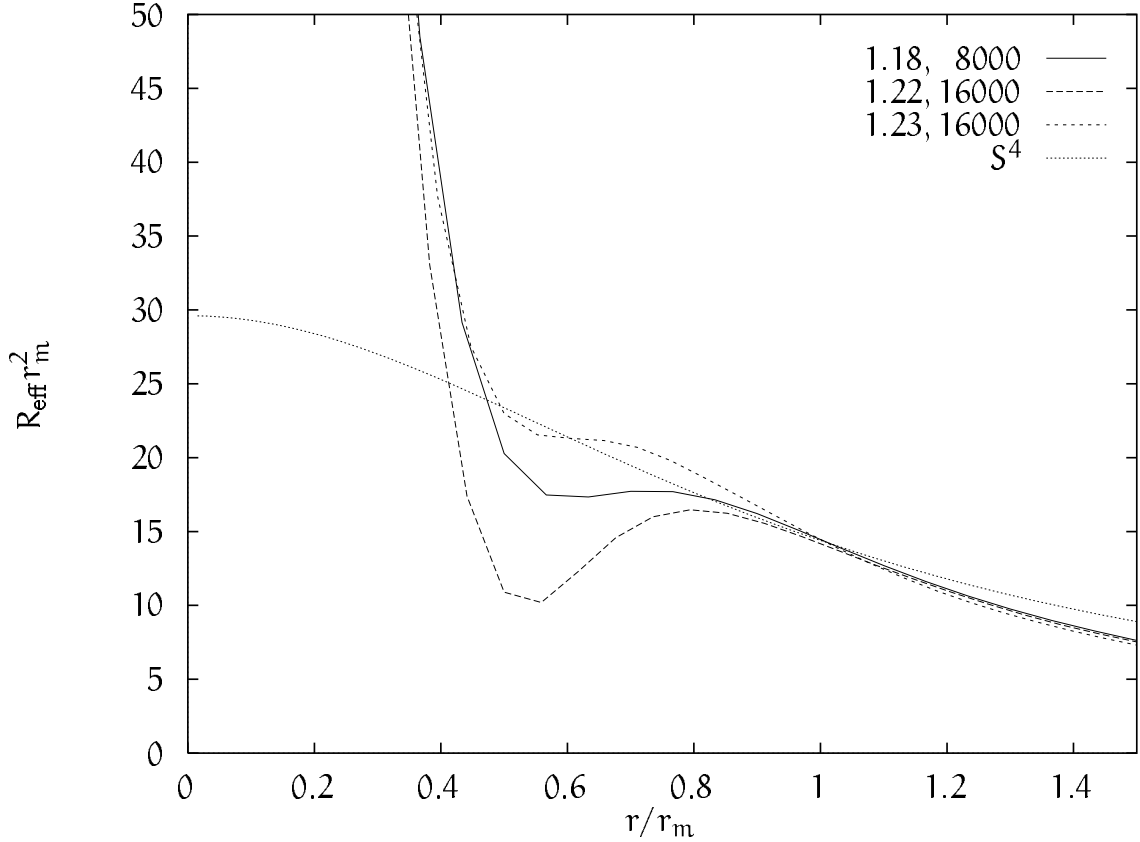

Since we are looking for classical behavior in the transition region it is instructive to compare with the classical form of $\widetilde{R}$ corresponding to the sphere $S^4$, for which $\rho = \rho_m \sin^3 \theta$, $\theta = \pi x/2$,

$$\widetilde{R} = \frac{18\pi^2}{\theta^2} \frac{\tan\theta - \theta}{5\tan\theta - 3\theta} = 3\pi^2 (1 - \frac{13\pi^2}{120} x^2 + \cdots). \qquad (3.40)$$

This function is also shown in figure 3.12. Our current data are evidently still far from the hypothetical classical limiting form (3.40).

## 3.5 Summary

The average number of simplices $N'(r)$ at geodesic distance $r$ gives us some basic information on the ensemble of Euclidean spacetimes described by the partition function (3.1). The function $N'(r)$ is maximal at $r = r_m$ and $r_m N'(r)/N$ shows scaling when plotted as a function of $r/r_m$. We explored a classical definition of curvature in the small to intermediate distance regime based on spacetime dimension four, the effective curvature $R_V$. We also explored a description at global distances by comparing $N'(r)$ with a sphere of effective dimension $d$. Judged by eye, the effective curvature fits and effective dimension fits give a reasonable description of $N'(r)$ in an appropriate distance regime (figures 3.2–3.4 on pages 50–52).

The resulting $R_V$ depends strongly on the fitting range, which led us to an explicitly distance dependent quantity, the 'running' curvature $R_{eff}(r)$. This dropped rapidly from lattice values of order of the Regge curvature at $r = 0$ to scaling values of order of the $R_V$ found in the effective curvature fits at intermediate distances. A preliminary analysis of scaling behavior then suggested the possibility of a classical regime with a precise definition of $r_m^2 R_V$ in the limit of large $N$. We shall now summarize the results further, keeping in mind the ambiguity in $R_V$ as derived from the fits to $N'(r)$.

For small $\kappa_2$ the effective curvature is negative. Furthermore the system resembles a $d$-sphere with very large dimension $d$ which increases with the volume. This suggests that no matter how large the volumes we use, there will never be a region of $r$ where the power law $V(r) \propto r^d$ gives good fits over large ranges of $r$. In other words, the curves for $d \ln V/d \ln r$ in figure 3.7 on page 55 will never have a plateau. This behavior is consistent with that of a space with constant negative curvature, where the volume rises exponentially with the geodesic distance for distances larger than the curvature radius and if we look at large enough scales the intrinsic fractal dimension equals infinity. The resulting euclidean spacetime cannot be completely described as a space with constant negative curvature as such a space with topology $S^4$ does not exist, and finite size effects take over at still larger distances.

At the transition the spacetime resembles a four-dimensional sphere with small positive effective curvature, up to intermediate distances.

For large $\kappa_2$ the system has dimension 2. In this region it appears to behave like a branched polymer, which has an intrinsic fractal dimension of 2 [David 1992]. Moving away from the transition, the curvature changes much more rapidly than in the small $\kappa_2$ phase and the effective curvature radius $r_V = \sqrt{12/R_V}$ soon becomes of the order of the lattice distance. A priori, two outcomes seem plausible. In the first, the system collapses and $r_V$ becomes of order 1 in lattice units, reflecting the unboundedness of the continuum action from below. In the second, the spacetime remains four-dimensional and $r_V$ can still be tuned to large values in lattice units, but very small compared to the global size $r_m$, for a sufficiently large system. So far the second outcome seems to be favored, for two reasons. Firstly, the function $N'(r)$ looks convex for $r \leqslant 6$ slightly above the transition, e.g. for $\kappa_2 = 1.24$ at 16000 simplices. In other words, the linear behavior as seen in figure 3.4 on page 52 does not set in immediately above the transition. Secondly, the system shows scaling and the scaling dimension $d_s$ as defined in (3.35) is approximately 4 even far into the elongated phase, indicating four-dimensional behavior at small scales.

High and low dimensions in the crumpled and elongated phases with the value four at the transition were reported earlier in [Agishtein & Migdal 1992b]. This dimension was apparently interpreted as a small scale dimension, whereas instead we find a small scale dimension of four in all phases. Similar results are also found in the Regge calculus approach to quantum gravity [Hamber 1993, Beirl *et al.* 1994a], where one also finds two phases, a strong (bare Newton) coupling phase with negative curvature and fractal dimension four, and (using an $R^2$ term in the action for stabilization) a weak coupling phase with fractal dimension around two.

## 3.6 Discussion

The evidence for scaling indicates continuum behavior. This brings up a number of issues which need to be addressed in a physical interpretation of the model.

One guideline in this work is the question whether there is a regime of distances where $V'(r) = V_{eff}N'(r)$ behaves classically for suitable bare Newton constant $G_0$. The connection with classical spacetime can be strengthened by identifying the geodesic distance $r$ with a cosmic time $t$ and $V'(r)^{1/3}$ with the scale factor $a(t)$ in a Euclidean Robertson-Walker metric. The classical action for $a(r)$ is given by

$$ S = -\frac{\pi}{8G} \int dr \, [a \left(\frac{da}{dr}\right)^2 + a] + \lambda \left(2\pi^2 \int dr \, a^3 - V\right), \qquad (3.41) $$

where $\lambda$ is a Lagrange multiplier enforcing a total spacetime volume $V$. It plays the role of a cosmological constant which is just right for getting volume $V$. For positive $G\lambda$ the solution of the equations of motion following from (3.41) is $a = r_0 \sin r/r_0$, which represents $S^4$ with $R = 12/r_0^2 = 192\pi G\lambda$. For negative $G\lambda$ the solution is $a = r_0 \sinh r/r_0$ which represents a space of constant negative curvature $R = -12/r_0^2 = 192\pi G\lambda$, cut off at a maximal radius to get total volume $V$.

The form (3.41) serves as a crude effective action for the system for intermediate distances around the maximum in $V'(r)$ at $r_m$, and couplings $\kappa_2 \propto G_0^{-1}$ in the transition region. At larger distances the fluctuations of the spacetimes grow, causing large baby universes and branching, and averaging over these may be the reason for the asymmetric shape of $V'(r)$. Because of this the Robertson-Walker metric cannot give a good description at these distances.

Intuitively one expects also strong deviations of classical behavior at distances of order of the Planck length $\sqrt{G}$, assuming that a Planck length exists in the model. A proposal for measuring it will be put forward in chapter five. The steep rise in the running curvature $R_{eff}(r)$ at smaller $r$ indeed suggests such a Planckian regime. It extends to rather large $r$ but it appears to shrink compared to $r_m$ as the lattice distance decreases, i.e. $N$ increases for a given scaling curve labelled by $\tau$. This suggests that the Planck length goes to zero with the lattice spacing, $G/r_m^2 \rightarrow 0$ as $N \rightarrow \infty$ at fixed $\tau$. This does not necessarily mean that the Planck length is of order of the lattice spacing, although this is of course quite possible. The theory, however, may also scale at Planckian distances and belong to a universality class. It might then be 'trivial'.

At this point it is instructive to recall other notorious models with a dimensionful coupling as in Einstein gravity, the four-dimensional nonlinear sigma models. The lattice models have been well studied, in particular the $O(4)$ model for low energy pion physics (see for example [Lüsher & Weisz 1988, Lüsher & Weisz 1989, Heller 1994]). It has one free parameter $\kappa = \ell^2 f_0^2$ which corresponds to the renormalized dimensionful coupling $f^2$; $f$ is the pion decay constant or the electroweak scale in the application to the Standard Model. With this one bare parameter it is possible to tune *two* quantities, $f/m$ and $\ell f$, where $m$ is the mass of the sigma particle or the Higgs particle. This trick is possible because the precise value of $\ell$ is unimportant, as long as it is sufficiently small.[*] In the continuum limit $\ell f \rightarrow 0$, however, triviality takes its toll: $m/f \rightarrow 0$ and the model becomes noninteracting.

The analogy $f^2 \leftrightarrow 1/G$, $m \leftrightarrow r_m^{-1}$ suggests that we may be lucky and there is a scaling region in $\kappa_2$–$N$ space, for a given scaling curve (given $\tau$), where the theory has universal properties and where we can tune $G/r_m^2$ to a whole range of desired values. Taking the scaling *limit* however might lead to a trivial theory with $G/r_m^2 = 0$. In case this scenario fails it is of course possible to introduce more parameters, e.g. as in $R^2$ gravity, to get more freedom in the value of $G/r_m^2$. This then raises the question of universality at the Planck scale.

---

[*]It is good to keep the numbers in perspective: for example in the Standard Model $f = 250$ GeV and for a Higgs mass $m = 100$ GeV or less, $\ell$ is 15 orders of magnitude smaller than the Planck length, or even much smaller.

We really would like to replace $1/r_m^2$ by $R_V$ in the reasoning in the previous paragraph, since we view $R_V$ as the local classical curvature, provided that a classical regime indeed develops as $N \to \infty$.

Next we discuss the nature of the elongated phase. Even deep in this phase we found evidence for scaling. Furthermore, for given scaling curve, increasing $N$ means decreasing $\kappa_2$. Hence, increasing $N$ at fixed $\kappa_2$ brings the system deeper in the elongated phase. This leads to the conclusion that there is nothing wrong with the elongated phase. It describes very large spacetimes which are two dimensional on the scale of $r_m$ but not necessarily at much smaller scales. It could be effectively classical at scales much larger than the Planck length but much smaller than $r_m$.

This reasoning further suggests that $N$ and $\kappa_2$ primarily serve to specify the 'shape' of the spacetime. The tuning $\kappa_2 \approx \kappa_2^c$ is apparently not needed for obtaining criticality but for obtaining a type of spacetime. The peak in the susceptibility of the Regge curvature could be very much a reflection of shape dependence. Most importantly, this suggests that the physical properties associated with general coordinate invariance will be recovered automatically in four-dimensional dynamical triangulation, as in two dimensions with fixed topology.[*]

---

[*]A field theory analogue is $\mathbb{Z}_n$ lattice gauge theory, which for $n \geqslant 5$ has been found to possess a Coulomb phase, a whole region in bare parameter space with massless photons; see for example [Frölich & Spencer 1982, Alessandrini 1983, Alessandrini & Boucaud 1983].

# Chapter Four

# Correlations

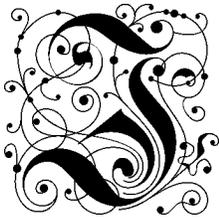 N THE dynamical triangulation model of 4-dimensional Euclidean quantum gravity we measure two-point functions of the scalar curvature as a function of the geodesic distance. To get the correlations it turns out that we need to subtract a squared one-point function which, although this seems paradoxical, depends on the distance. At the transition and in the elongated phase we observe a power law behaviour, while in the crumpled phase we cannot find a simple function to describe it.

## 4.1 Introduction

In the dynamical triangulation model of four dimensional euclidean quantum gravity the path integral over metrics on a fixed manifold is defined by a weighted sum over all ways to glue four-simplices together at the faces [Ambjørn & Jurkiewicz 1992, Agishtein & Migdal 1992a]. This idea was first formulated in [Weingarten 1982], using hypercubes instead of simplices.

The partition function of the model at some fixed volume of $N$ simplices is

$$Z(N, \kappa_2) = \sum_{\mathcal{T}(N_4 = N)} \exp(\kappa_2 N_2). \qquad (4.1)$$

The sum is over all ways to glue $N$ four-simplices together, such that the resulting complex has some fixed topology which is usually (as well as in this chapter) taken to be $S^4$. The $N_i$ are the number of $i$-simplices in this complex. $\kappa_2$ is a coupling constant which is proportional to the inverse of the bare Newton constant: $\kappa_2 \propto G_0^{-1}$.

It turns out that the model has two phases. For low $\kappa_2$ the system is in a crumpled phase, where the average number of simplices around a vertex is large and the average distance between two simplices is small. In this phase the volume within a distance $r$ appears to increase exponentially with $r$, a behaviour like that of a space with constant negative curvature. At high $\kappa_2$ the system is in an elongated phase and resembles a branched polymer. As is the case with a branched polymer, the (large scale) internal fractal dimension is two.

The transition between the two phases occurs at a critical value $\kappa_2^c$ which depends somewhat on N. This transition appears to be a continuous one, making a continuum limit possible [Agishtein & Migdal 1992b, Ambjørn *et al.* 1993c, Catterall *et al.* 1994a] (see chapter two). At the transition, the space behaves in several respects like the four dimensional sphere (see chapter three).

## 4.2  Curvature and volume

In the Regge discretization of general relativity, all the simplices are pieces of flat space. The curvature is concentrated on the subsimplices of codimension two, in our case the triangles. On these triangles it is proportional to a two-dimensional delta function. From the definition of curvature as the rotation of a parallel transported vector, one can find the integrated curvature over a small region $V_\epsilon(\triangle)$ around such a triangle

$$\int_{V_\epsilon(\triangle)} R\sqrt{g}\, dx = 2A_\triangle \delta_\triangle, \qquad (4.2)$$

where $A_\triangle$ is the area of the triangle and $\delta_\triangle$ is the deficit angle around the triangle (see e.g. [Hamber 1986]). The deficit angle around a triangle is the angle which is missing from $2\pi$

$$\delta_\triangle = 2\pi - \sum_{d \in \{S(\triangle)\}} \theta_d \qquad (4.3)$$

where $\{S(\triangle)\}$ are the simplices around the triangle and $\theta_d$ is the angle between those two faces of the simplex that border the triangle. The deficit angle $\delta_\triangle$ can be negative.

In dynamical triangulation, all the simplices have the same size and shape and the deficit angle is a simple function of the number $n_\triangle$ of simplices around the triangle. Then expression (4.2) reduces to

$$\int_{V_\epsilon(\triangle)} R\sqrt{g}\, dx = 2V_2(2\pi - \theta n_\triangle), \qquad (4.4)$$

where $\theta$ is the angle between two faces of a simplex, which for $D$ dimensions equals $\arccos(1/D)$, and $V_2 = A_\triangle$ is the now constant area of a two-simplex (that is a triangle).

For each triangle we can now define a local four-volume that belongs to the triangle by assigning that part of each adjoining simplex to it which is closer to the triangle than to any other. For our equal and equilateral simplices, this just results in $V_4/10$ per adjoining simplex with $V_4$ the volume of a four simplex. In other words this local volume is

$$V_\triangle = \int_{\Omega(\triangle)} \sqrt{g}\, dx = \frac{V_4}{10} n_\triangle, \tag{4.5}$$

where $\Omega(\triangle)$ is the region of space associated to that triangle. It is not clear what $V_\triangle$ would mean in the continuum limit. We define it here mainly to compare our results with other work on simplicial quantum gravity.

If we view the delta function curvature as the average of a constant curvature over the region $\Omega(\triangle)$, this constant curvature would be equal to

$$R_\triangle = \frac{20V_2}{V_4} \frac{2\pi - \theta n_\triangle}{n_\triangle}. \tag{4.6}$$

Because neither a constant term nor a constant factor is important for the behaviour of correlation functions, we will in the rest of this chapter for convenience define the curvature as

$$R_\triangle \equiv n_\triangle^{-1}, \tag{4.7}$$

and the local volume as

$$V_\triangle \equiv n_\triangle. \tag{4.8}$$

## 4.3 Two-point functions

One of the interesting aspects of the dynamical triangulation model one can investigate is the behaviour of two-point functions of local observables. In continuum language, such a correlation function of a local observable $O(x)$ at a distance $d$ would be defined by

$$\langle OO \rangle(d) = \left\langle \frac{\int dx\, \sqrt{g(x)} \int dy\, \sqrt{g(y)}\, O(x)O(y)\delta(d(x,y)-d)}{\int dx\, \sqrt{g(x)} \int dy\, \sqrt{g(y)}\, \delta(d(x,y)-d)} \right\rangle, \tag{4.9}$$

where the average is

$$\langle A \rangle = \frac{\int \mathcal{D} g_{\mu\nu} \, A \exp(-S[g_{\mu\nu}])}{\int \mathcal{D} g_{\mu\nu} \, \exp(-S[g_{\mu\nu}])}, \tag{4.10}$$

where $d(x, y)$ is the geodesic distance between the points $x$ and $y$ for a given metric $g_{\mu\nu}$. In other words, for each configuration (i.e. for each metric) we average over all pairs of points that have geodesic distance $d$. Obviously, it makes little sense to define such a correlation function for two fixed points $x$ and $y$. Because of general coordinate invariance, such a correlation could only depend on whether $x$ and $y$ coincide or not.

The discrete version of equations (4.9) and (4.10) becomes

$$\langle OO \rangle (d) = \left\langle \frac{\sum_{x,y} O(x)O(y)\delta_{d(x,y),d}}{\sum_{x,y} \delta_{d(x,y),d}} \right\rangle \tag{4.11}$$

with the average

$$\langle A \rangle = \frac{\sum_{\mathcal{T}} A \exp(\kappa_2 N_2(\mathcal{T}))}{\sum_{\mathcal{T}} \exp(\kappa_2 N_2(\mathcal{T}))}. \tag{4.12}$$

In figure 4.1 on the facing page we have plotted the correlation function of the curvature, with the square of its expectation value subtracted. Most of the data in this paper are for a volume $N = 16000$ simplices. The values of $\kappa_2$ correspond to a system in the crumpled phase ($\kappa_2 = 0.8$), near (but slightly below) the transition ($\kappa_2 = 1.22$) and in the elongated phase ($\kappa_2 = 1.5$).

Configurations were recorded every 10000 sweeps, where a sweep is defined as a number of accepted moves equal to the number of simplices $N$. For $\kappa_2 = 0.8$, 1.22 and 1.5 we used 16, 51 and 21 configurations respectively.

We define the geodesic distance between two triangles as the smallest number of steps between neighbouring triangles needed to get from one to the other. For this purpose, we define two triangles to be neighbours if they are subsimplices of the same four-simplex and share an edge. Other definitions of neighbour are conceivable. One such a definition would be to define two triangles to be neighbours if they share an edge, irrespective of whether they are in the same simplex. The one we use has the advantage that it is quite narrow and therefore results in larger distances.

Figure 4.1. The correlation function $\langle RR \rangle(d) - \langle R \rangle^2$ for various values of $\kappa_2$.

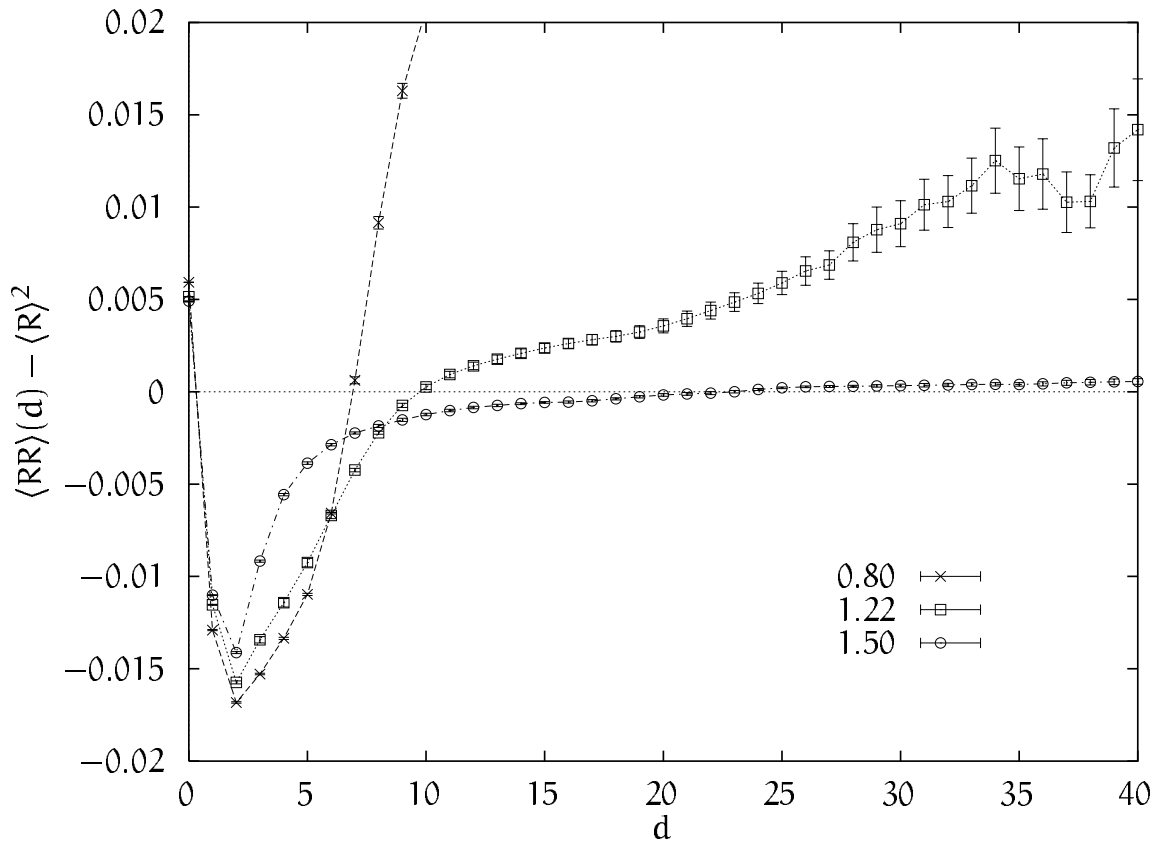

One thing is immediately striking, the correlation functions do not go to zero at long distances. To keep the short distance behaviour visible, the full range in the elongated phase has not been plotted, but we already see that also in this phase it crosses the zero axis and indeed this curve does eventually go to large ($\approx 0.02$) positive values.

The local volume $V_\triangle$ is proportional to the number of simplices $n_i$ around a triangle i. We see that in this model, this observable $V_\triangle$ is essentially the same as the scalar curvature. At first sight, one would therefore expect them to have the same behaviour. If one is positively correlated, the other one would also be positively correlated. Figure 4.2 on the next page shows the correlation of n. We see that quite the opposite is true. With few exceptions, n is positively correlated where $R = n^{-1}$ is negatively correlated and vice versa.

This behaviour is similar to that reported for the Regge calculus formulation of simplicial quantum gravity in [Hamber 1994]. There it is also found that the curvature correlations are positive and the volume correlations negative at large distances.

Figure 4.2. The correlation function $\langle nn \rangle(d) - \langle n \rangle^2$ for various values of $\kappa_2$.

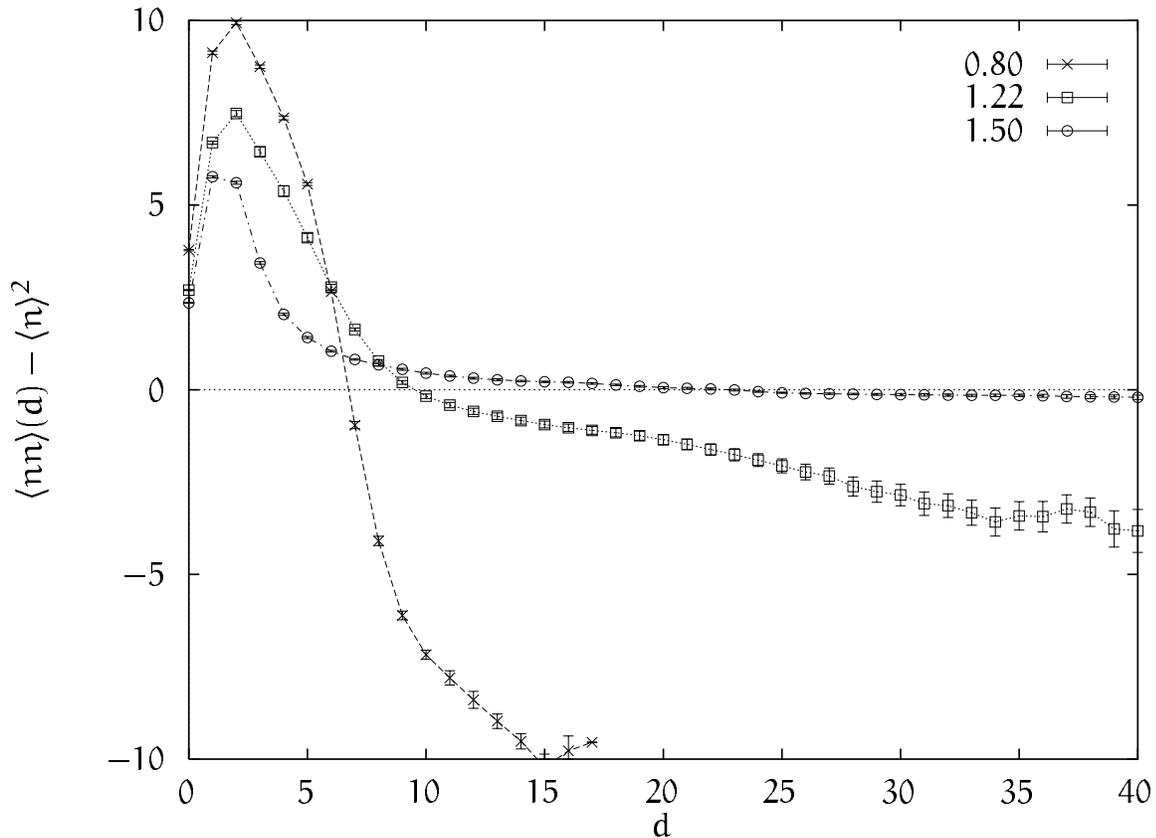

This difference in behaviour can be explained intuitively as follows. Because triangles with large $n$ have more neighbours, any random triangle will have a large chance to be close to a point with large $n$ and a small change to be close to a point with small $n$. So whatever the value of $n$ at the origin, the points nearby have large $n$ and the points far away have small $n$. The average $\langle nn \rangle$ will then be large at small distances and small at large distances. Because large $n$ means small R, the situation is reversed if we substitute R for $n$ in this discussion, qualitatively explaining figure 4.1 on the preceding page and figure 4.2.

At first sight one might conclude from this explanation that a point with large $n$ having many neighbours is just an artefact of the model. This is not true, however. Large $n$ corresponds to large negative curvature and also in the continuum a point with large negative curvature has a larger neighbourhood. To be more precise, the volume of $d$-dimensional space within a radius $r$ around a

Figure 4.3. The curvature as a function of the distance $\langle R \rangle(d)$ for various values of $\kappa_2$.

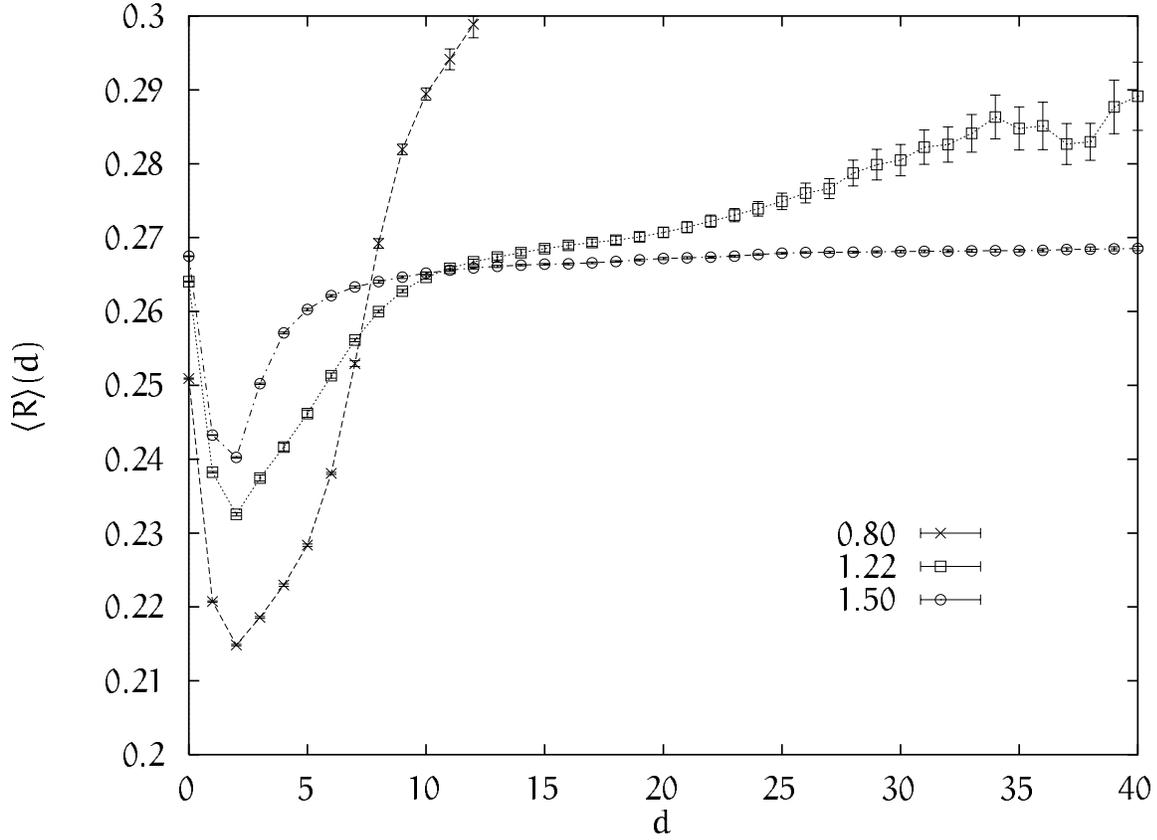

point with scalar curvature R equals

$$V(R) = C_d r^d (1 - \frac{R}{6(d+2)} r^2 + O(r^4)). \qquad (4.13)$$

## 4.4 Connected part

The above reasoning leads us to the somewhat unusual concept of a correlation function which does not depend on some observable at the origin. We define such a correlation as

$$\langle R \rangle(d) = \left\langle \frac{\sum\limits_{x,y} R_x \delta_{d(x,y),d}}{\sum\limits_{x,y} \delta_{d(x,y),d}} \right\rangle, \qquad (4.14)$$

Figure 4.4. Comparison between the correlation function $\langle RR\rangle(d)$ (upper curves of each pair) and the squared one-point function $\langle R\rangle(d)^2$ (lower curves) at various values of $\kappa_2$.

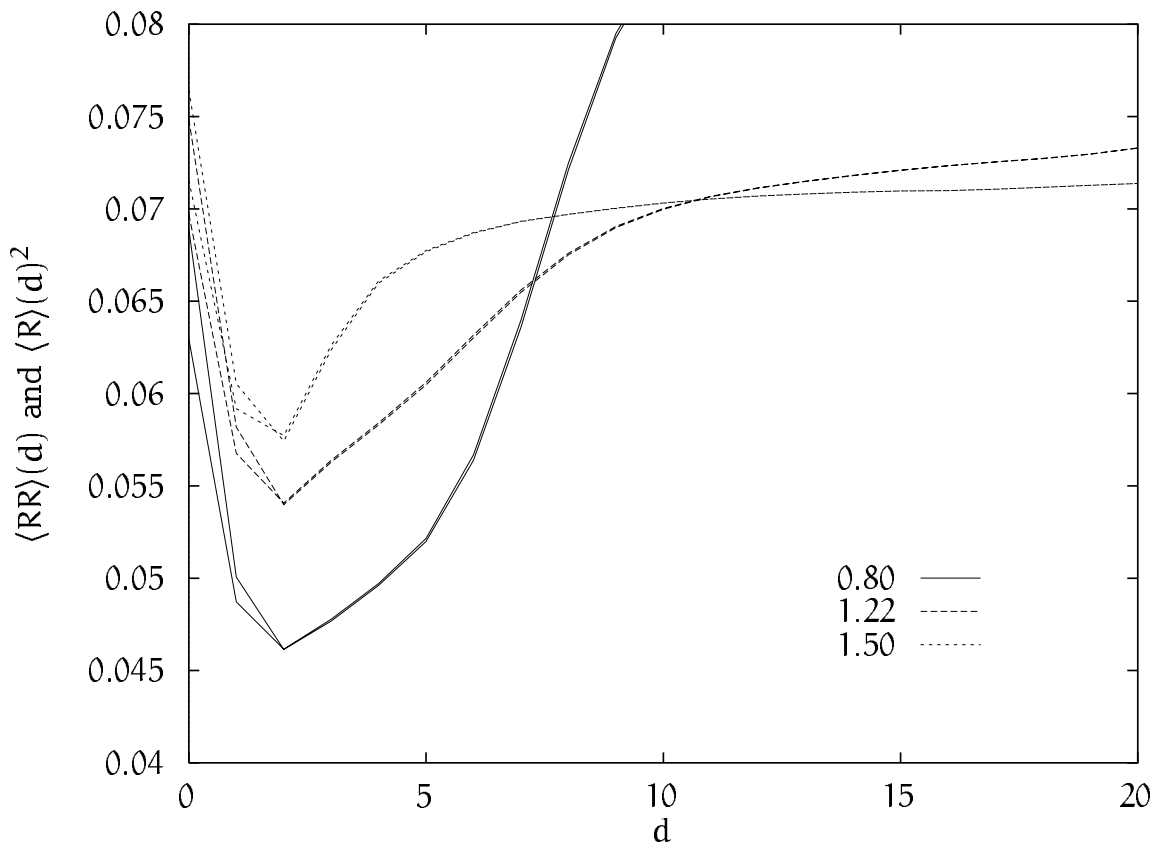

where $x$ and $y$ denote a triangle. In the more usual case of a quantum field theory on flat space this could never depend on the distance, but here it does. The reason is that we correlate functions of the geometry with the distance, which is itself a function of the geometry.

Figure 4.3 on the preceding page shows this correlation function. No average has been subtracted. The behaviour of this one-point function turns out to be very similar to that of the curvature correlation in figure 4.1 on page 71. This correlation function again shows that any particular point has a large chance to be in the neighbourhood of a point with low curvature, which can be simply explained with the fact that points with low curvature have more neighbourhood.

The same plot for $n$ (not shown) shows the opposite behaviour. At small distances it is larger than average, while at large distances it is smaller than average. This is rather obvious, because where $n$ is large, its inverse is small and vice versa.

Figure 4.5. Corrected correlation function $C_R(d)$ at various values of $\kappa_2$.

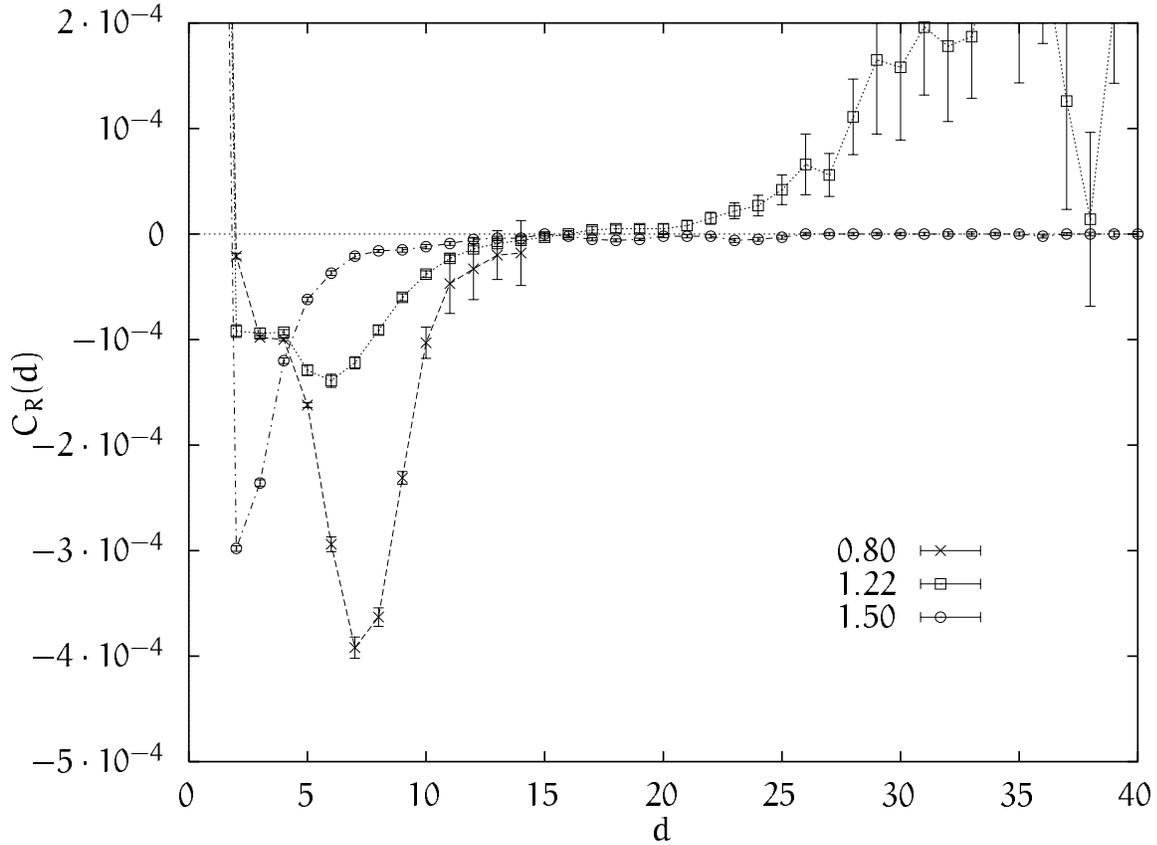

We can now investigate how much of the curvature correlation shown of figure 4.1 on page 71 is due to this effect. Figure 4.4 on the preceding page compares the curvature correlation $\langle RR \rangle(d)$ with the square of this one-point function. We see that, except at small distances, the two are indistinguishable on this scale. In other words, we have not been measuring any curvature correlations. All we have measured are correlations between the curvature and the geodesic distance.

It is now easy to explain the difference in behaviour between the curvature and the volume correlations. Because they are almost equal to the square of $\langle R \rangle(d)$ and $\langle n \rangle(d)$ respectively, they behave just like them. And as we just mentioned it is easy to understand that these have opposite behaviours.

The way to go now is to subtract the two things and see what real curvature correlations are left. This is similar to subtracting a disconnected diagram and keeping the connected part. We get the corrected correlation functions for the curvature

$$C_R(d) = \langle RR \rangle(d) - \langle R \rangle(d)^2, \tag{4.15}$$

Figure 4.6. Corrected correlation function $C_V(d)$ at various values of $\kappa_2$.

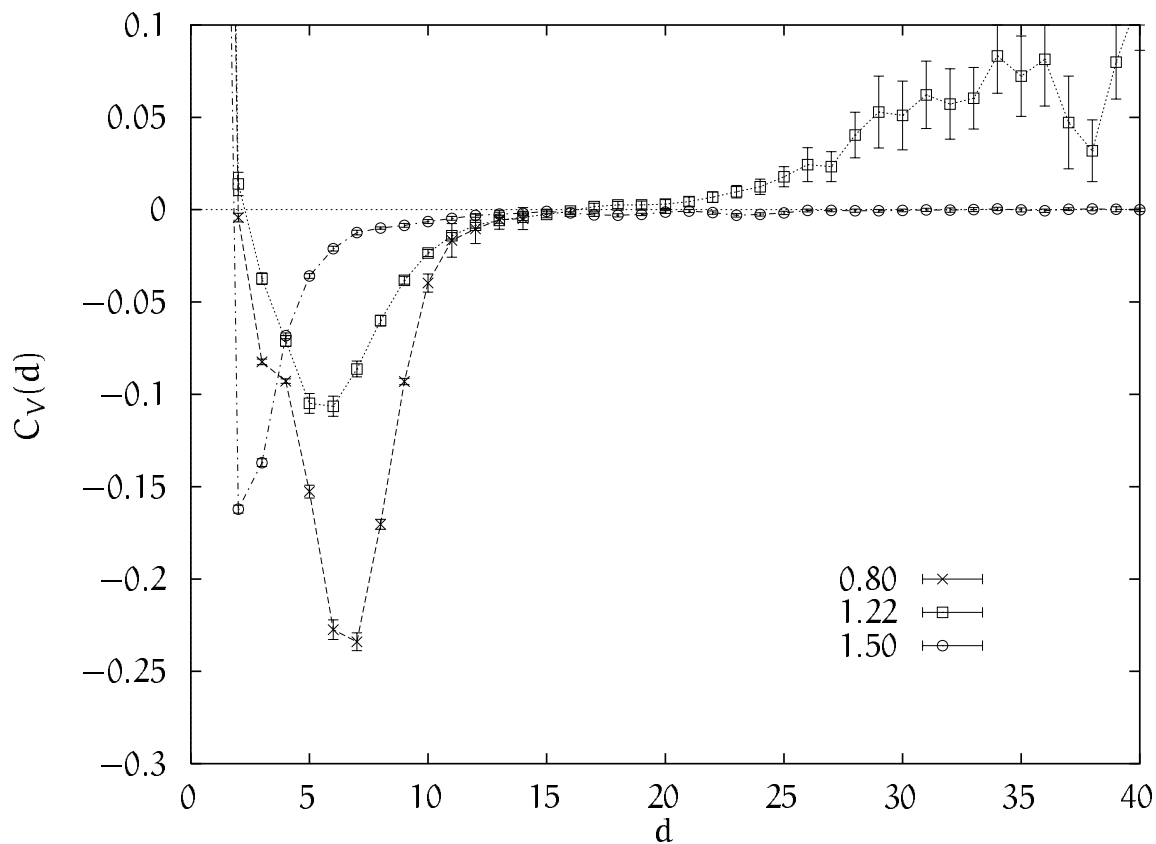

and the volume

$$C_V(d) = \langle nn \rangle(d) - \langle n \rangle(d)^2. \tag{4.16}$$

The results for the curvature are plotted in figure 4.5 on the preceding page and those for the volume in figure 4.6. The error bars were found by a jackknife method, each time leaving out one of the configurations in the calculation of $C_R(d)$ and $C_V(d)$. Now both correlations behave almost exactly the same. Note the large difference in scale between these figures and figures 4.1 and 4.2.

In the crumpled phase we were not able to fit $C_R(d)$ to a simple function. This is probably due to the fact that we cannot reach very large distances in this phase. Near the transition however it is possible to fit the correlation function to a power law decay, at not too small distances. This is shown in figure 4.7 on the facing page. In the region $9 \leqslant d \leqslant 18$ it fits nicely to $ad^b$ with the result

$$a = -0.5(2) \tag{4.17}$$

$$b = -4.0(2) \tag{4.18}$$

Figure 4.7. Power law fit to curvature correlation $C_R(d)$ near the phase transition at $N_4 = 32000$ and $\kappa_2 = 1.255$.

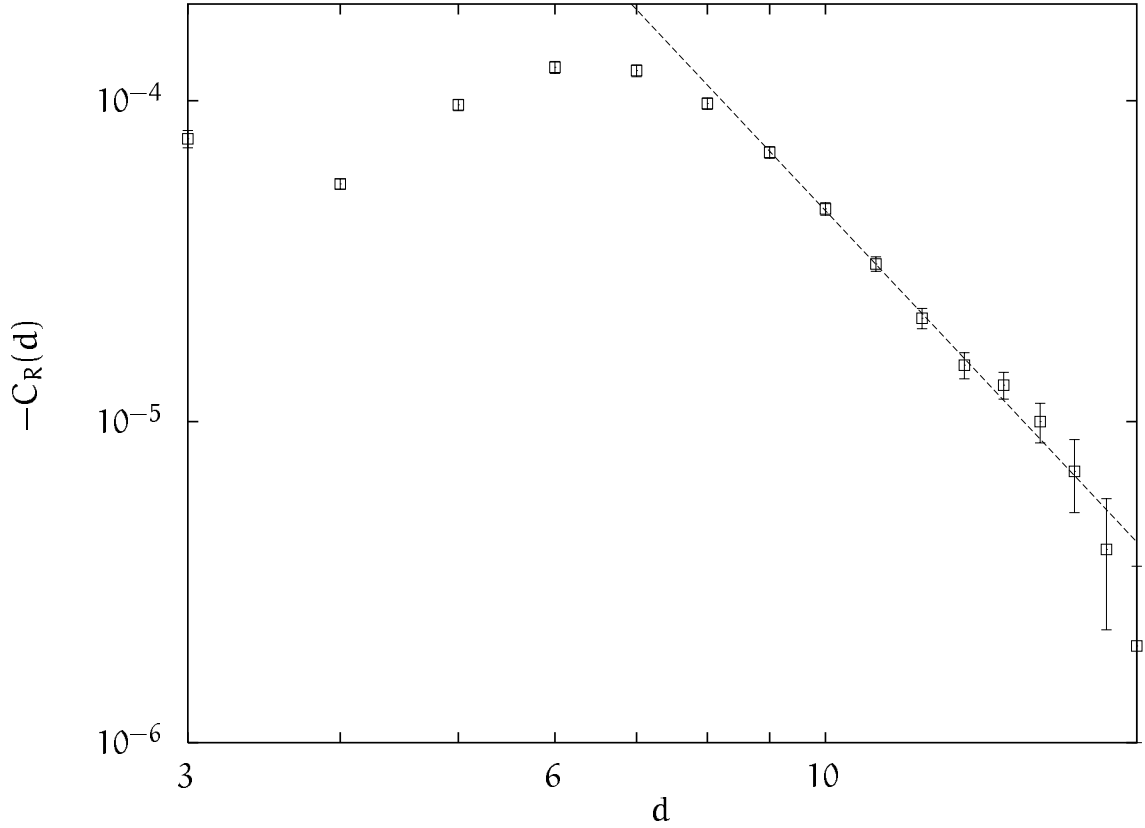

$$\chi^2 = 5 \text{ at } 8 \text{ d.o.f.} \tag{4.19}$$

This data was made at a volume of 32000 simplices, with $\kappa_2 = 1.255$. We used 65 configurations, which were recorded every 5000 sweeps.

This result should be taken with caution, however. One would really like to have a good fit over a larger range. To get some idea of the typical ranges involved, we consider the number of triangles at distance $d$

$$N'(d) = \left\langle \frac{\sum\limits_{x,y} \delta_{d(x,y),d}}{N_2} \right\rangle, \tag{4.20}$$

where $N_2$ is the number of triangles of the configuration. The corresponding quantity with 'triangles' replaced by 'four-simplices' was studied more closely in chapter three. The value of $d_m$, which is that $d$ for which $N'(d)$ has its maximum, is an indication of the distance at which finite size effects might become important.

Figure 4.8. Power law fit to curvature correlation $C_R(d)$ in the elongated phase at $\kappa_2 = 1.5$.

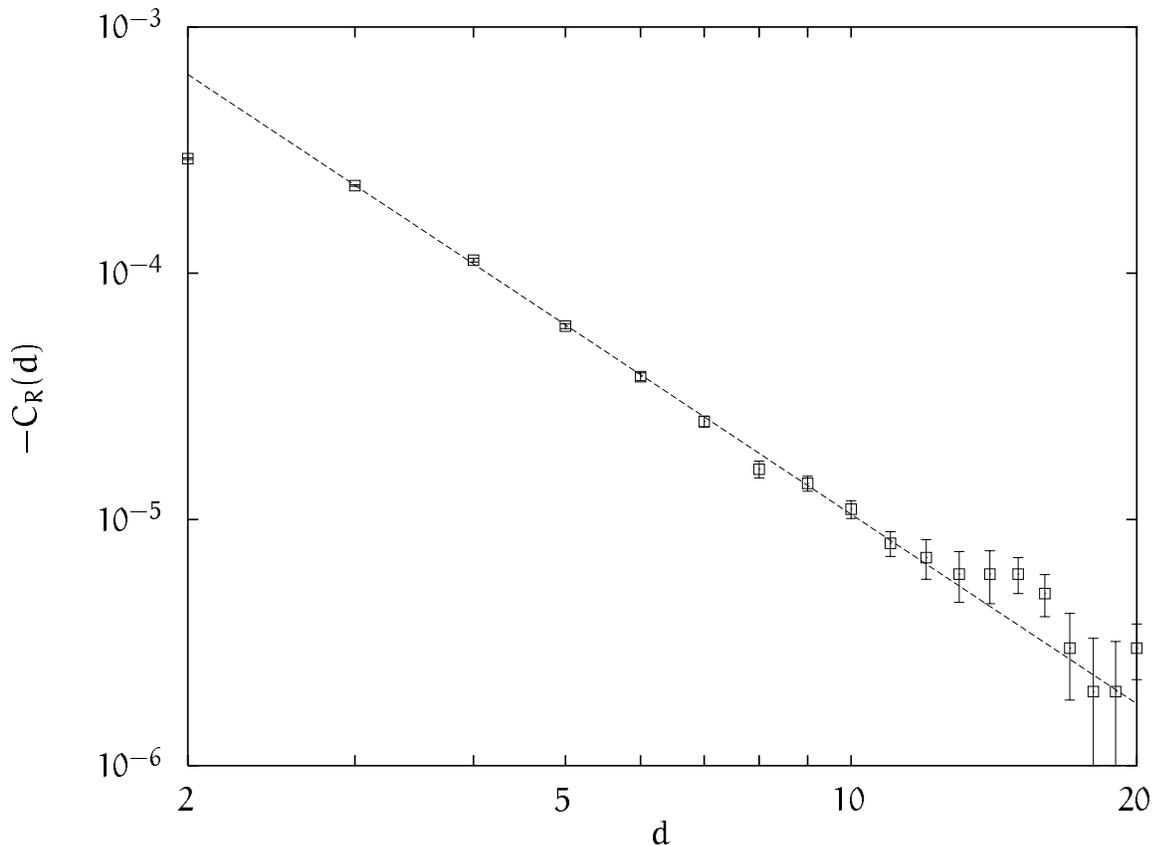

At $\kappa_2 = 1.255$, this $d_m$ is only 11, indicating that finite size effects might play a role in the power that was measured.

The situation is even better far in the elongated phase. Here a power law fits well, as can be seen in figure 4.8. This fit was done to the points $3 \leqslant d \leqslant 15$ and the parameters of this fit are

$$a = -0.0038(1) \tag{4.21}$$

$$b = -2.56(3) \tag{4.22}$$

$$\chi^2 = 17 \text{ at } 11 \text{ d.o.f.} \tag{4.23}$$

The value of d with maximum number of triangles was 32, so in this case we are in a region of small distances compared to the system size. This data was made from 23 configurations of 32000 simplices. We have also fitted the connected RR-correlation at other points in the elongated phase and at 16000 simplices. The power that emerged was within the errors equal to the one given above.

## 4.5 Discussion

We have investigated the behaviour of the curvature and volume correlation functions. It turned out that the naive correlation functions could be almost entirely described by a "disconnected part", which we therefore subtracted. The difference turns out to behave according to a power law in the elongated phase and near the transition. In the latter case the power is close to four, which is reminiscent of two graviton exchange.

In chapter three we explored the possibility of a semiclassical region near the transition, in which the system behaves like a four-sphere for not too small or large distances. To this end, we defined a scale dependent effective curvature. See figure 3.12 on page 62. For $\kappa_2$ near the transition the following picture emerged. At small distances, this effective curvature is large, indicating a Planckian regime. At intermediate distances there seems to be a semiclassical regime, where the space behaves like a four-sphere. The fluctuations around this approximate $S^4$ might then correspond to gravitons. We consider it therefore encouraging that the power b in (4.18) is compatible with four.

For the volumes in current use, the effective curvature shows that the semiclassical regime sets in at a distance roughly $2/3$ of $r_m$. Here, $r_m$ is the geodesic distance through the simplices where the number of simplices $N'(r)$ has its maximum. Similarly, $2/3$ of $d_m$ turns out to be the distance where the curvature correlations start to behave like $d^{-4}$. We like to think of this as a confirmation of the point of view sketched above.

Two-point functions of curvature and volume have been studied in the Regge calculus formulation of simplicial quantum gravity in references [Hamber 1994, Beirl *et al.* 1994b]. In these studies there are only results in what is called the well-defined phase of the Regge calculus approach, which corresponds to our crumpled phase. This makes it hard to do more than the qualitative comparison which was done in section 4.3 on page 69.

The curvature correlations have also been investigated in the continuum. In [Antoniadis & Mottola 1992] a theory is developed for the conformal factor in four-dimensional quantum gravity and from this the curvature correlation is calculated. The conformally invariant phase discussed in [Antoniadis & Mottola 1992] seems to correspond to the elongated phase in the dynamical triangulation model. Intuitively, this can be understood by visualizing large fluctuations in the conformal factor as generating many baby universes. Many baby universes is also a feature of the branched polymer like elongated phase of simplicial quantum gravity [Ambjørn *et al.* 1993b]. Furthermore, the conformally invariant phase is supposed to occur at very large distance scales. In chapter three we argued that the elongated phase also describes scales which are large compared to a typical physical curvature scale. In this conformally invariant phase a power law is predicted for the curvature correlations (see also [Antoniadis *et al.* 1992]), with a power of 0.7. Unfortunately, a direct comparison with [Antoniadis & Mottola 1992, Antoniadis *et al.* 1992] is not possible because in the continuum the correlation function is defined as a function of the distance in a fixed fiducial metric, a quantity that is not yet defined in our model.

# Chapter Five

# Binding

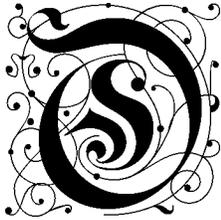NE OF the most obvious features of gravity is that it attracts objects to each other. A model of quantum gravity should therefore be able to show gravitational attraction. In this chapter we investigate this in the dynamical triangulation model by putting a scalar field on the configurations and looking for a non-zero binding energy. In the end we will see that attraction is indeed present.

## 5.1 Description of the model

We look at the behaviour of a free scalar field $\phi$ with bare mass $m_0$ in a quantum gravity background. The Euclidean action of this system in continuum language is a sum of a gravitational and a matter part

$$S = S[g] + S[g, \phi], \qquad (5.1)$$

$$S[g] = \frac{1}{16\pi G_0} \int d^D x \sqrt{g} \left(2\Lambda_0 - R\right), \qquad (5.2)$$

$$S[g, \phi] = \int d^D x \sqrt{g} \left(\frac{1}{2} g^{\mu\nu} \partial_\mu \phi \partial_\nu \phi + \frac{1}{2} m_0^2 \phi^2\right), \qquad (5.3)$$

where $\Lambda_0$ is the bare cosmological constant, $R$ is the scalar curvature and $G_0$ is the bare Newton constant.

We take $\phi$ as a test particle here, i.e. the back reaction of the field $\phi$ on the metric is not taken into account. This approach is often called the quenched approximation. In QCD this approximation turns out to give good results (see e.g. [Sharpe 1994]). In that case it is also called the valence quark approximation, because it neglects diagrams with internal quark (in our case $\phi$) loops. A continuum calculation of the gravitational attraction of a scalar field in this same quenched approximation was done in [Modanese 1995]. It is seen in other simulations [Ambjørn *et al.* 1993a] that including matter has little influence on the gravity sector of the theory.

We will use the following notation for expectation values of an observable A. On a fixed background geometry we can average over configurations of the matter field

$$\langle A \rangle_\phi = \frac{\int \mathcal{D}\phi \, A \exp(-S[g, \phi])}{\int \mathcal{D}\phi \, \exp(-S[g, \phi])}, \tag{5.4}$$

and we can average over metrics

$$\langle A \rangle_g = \frac{\int \mathcal{D}g \, A \exp(-S[g])}{\int \mathcal{D}g \, \exp(-S[g])}, \tag{5.5}$$

The quenched expectation value is then

$$\langle A \rangle = \langle \langle A \rangle_\phi \rangle_g. \tag{5.6}$$

We can now look at propagators in a fixed geometry. The one particle propagator, denoted by $G(x)$, is defined as

$$G(x) = \langle \phi_x \phi_0 \rangle_\phi, \tag{5.7}$$

where $0$ is an arbitrary point. The connected two particle propagator will then be the square of the one particle propagator

$$\langle \phi_x \phi_x \phi_0 \phi_0 \rangle_{\phi, \text{conn}} = G(x)^2. \tag{5.8}$$

Letting the metric fluctuate, we take the average of the propagators over the different metrics. Because of reparametrization invariance, the average $\langle G(x) \rangle_g$ will not depend on the place x. Therefore, we look at averages at fixed geodesic distance d

$$\langle G(d) \rangle_g = \frac{1}{Z} \int \mathcal{D}g \, \exp(-S[g]) \frac{\int d^D x \sqrt{g} \, G(x) \, \delta(d(x, 0) - d)}{\int d^D x \sqrt{g} \, \delta(d(x, 0) - d)}, \tag{5.9}$$

where $d(x, y)$ is the minimal geodesic distance between $x$ and $y$. For a massive particle, we expect the propagator (5.9) to fall off exponentially as

$$\langle G(d) \rangle_g \propto d^\alpha \exp(-md), \tag{5.10}$$

$$\langle G(d) \rangle_g^2 \propto d^{2\alpha} \exp(-2md), \tag{5.11}$$

with some power $\alpha$ and the renormalized mass $m$, which in general will not equal the bare mass $m_0$. These expressions neglect finite size effects and should probably be modified when looking at distances comparable to the size of the system. We will only use the data at relatively short distances, so this should not be important. The two particle propagator will behave similarly as

$$\langle G(d)^2 \rangle_g \propto d^\beta \exp(-E_c d), \tag{5.12}$$

where $E_c$ is the energy of the two particle compound. If this energy turns out to be less than two times the mass of a single particle, the difference can be interpreted as a binding energy between the particles. This would show gravitational attraction between them. Because the average of the square of a fluctuating quantity is always greater than the square of its average, it is obvious that $\langle G(d)^2 \rangle_g > \langle G(d) \rangle_g^2$. This does not yet imply anything about the way they fall off. In particular it is not guaranteed that $E_c < 2m$.

## 5.2   Implementation

We have run numerical simulations with both two and four dimensional dynamical triangulations.

In two dimensions the volume can be kept constant, and for fixed topology no parameters will be left. We used systems of 32000 and 64000 triangles with the topology of the two-sphere.

In four dimensions the analogue of the continuum gravitational action is

$$S[g] = \frac{1}{16\pi G_0} \int d^4 x \sqrt{g} \, (2\Lambda_0 - R) \tag{5.13}$$

$$\rightarrow \kappa_4 N_4 - \kappa_2 N_2, \tag{5.14}$$

where $N_2$ and $N_4$ are the number of triangles and four-simplices respectively. We used systems of about 32000 simplices and the topology of the four-sphere. To keep the number of simplices around the desired value, we added a quadratic term to the action as was described in section 2.2.

Figure 5.1. The two particle propagator and the square of the one particle propagator versus the geodesic distance for three different bare masses $m_0$ in two dimensions. The vertical scale is logarithmic.

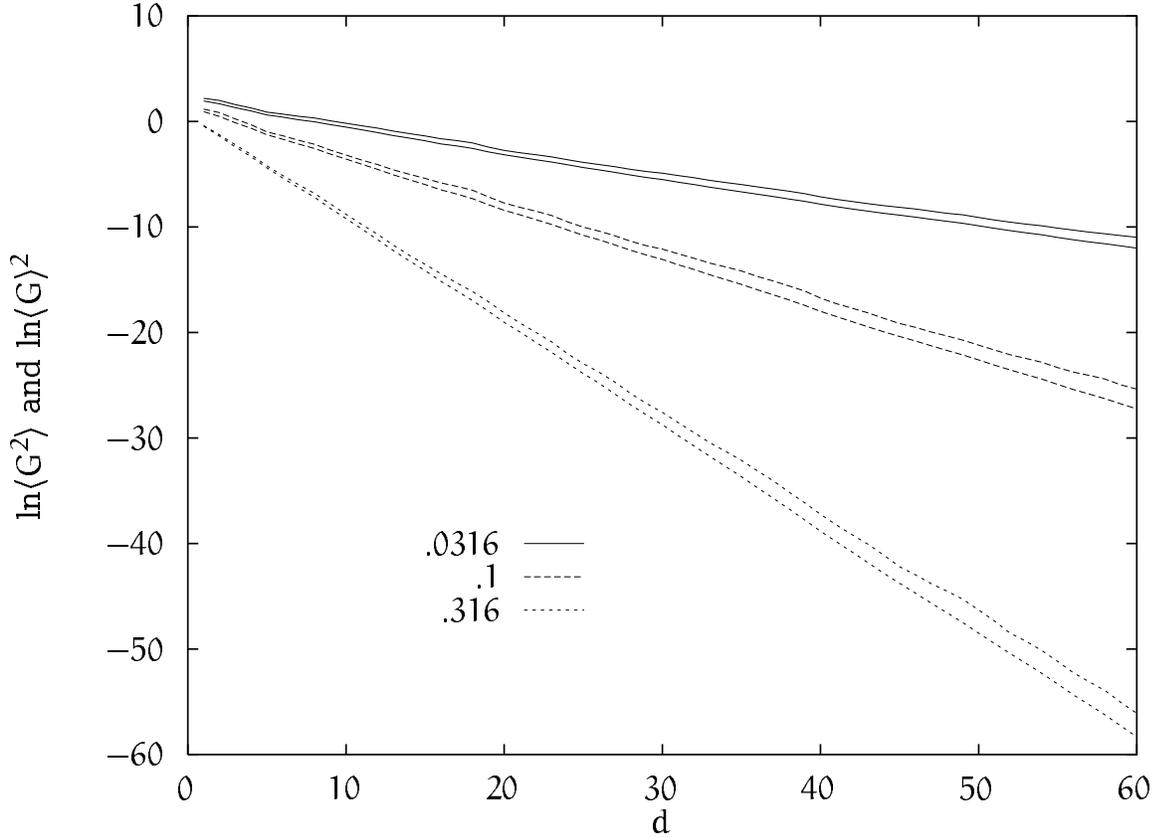

The parameter $\kappa_2$ is inversely proportional to the bare gravitational constant $G_0$. As we have shown in chapter two, we see indications of a second order phase transition as $\kappa_2$ varies.

On each dynamical triangulation configuration we then calculated the propagator

$$G(x) = (\Box^2 + m_0^2)_{0x}^{-1} \qquad (5.15)$$

of the scalar field, using the algebraic multigrid routine AMG1R5. The discrete Laplacian is defined as

$$(\Box^2)_{xy} = \begin{cases} d+1 & \text{if } x = y, \\ -1 & \text{if } x \text{ and } y \text{ are nearest neighbours}, \\ 0 & \text{otherwise}. \end{cases} \qquad (5.16)$$

Figure 5.2. The effective binding energy $E_b$ as a function of the geodesic distance for three different bare masses in two dimensions.

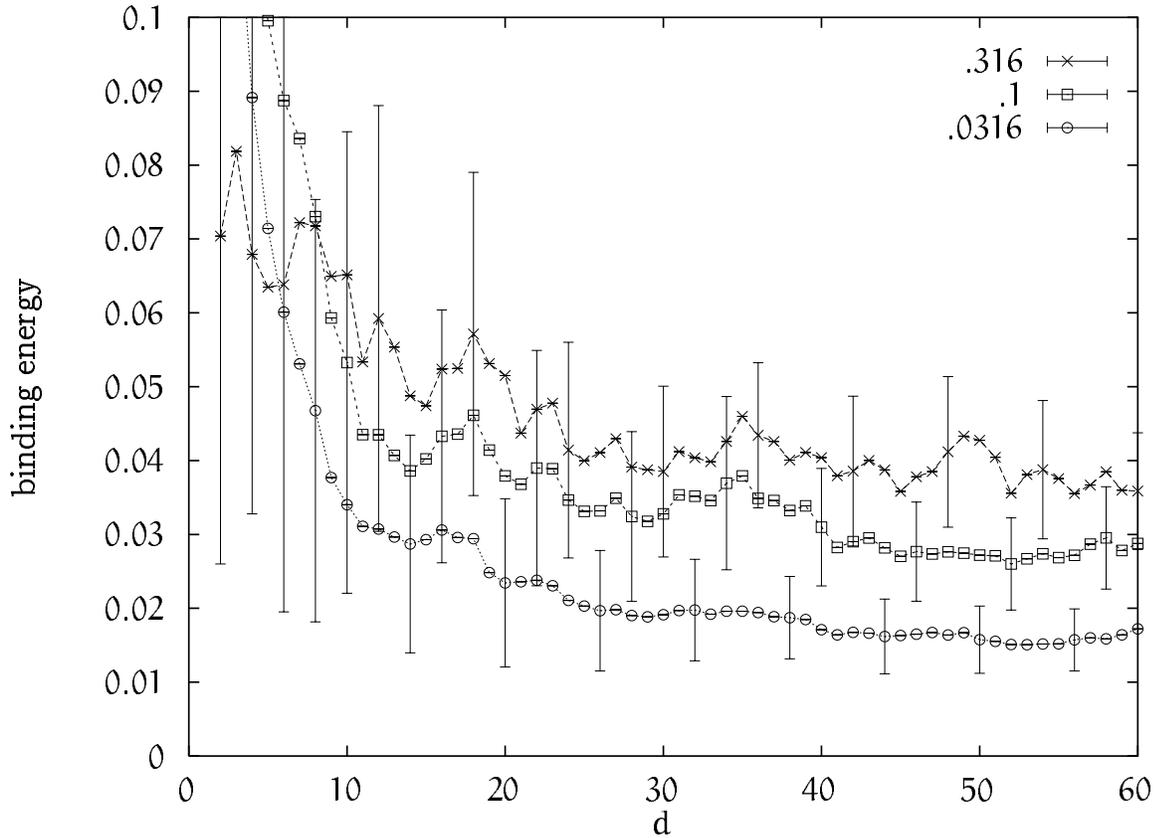

Eventually we hope to be able to extract the renormalized Newton constant $G_R$ near the critical value of the (inverse) bare Newton constant $\kappa_2$, for example according to the non-relativistic formula

$$E_b \equiv 2m - E_c = \frac{1}{4} G_R^2 m^5. \tag{5.17}$$

This formula is just the familiar energy $\alpha^2 m_e / 2$ of the hydrogen atom in the ground state, but with the gravitational parameters substituted as $\alpha \to G_R m^2$ and taking into account that we now have two particles of equal mass. As it is non-relativistic it may not suffice to fit the data.

## 5.3 Results

In figure 5.1 on the preceding page we see the results in two dimensions for three different bare masses. Each pair of lines corresponds to one bare mass. In each pair the upper line is $\ln\langle G(d)^2\rangle$ (the two particle propagator) and the lower line is $\ln\langle G(d)\rangle^2$ (the product of two single particle propagators).

Figure 5.3. As in figure 5.1, but in four dimensions. $\kappa_2 = 1.255$, which is very close to the transition.

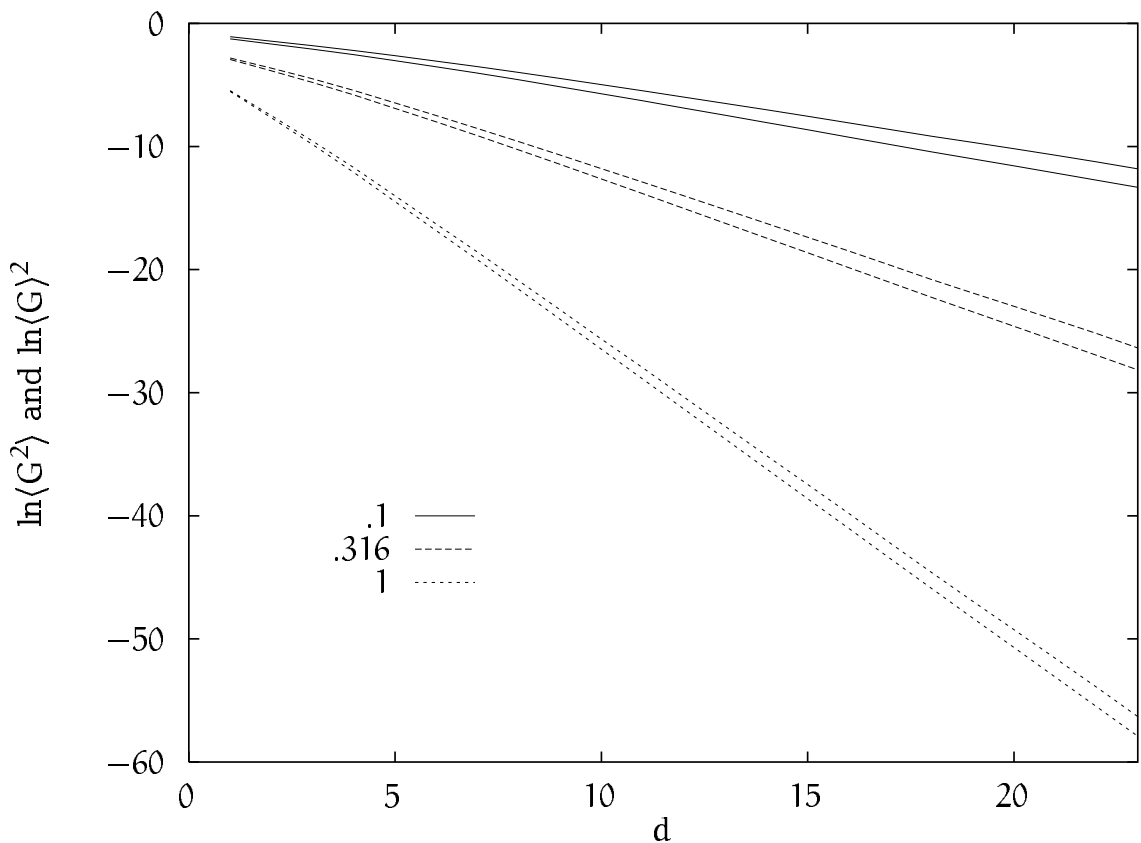

There is clearly a difference in slope between the lines in each pair. This shows that the energy of the two particle compound is less than two times the mass of a single particle and consequently that there is a positive binding energy between the particles.

In the two dimensional case, one might expect not to see any attraction, because of the absence of dynamics in the classical system. In the quantum case, however, a non-trivial theory results due to the conformal anomaly [Polyakov 1981].

Figure 5.3 shows similar data in four dimensions, with a coupling constant $\kappa_2 = 1.255$. This is very close to the phase transition. We used 144 configurations recorded every 5000 sweeps. For the three masses from lowest to highest we used 89, 120 and 34 origins respectively. We calculated $\langle G(d)\rangle_\phi$ by averaging it over all points at distance d from the origin. This corresponds to taking the propagator from a source that is not a single point, but a complete shell around the origin.

Figure 5.4. As in figure 5.2, but in four dimensions. Again, $\kappa_2 = 1.255$.

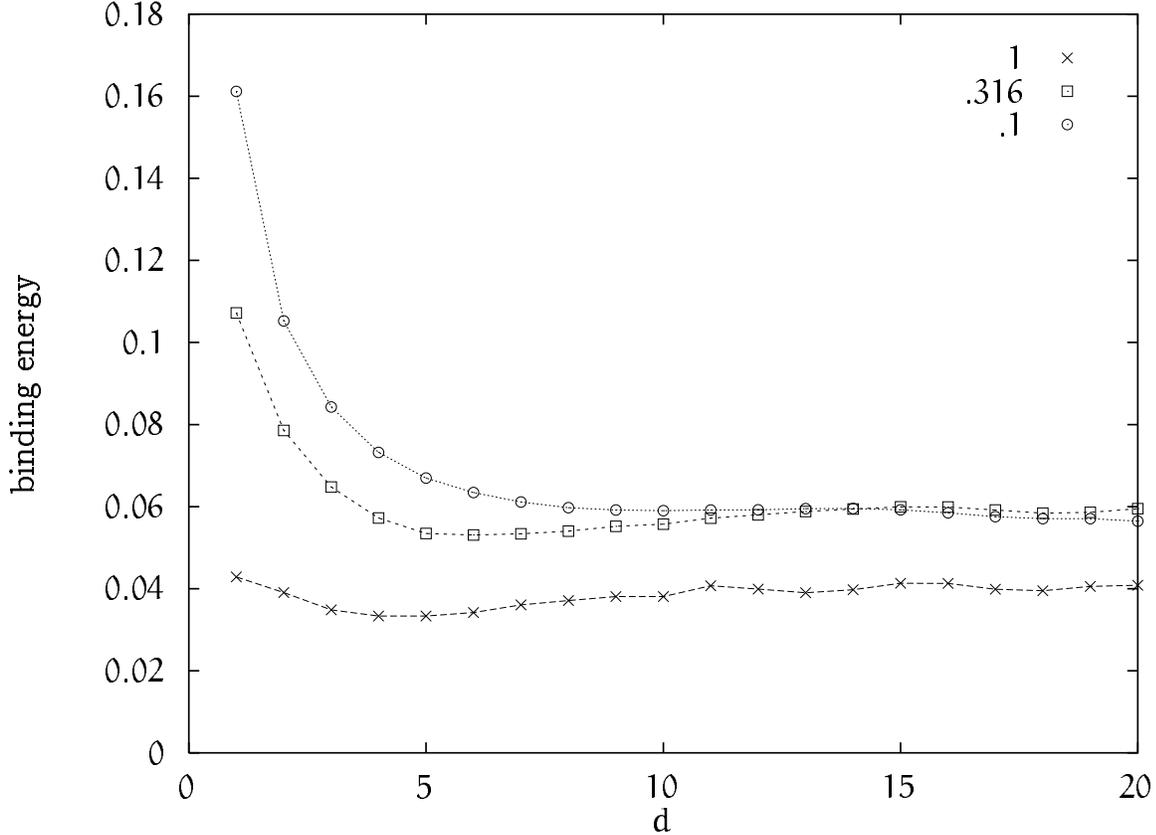

The use of what are called "smeared sources" can improve the data by increasing the contribution of the ground state and decreasing the contribution of the excited states.

As in the two-dimensional case there is a clear difference in slope. Using this data we can measure the renormalized mass $\mathfrak{m}$. The results are

| $\mathfrak{m}_0$ | $\mathfrak{m}$ | $\mathfrak{m}/\mathfrak{m}_0$ |
|---|---|---|
| 0.1 | 0.27 | 2.7 |
| 0.316 | 0.57 | 1.8 |
| 1 | 1.19 | 1.19 |

$$(5.18)$$

It was argued in [Agishtein & Migdal 1992b] that the physical mass should vanish at zero bare mass and that therefore the renormalization would be only multiplicative. Our data seem to show that the relation is more complicated. Increasing $\mathfrak{m}_0$ by a factor of $\sqrt{10} \approx 3.16$ increases $\mathfrak{m}$ by a factor of about 2.1.

The propagators curve downward towards the origin. This is somewhat unexpected, as it is interpreted as a sum over decaying exponentials. We do not know how to explain this phenomenon.

Using these data, we can now estimate the binding energy of the particles. From (5.11) and (5.12) we have

$$E_b \equiv 2m - E_c \qquad\qquad (5.19)$$

$$= d^{-1} \ln \frac{\langle G(d)^2 \rangle_g}{\langle G(d) \rangle_g^2}, \qquad d \to \infty. \qquad (5.20)$$

As we cannot use infinite distances, we will consider this expression at finite d and look whether the effective binding energy $E_b(d)$ becomes constant.

Figure 5.2 on page 85 shows this quantity as a function of the geodesic distance. The three curves again correspond to the three different bare masses. Although the result is not yet very accurate, it is clear that the binding energy goes to a non-zero value.

Figure 5.4 on the preceding page shows the data in the four dimensional simulations. These data are still somewhat preliminary. Nevertheless, this figure also clearly indicates a non-zero binding energy. Unfortunately, unlike the two-dimensional case, the correlation between the mass and the binding energy does not appear to be strictly positive. The lowest binding energy belongs to the highest mass.

Considering that the space looks like a four-sphere as argued in chapter three, it might be interesting to compare the propagators to those on a real four-sphere. This could perhaps reduce the finite size effects in the effective binding energies at large distances.

# Chapter Six

# Non-computability

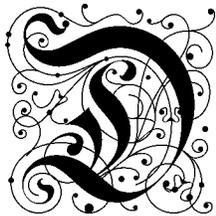 UE TO the unrecognizability of certain manifolds there must exist pairs of triangulations of these manifolds that can only be reached from each other by going through an intermediate state that is very large. This might reduce the reliability of dynamical triangulation, because there will be states that will not be reached in practice. This problem was investigated numerically for the manifold $S^4$ in [Ambjørn & Jurkiewicz 1995a], but it is not known whether $S^4$ is recognizable. We perform a similar investigation for the manifold $S^5$, which is known to be unrecognizable, but see no sign of any unreachable states.

## 6.1   Non-computability

To generate all the configurations of dynamical triangulation we need an algorithm that is ergodic, i.e. a set of moves that can transform any triangulation into any other triangulation with the same topology. A well known set of moves that satisfy this condition are the so-called $(k, l)$ moves, whose ergodicity was shown in [Pachner 1986, Gross & Varsted 1992]. These moves are extensively discussed in chapter eight.

Unfortunately, the number of moves we need to get from one configuration to another can be very large. To be more precise, the following theorem holds: if the manifold under consideration is unrecognizable, then for any finite set of elementary moves the number of moves needed to get from one configuration of N simplices to another such configuration is not bounded by a computable function of N. This was shown in reference [Nabutovsky & Ben-Av 1993]. We will explain some of the terms in this theorem in a way that is not mathematically precise, but hopefully intuitively clear. See [Nabutovsky & Ben-Av 1993] for details.

A manifold A is unrecognizable if, given a triangulation $\mathcal{T}_0(A)$ of this manifold, there does not exist an algorithm that, given as input an arbitrary triangulation $\mathcal{T}_0(B)$ of a manifold B, can decide whether A and B are homeomorphic. The definition of unrecognizability is not important for the rest of this article, it is only important to know that for some manifolds the above theorem holds. Certain four-dimensional manifolds are unrecognizable, but for the sphere $S^4$, which is usually used in dynamical triangulation, this is not known. It is known, however, that the five dimensional sphere $S^5$ is unrecognizable.

A computable function is a function from $\mathbb{N}$ to $\mathbb{N}$ that can be computed by a large enough computer. Although the computable functions are only an infinitesimally small fraction of all the functions from $\mathbb{N}$ to $\mathbb{N}$, most functions one can think of are computable. A fast-growing example of a computable function would be N!!$\cdots$! with N factorial signs.

Elementary moves can be any type of moves that are computable, i.e. any type that we can actually do in our computer in finite time. The authors of [Ambjørn & Jurkiewicz 1995a] restrict the discussion to local moves. Local moves are moves that involve a number of simplices that is bounded by a constant, in other words a number that does not grow with the volume of the configuration. But their reasoning extends to any computable set of moves. Because having a computable bound on the necessary number of computable moves allows you to compute all possible reachable configurations and in that way recognize unrecognizable manifolds. The conclusion is that implementing non-local moves, like the baby universe surgery of [Ambjørn & Jurkiewicz 1995b], will not allow you to get around this problem.

The proof of the theorem goes along the following lines. Suppose that such a computable bound exists. We could then try all possible sequences of moves which are not longer than this bound, starting from the triangulation $\mathcal{T}_0(A)$. For each triangulation $\mathcal{T}_i(A)$ of A that we get this way, we can check whether it is equal to $\mathcal{T}_0(B)$. As this would give us all possible triangulations of A, this allows us to check whether A and B are homeomorphic and hence recognize B as being equal to A. Given that A is not recognizable, this is impossible.

The above theorem might seem a terrible obstacle for numerical simulation, but the theorem says nothing about the number of moves needed to generate all configurations of any particular size. It is therefore far from clear whether it will have implications for the present simulations.

## 6.2 Barriers

From the theorem stated above it follows that for an unrecognizable manifold the maximum size $N_{int}(N)$ of the intermediate configurations needed to interpolate between any two configurations of size N is also not bounded by a computable function of N. If $N_{int}(N)$ did have such a bound, a bound on the number of possible configurations of size less than or equal to $N_{int}(N)$ would be a bound on the number of moves needed, which would violate the theorem. As I explained on page 28 in chapter two, a simple computable bound on the number of configurations of size N is $(5N_4 - 1)!! \times 12^{5N_4/2}$, where d is the dimension of the simplices.

It was pointed out in [Ambjørn & Jurkiewicz 1995a] that this means that for such a manifold there must exist barriers of very high sizes between certain points in configuration space. Although the situation is not clear from the theorem, it seems natural that these barriers occur at all volume scales. We can then apply the following method, which was formulated in [Ambjørn & Jurkiewicz 1995a]. We start from an initial configuration with minimum size. For $S^4$ and $S^5$, there is a unique configuration of minimum size with six and seven simplices respectively. This is just the boundary of one five- respectively six-dimensional simplex. We increase the volume to some large number and let the system evolve for a while, which might take it over a large barrier. Next, we rapidly decrease the volume, hoping to trap the configuration on the other side of this barrier.

We can check whether this has happened by trying to decrease the volume even more. If this brings us back to the initial configuration, we have gone full circle and cannot have been trapped at the other side of a barrier. Conversely, if we get stuck we are apparently in a metastable state, i.e. at a point in configuration space where the volume has a local minimum.

This was tried in [Ambjørn & Jurkiewicz 1995a] for $S^4$, but no metastable states were found. To judge the significance of this, it is useful to investigate the situation for a manifold which is known to be unrecognizable. It is rather difficult to construct a four dimensional manifold for which this is known, but if we go to five dimensions this is easy, because already the sphere $S^5$ is not recognizable.

## 6.3 Results

Because my program for dynamical triangulation was written for any dimension (see chapter eight), it was not difficult to investigate $S^5$. The Regge-Einstein action in the five dimensional model is

$$S = \kappa_5 N_5 - \kappa_3 N_3, \tag{6.1}$$

where $N_i$ is the number of simplices of dimension i. This is not the most general action linear in $N_i$ in five dimensions as this would take three parameters. This follows straightforwardly from the Dehn-Sommerville relations (1.25), which imply

$$N_4 = 3N_5, \tag{6.2}$$

$$N_2 = 2N_3 - 5N_5, \tag{6.3}$$

$$N_1 = N_0 + N_3 - 3N_5. \tag{6.4}$$

For the purposes of this chapter this is not relevant, however.

I generated 26, 24 and 8 configurations at $N_5 = 8000$, $16000$ and $32000$ simplices respectively. These were recorded each 1000 sweeps, starting already after the first 1000 sweeps, where a sweep is $N_5$ accepted moves. All configurations were made at curvature coupling $\kappa_3 = 0$, making all configuration of the same volume contribute equally to the partition function, in other words making them appear equally likely in the simulation. The volume was kept around the desired value by adding a quadratic term to the action as I have explained in section 2.2. Looking at the number of hinges $N_3$, the system seemed to be thermalized after about 6000 sweeps, irrespective of the volume.

The critical value of $\kappa_5$ (the bare cosmological constant) below which the volume diverges was measured as explained for $\kappa_4^c$ in chapter two. See equation (2.11) on page 29. The measured values of $\langle N_5 \rangle$ were not exactly the values I aimed for. In the modified action (2.26) I just set $\kappa_5$ to zero and V to the volumes mentioned. The results were

$$
\begin{array}{c|c|c}
V & \langle N_5 \rangle & \kappa_5^c \\
\hline
8000 & 8825.1(4) & 0.8251(4) \\
16000 & 16836.6(5) & 0.8366(5) \\
32000 & 32844.6(8) & 0.8446(8)
\end{array}
\tag{6.5}
$$

The errors at the largest volume cannot be trusted, because of the low statistics at this volume.

Starting with these configurations, the volume was decreased by setting $\gamma$ (see equation (2.26)) to zero and $\kappa_5$ to a fixed number larger than the critical value. We call this process cooling, because it attempts to reach a configuration of minimum volume and thereby minimum action.

Figure 6.1. A typical cooling run starting at $N_5 = 32000$, using $\kappa_5 = 2$. The horizontal units are 1000 accepted moves. The vertical axis is the number of five-simplices. The inset is a blowup of a small part of the curve.

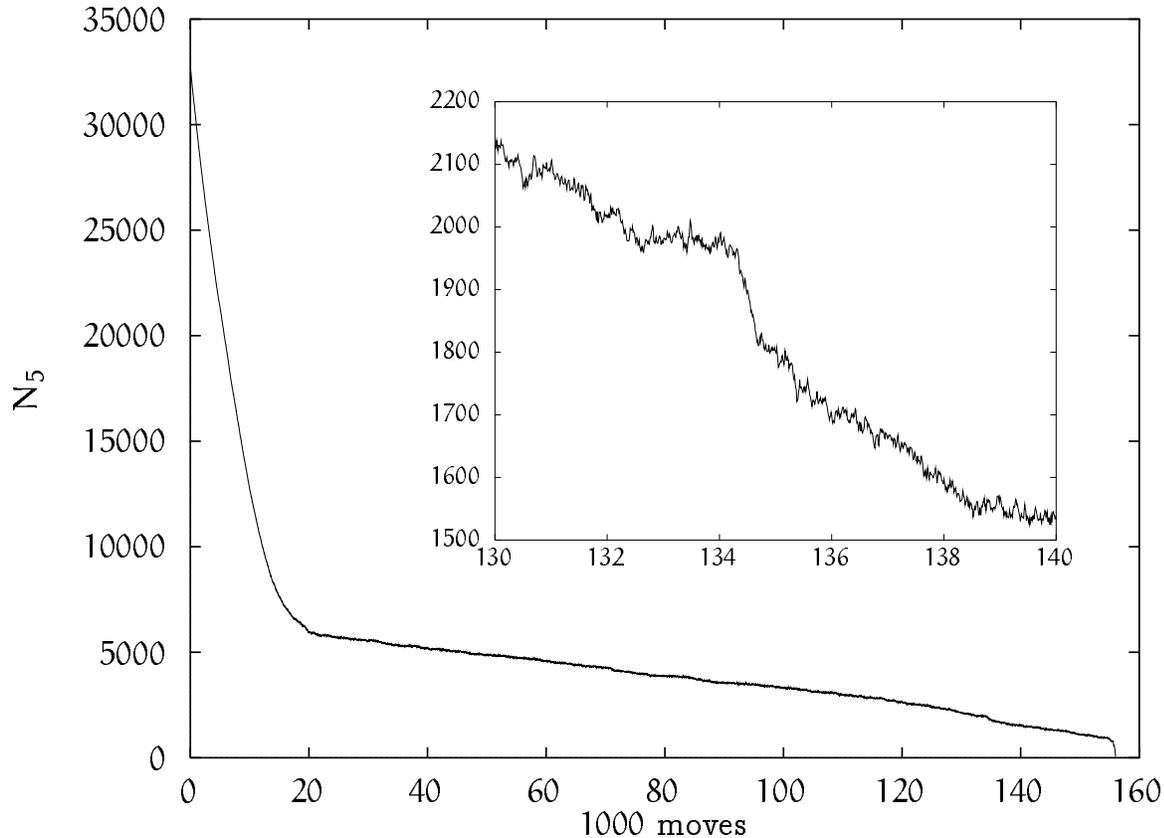

For each configuration we cooled four times with $\kappa_5 = 2$ and two times with $\kappa_5 = 3$. For both values of $\kappa_5$ used one of the runs is shown in figures 6.1 and 6.2. In the insets we can see the typical volume fluctuations that occurred. These were of the order 30 at $\kappa_5 = 2$ and 6 at $\kappa_5 = 3$. The volume would first decrease very quickly until it reached roughly a quarter of the starting value and then started to decrease much slower. In all cases the initial configuration of seven simplices was reached. We also tried to use $\kappa_5 = 4$. The same behaviour of fast and slow cooling was seen, but the latter was so slow that due to CPU constraints these had to be stopped before either a stable situation or the minimal volume was reached.

There is an important difference between four and five dimensions. In four dimensions there is a move that leaves the volume constant. Therefore the system can evolve at constant volume. In five dimensions this is not possible, because all moves change the volume. In this case the volume has to fluctuate for the con-figuration to change. This is why much larger values of the cosmological constant (such as were used in [Ambjørn & Jurkiewicz 1995a]) would effectively freeze the system.

Figure 6.2. As figure 6.1, but with $\kappa_5 = 3$.

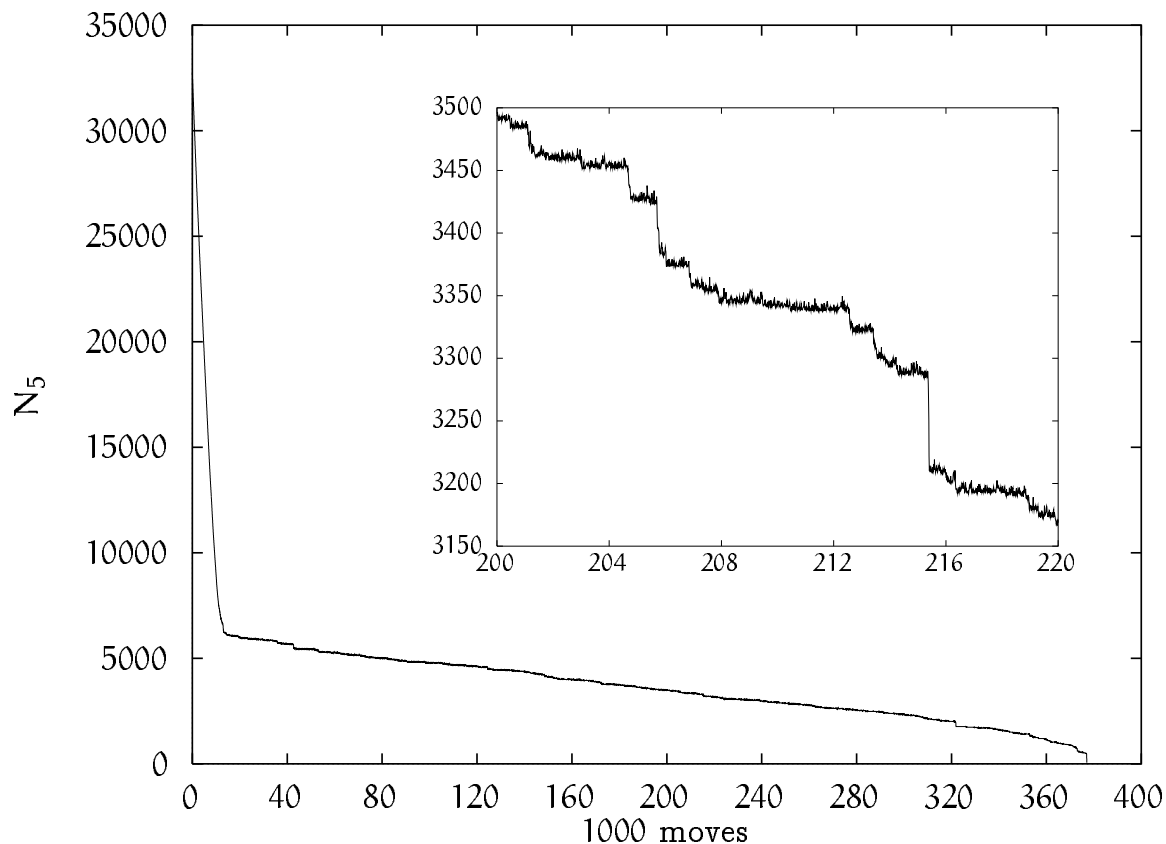

Initially, before the system was thermalized, there was a strong positive correlation between the time used to evolve the system at large volume and the time needed to cool the system back to the initial configuration. The rates of fast and slow cooling did not change, but the volume at which the slow cooling set in became larger. Some of these times are shown in figure 6.3 on the next page for the case of 16000 simplices. This trend does not continue and the cooling times seem to converge.

## 6.4 Discussion

So, contrary to expectation, no metastable states were seen in any of the 384 cooling runs. Small volume fluctuations were necessary, but these gave no indication of the high barriers expected.

Figure 6.3. Number of cooling moves needed at $\kappa_5 = 2$ as a function of the number of equilibrium sweeps at the large volume of 16000 simplices.

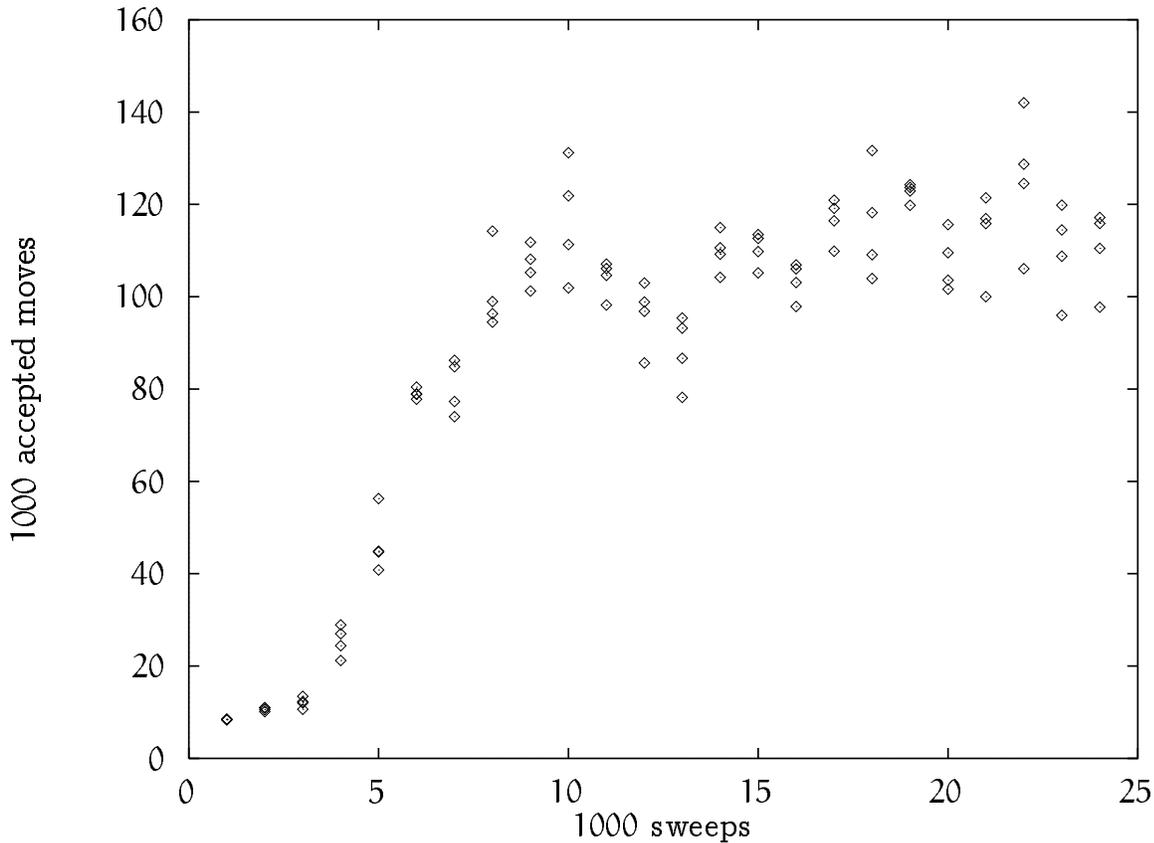

It is not clear why we don't see any metastable states. There are several possibilities. First the barriers might be much larger than 32000 and we need to go to extremely large volumes before cooling. Second, there might be no barriers much larger than the volume for the volumes we looked at and the size of the intermediate configurations needed still grows very slowly for the volumes considered, even though this size is not bounded by a computable function. Third, the metastable regions in configuration space might be very small and the chance that we see one is therefore also very small.

It was speculated in [Ambjørn & Jurkiewicz 1995a] that the absence of visible metastable states might indicate that $S^4$ is indeed recognizable. The results shown in this paper for $S^5$ (which is known to be unrecognizable) indicate that, unfortunately, the results for $S^4$ say nothing about its recognizability. This, of course, in no way invalidates the conclusion of [Ambjørn & Jurkiewicz 1995a] that the number of unreachable configurations of $S^4$ seems to be very small.

Recognizability is not the only thing that matters. Even if $S^4$ is recognizable,

this says little about the actual number of $(k, l)$ moves needed to interpolate between configurations, except that this number might be (but does not have to be) bounded by a computable function.

# Chapter Seven

# Fluctuating topology

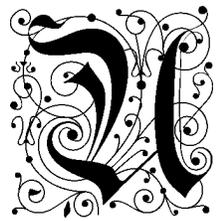 SUALLY one restricts the dynamical triangulation model of quantum gravity to systems where the spacetime topology is fixed (although the topology of space is allowed to change with time). In a path integral formulation of quantum gravity, summing over the topologies seems a natural thing to do. In this chapter I define a model where we take this sum over topologies and describe some preliminary Monte Carlo results.

## 7.1  Introduction

The previous chapters of this thesis considered a partition function which was a sum over triangulations with fixed four-topology. This was done mainly for practical reasons. Fixing the topology seems somewhat unnatural to me. To see why, take a look at figure 7.1 on the following page. It depicts the splitting and converging of space. Both processes are allowed with fixed spacetime topology. But the combination, a space that splits first and then reconverges, is not. Somehow the space has to remember whether it split or not to see if it can converge. To me, this seems strange.

A third possibility is that the topology of spacetime is restricted to $\mathcal{M}^3 \times \mathbb{R}$, where $\mathcal{M}^3$ is fixed. This would not raise the objection expressed in the previous paragraph. A mechanism that imposes such a restriction dynamically is described in [Anderson & DeWitt 1988].

Because the typical curvature fluctuations become larger at smaller scales, allowing an arbitrary topology will result in a very complicated structure at the Planck scale [Wheeler 1964]. Such a spacetime which is full of holes is commonly called spacetime foam. It was estimated in [Hawking 1978a] that the dominant contribution to the path integral would come from spacetimes where the Euler characteristic $\chi$ is of the order of the volume of the spacetime in Planck units.

Figure 7.1. Splitting and converging universe is allowed with fixed spacetime topology. Why not both?

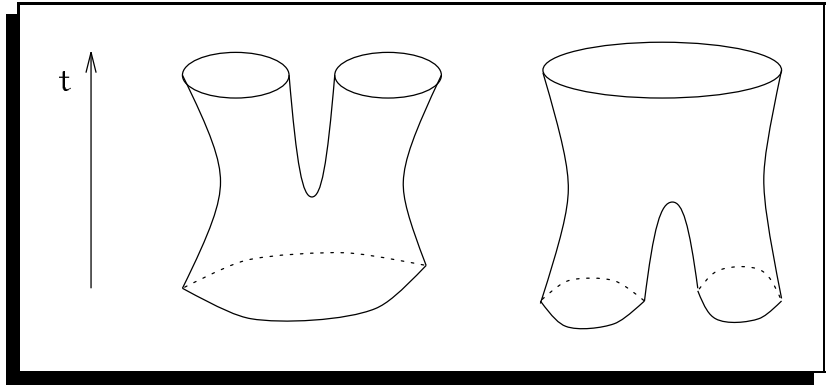

## 7.2   Definition of the model

The partition function of the model in d dimensions is

$$Z(\kappa, N_d) = \sum_{\mathcal{T}(N_d)} \exp(\kappa N_{d-2}), \qquad (7.1)$$

This expression is the same as for a fixed topology, but here the sum is over all possible ways to glue a fixed number $N_d$ of d-simplices together, maintaining orientation (i.e. only identify $d - 1$ dimensional faces with opposite orientation). Because the faces have to be glued in pairs, the volume $N_d$ must be even if the number of spacetime dimensions is even.

At this stage we include disconnected configurations. Because the action is a sum of the actions of the connected components, the Boltzmann weight factorizes, which means that the local physics will not change. Including these configurations, however, will tell us something about the probability to obtain a particular size of connected component. A universe which is most likely to be split up into many small parts seems an unlikely candidate for the real world.

Naively gluing together simplices will not result in manifolds, but only in pseudomanifolds. Several ways to deal with this problem are conceivable. First, as nobody knows what spacetime looks like at the Planck scale, one could argue that this is not a problem. The theory could be universal, such that the inclusion of pseudomanifolds does not change the continuum behaviour.

Second, one can (at least for $d \leqslant 4$) locally deform the resulting pseudomanifold to turn it back into a manifold by removing a small (compared to the lattice spacing) region around the singular points and pasting in a regular region. Taking out the region around the singular points will generate a boundary which is an oriented $(d-1)$-manifold. A well known result from cobordism theory says that for $d = 2$, 3 or 4, we can always find an oriented $d$-manifold with that $(d-1)$-manifold as its boundary. By scaling this region down by a factor $\lambda$, the curvature will increase like $\lambda^2$, but the volume will decrease like $\lambda^{-d}$. Therefore in three and four dimensions this can be done while changing the total curvature only by an arbitrarily small amount. In two dimensions there was no problem to begin with, because the manifold condition is always satisfied as the only connected compact one-dimensional manifold is the circle $S^1$.

Third, these configurations might be unimportant in the limit $\kappa \to \infty$ (which will have to be taken, see below). E.g. in three dimensions for each fixed $N_0$ (and volume) the number of edges (which couples to $\kappa$) is maximal if and only if the configuration is a real (i.e. non-pseudo) manifold. The explanation is analogous to the derivation of (1.24). Around each vertex we have the relation

$$M_0 - M_1 + M_2 = \chi^{(2)} = 2(1 - h), \tag{7.2}$$

where $h$ is the number of handles of the two-dimensional manifold around the vertex. Summing over the vertices $x$ gives us

$$2N_1 - 3N_2 + 4N_3 = 2N_0 - 2\sum_x h_x. \tag{7.3}$$

The sum over $h_x$ vanishes if and only if all the vertices are surrounded by a simplicial ball. Otherwise it is positive. This proves the claim that $N_1$ is maximal if and only if all the $h_x$ vanish and the complex is a simplicial manifold. Using the relation $2N_2 = 4N_3$ (the analog of (1.21)), it also follows that

$$\chi = N_0 - N_1 + N_2 - N_3 \geqslant 0, \tag{7.4}$$

where the equality holds if and only if the pseudomanifold is a manifold. The situation is less clear in four dimensions, though, because a similar reasoning only results in the neighbourhood of a point being a non-pseudo manifold, but not necessarily a ball. Taking the relation (7.4) as applying to the neighbourhood of a vertex in the four-dimensional complex and summing it over all vertices gives us the generalization of (1.24)

$$2N_1 - 3N_2 + 4N_3 - 5N_4 \geqslant 0, \qquad (7.5)$$

where the equality holds if and only if it holds in (7.4) for all vertices.

At low $\kappa$ the connected configurations will contribute most as they have the highest entropy. This is due to the fact that the total number of configurations increases factorially with the number of simplices, while the entropy of the disconnected ones is extensive and therefore their number only rises exponentially. The number of disconnected configurations will also pick up a factor $P(N)$, the number of partitions of $N$, but this behaves only like $\exp(\sqrt{N})$.

At high $\kappa$ the disconnected configurations will contribute most as they have the lowest action. Because lowest action means highest number of hinges ($(d-2)$-dimensional subsimplices where the curvature in concentrated), these hinges will have very few d-simplices around them. In fact, the configuration with the lowest action will have every hinge contained in only one simplex, making no connections between the simplices and therefore a completely disconnected configuration. In an even dimension d this is not possible and the lowest action will occur in a configuration which completely consists of connected components of only two simplices.

It is a priori not clear whether this change between domination of connected and disconnected configurations occurs gradually or whether there is a phase transition. Suppose for a moment that there is a sharp crossover at some $\kappa^c$ depending on the volume. Because, as explained above, the number of connected configurations rises faster with the volume than the number of disconnected ones, the value of $\kappa^c$ will increase with the number of simplices. This means that in a possible continuum limit, the value of $\kappa$ will have to be taken to infinity. Because the number of connected configurations rises like $N! \approx \exp(N \ln N)$, the value of $\kappa^c$ will diverge logarithmically with $N$.

## 7.3   Tensor model

We can formally write down a tensor model which is a generalization of the well known one matrix model of two dimensional quantum gravity (see e.g. [Ginsparg & Moore 1993] and references therein). The partition function of this tensor model is written in terms of an n-dimensional tensor $M$ of rank d. The tensor $M$ is invariant under an even permutation of its indices and goes to its complex conjugate under an odd permutation of its indices. The partition function for three dimensions is

$$Z = \int dM_{abc} \exp\left(-\frac{1}{6}M_{abc}M_{abc}^* + gM_{abc}M_{ade}^*M_{bdf}M_{cef}^*\right), \qquad (7.6)$$

and for four dimensions it is

$$Z = \int dM_{abcd} \exp\left(-\frac{1}{24}M_{abcd}M_{abcd}^* + gM_{abcd}M_{aefg}^*M_{behi}M_{cfhj}^*M_{dgij}\right), \qquad (7.7)$$

$$a, b, \ldots = 1, \ldots, n \qquad (7.8)$$

Using the properties of $M$, the action can easily be seen to be real. The generalization to more dimensions should be obvious from these expressions.

These expressions are only formal, because the interaction term can be negative for any $g$ and is of higher order than the gaussian term. Therefore, these integrals will not converge. Nevertheless, if we expand these expressions in $g$, each term of the expansion is well-defined and the dual of each of its Feynman diagrams is a dynamical triangulation configuration of size $N_d$ equal to the order of $g$ used. The propagators carry $d$ indices, corresponding to the $d-2$ dimensional hinges of the simplices. To be precise,

$$\langle M_{a_1 \ldots a_d} M_{b_1 \ldots b_d}^* \rangle = \sum_{\text{even } \pi} \delta_{a_1 \pi(b_1)} \ldots \delta_{a_d \pi(b_d)}, \qquad (7.9)$$

where the sum is over the even permutations $\pi$. As an example, in three dimensions this becomes

$$\langle M_{abc} M_{def} \rangle = \begin{cases} 3 & \text{if all indices are equal,} \\ 1 & \text{if } abc \text{ is an odd permutation of } def, \\ 0 & \text{otherwise.} \end{cases} \qquad (7.10)$$

The second clause should only be considered if the first one does not apply.

Each vertex of the tensor model corresponds to a $d$-simplex, and each propagator between them corresponds to the identification of two $d-1$ dimensional faces of simplices. For the propagator not to vanish, the sets of indices at each end of a propagator must be an odd permutation of each other, making sure that the simplicial complex has an orientation. As in the matrix model, each hinge corresponds to an index loop in the diagram, giving a factor $n$, while each simplex corresponds to a vertex, giving a factor $g$. The contribution of a particular

diagram will be

$$g^{N_d} n^{N_{d-2}} = \exp(\ln(g)N_d + \ln(n)N_{d-2}) \qquad (7.11)$$

which is precisely proportional to the Boltzmann weight of the dynamical triangulation configuration according to the Regge-Einstein action, with the identification $\kappa = \ln n$.

This model has been discussed for three dimensions in [Ambjørn *et al.* 1991]. A different generalization of the matrix model where the dimension of the matrix couples to the number of points in the dynamical triangulation configuration (which means one does not get the Regge-Einstein action in more that 3 dimensions) has been discussed in [Gross 1991].

## 7.4 Monte Carlo simulation

A move can most easily be described in the tensor model formulation. It consists of first cutting two propagators and then randomly reconnecting them. Due to the orientability, there are $d!/2$ ways to connect two propagators. In the dynamical triangulation model, this corresponds to cutting apart the simplices at each side of two of the $d - 1$ dimensional faces and pasting these four faces together in two different pairs.

Unlike the case of fixed topology with $(k, l)$ moves, one can easily see that these moves are ergodic. We can always get one propagator correct in a single move. The number of moves needed to get from one configuration to another is $O(N_d)$, raising none of the computability problems associated with the fixed topology case [Nabutovsky & Ben-Av 1993] (see chapter six). The non-existence of a classification of four-topologies and their unrecognizability is usually mentioned as a problem for the summation over topologies. In dynamical triangulation, however, it seems to be more a problem for fixing the topology.

We use the standard Metropolis test to accept or reject the moves. This means that we accept any move that lowers the action and accept a move that raises the action with a probability $\exp(-\Delta S)$. Because (again unlike the fixed topology case) the number of possible moves does not depend on the configuration, detailed balance (see section 8.2) is easily obtained.

One could restrict the simulation to connected configurations by checking connectedness for each move accepted by the Metropolis test. This might be rather slow, because this is not a local test. Also, although it seems very plausible, it is not clear whether this would be ergodic in the space of connected configurations.

Figure 7.2.  The number of connected components as a function of κ in the four-dimensional model.

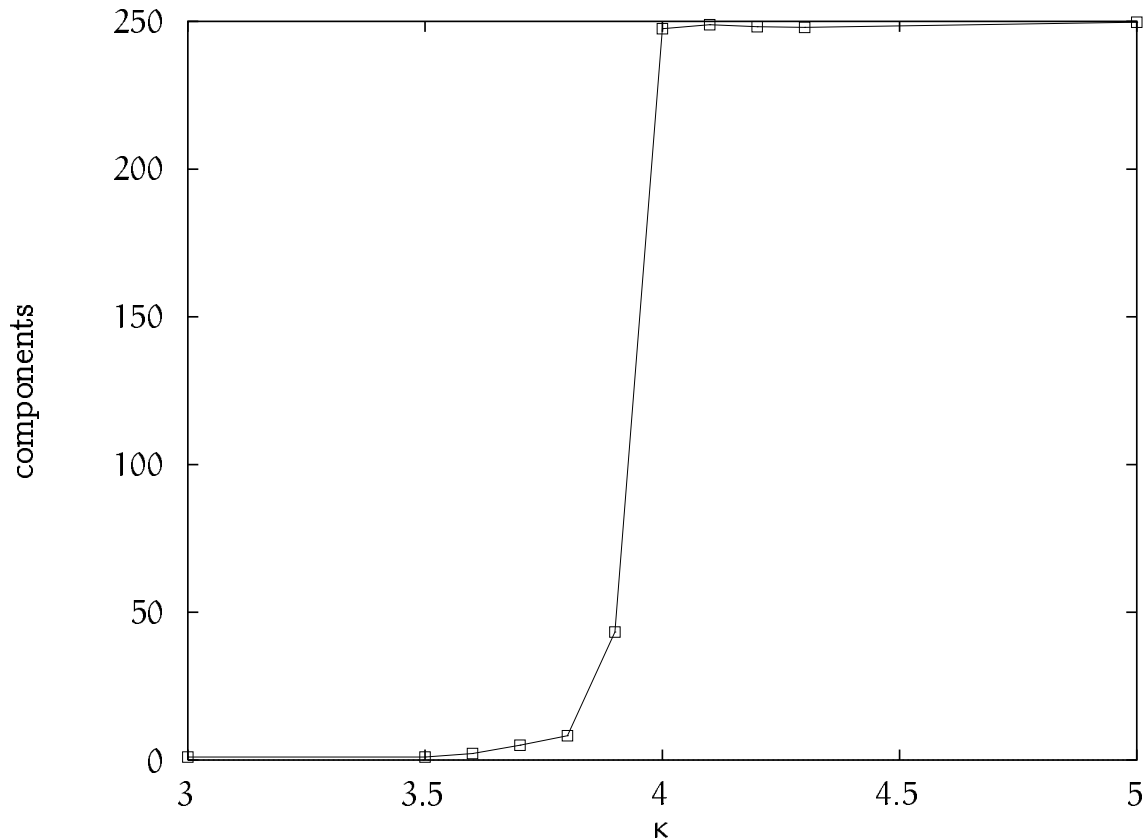

## 7.5   Results

We have simulated this model in four dimensions at a volume $N_4 = 500$. We used a hot start, that is a configuration with completely random connections. This would be an equilibrium configuration at κ = 0.

The number of connected components is plotted in figure 7.2. We see that at low κ the average number of components is almost one, while at high κ the average number of components is almost equal to the maximum number of $N_d/2$. Already at this low number of simplices, the change between few and many components looks quite sharp.

Unfortunately, the acceptance rates for these moves are quite low for κ near $κ^c$, of the order of 0.1%. One of the reasons for this is that the proposed moves are not local in the sense that they can connect any two points in the complex. For κ near the transition thermalization took much longer than for κ far away from it in either direction. To illustrate this phenomenon, I have plotted the number of connected components as a function of computer time in figure 7.3. As usual, a sweep is defined as $N_4$ succeeded moves. The curve at $\kappa = 4.0$ eventually ends up near 250, but takes much longer than the one at $\kappa = 4.3$. Both effects mean that simulating much larger systems will probably not be feasible with only these moves.

Figure 7.3. The number of connected components as a function of computer time for various values of κ.

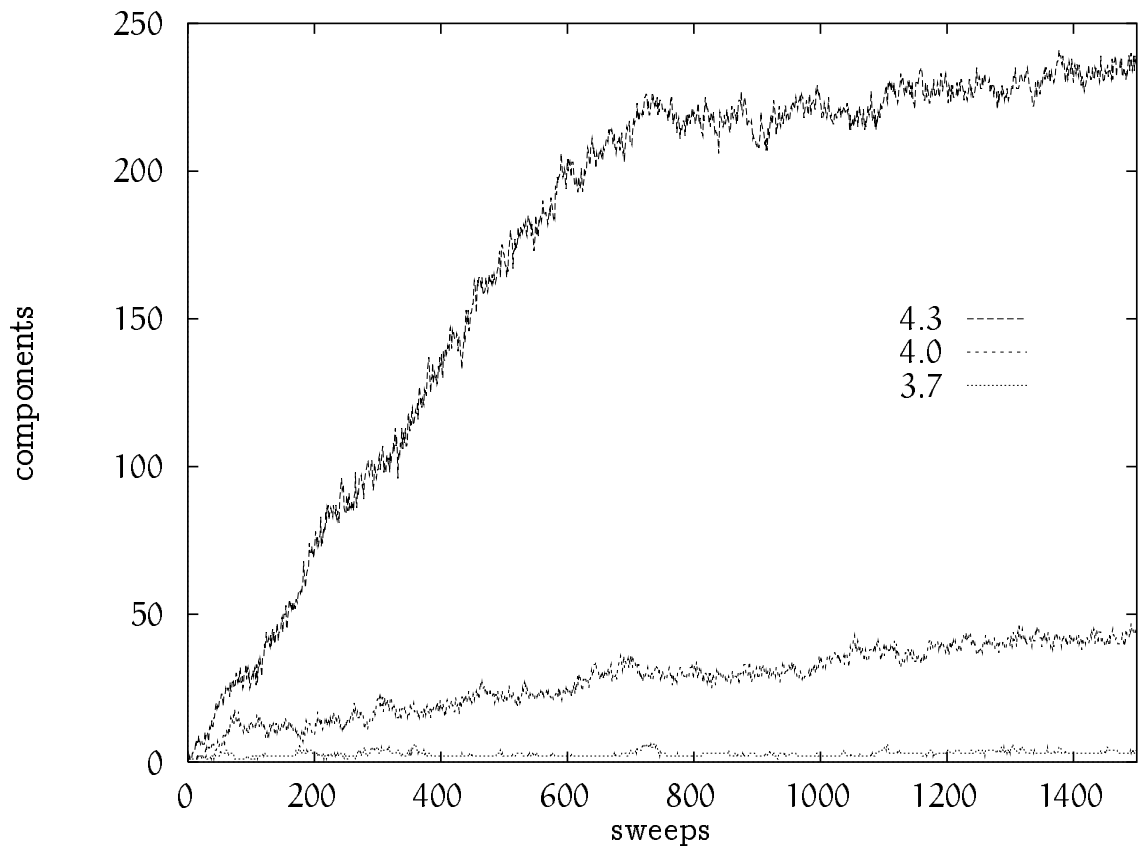

## 7.6 Discussion

The main problem to investigate is the existence of a sensible continuum limit. Although, due to the factorially increasing number of configurations, the grand canonical partition function in the normal definition does not exist, this does not exclude that the local behaviour of the system might show scaling for large volumes.

In two dimensions a limit of the grand canonical partition function, the so-called double scaling limit, is known [Brézin & Kazakov 1990, Douglas & Shenker 1990, Gross & Migdal 1990]. In this limit we also see that the coupling $\kappa$ has to be taken to infinity. We can similarly investigate the problem in two dimensions with fixed volume. For simplicity we will now restrict ourselves to connected configurations. The partition sum with fluctuating topology can be written as a sum over the number of handles h. Using (1.33) this becomes

$$Z(N_2) = \sum_h Z(N_2, h) = f(N_2) \sum_h \exp\left[2h\left(\frac{5}{4}\ln N_2 - \kappa\right)\right], \qquad (7.12)$$

where we have put the factors that do not depend on h in the function $f(N_2)$. This function is not relevant for this discussion. We see that by increasing $\kappa$ with the volume $N_2$ as

$$\kappa = \frac{5}{4}\ln N_2 + \lambda, \qquad (7.13)$$

where $\lambda$ is some positive constant, it is possible to keep the relative importance of each topology constant in the thermodynamic limit.

I think that the above discussion shows that a factorially increasing number of configurations does not by itself preclude a continuum limit. A similar mechanism could apply in four dimensions. It would be nice if we had any idea how the function $f(N_4, \kappa_2)$ as defined in (2.28) behaves as a function of the topology. This would allow a similar analysis in four dimensions. Perhaps it is necessary to introduce another coupling constant, like $\kappa_0$, to do this in a meaningful way.

# Chapter Eight

# Simulations

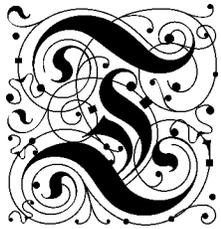HIS CHAPTER contains a description of the computer program that was used to generate the configurations of the dynamical triangulation model of quantum gravity. It consists of three sections: the first is a description of the moves that are used to generate the configurations. In the second section we calculate the relative probabilities with which these moves must be used. The last section describes in detail the code used to implement the moves.

Of course, generating appropriately weighted configurations is not all that is needed. We need to perform various measurements on these configurations. These measurements are described, although not in the same detail, in the chapters that deal with the results.

Unless stated otherwise, in this chapter the word *simplex* will strictly refer to a simplex of the dimension d we are working in. A simplex of dimension $d' \leqslant d$ will be called a subsimplex.

## 8.1   Moves

As explained in chapter one we need to generate a set of simplicial complexes of some fixed dimension and topology. This can be done by starting with one of those complexes and randomly performing a sequence of moves that take one such complex into another. These moves have to be ergodic, which means that one must be able to reach all possible complexes using these moves.

Two sets of moves are known that satisfy this condition of ergodicity. These are called respectively the Alexander moves [Alexander 1930] and the (k, l) moves. These have been proven to be ergodic in respectively [Alexander 1930] and [Pach-ner 1986, Gross & Varsted 1992], in the latter case by proving them to be equivalent to the Alexander moves. The $(k, l)$ moves are more convenient to implement on a computer and I will therefore restrict the discussion to these.

Figure 8.1. The possible $(k, l)$ moves in two dimensions

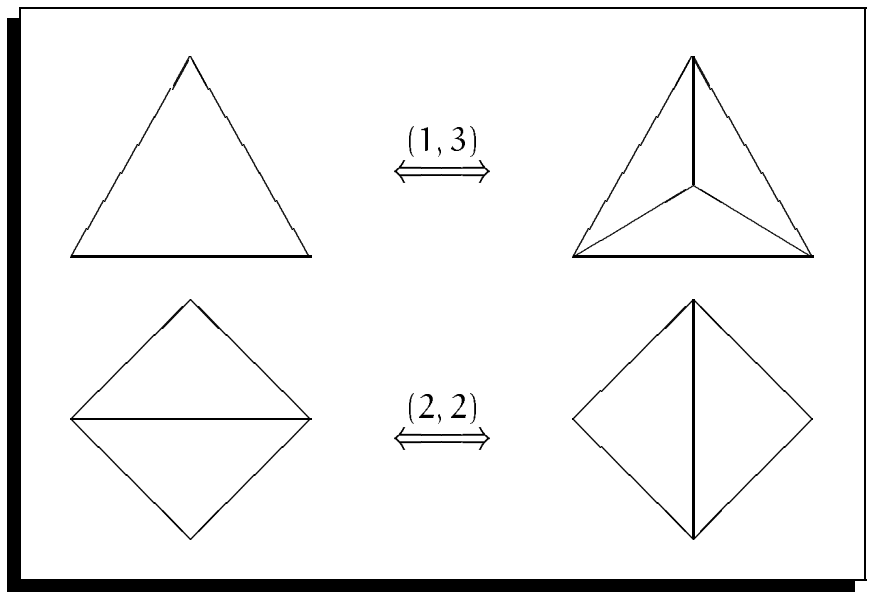

The $(k, l)$ moves in d dimensions can be defined by taking the boundary of a $d + 1$ dimensional simplex, which is itself a d dimensional complex. A $(k, l)$ move consists of replacing a subcomplex which is equal to a part of this boundary by the other part of this boundary.

Let me illustrate this in two dimensions. Figure 8.1 shows the two possible moves in two dimensions. One should imagine these configurations to be embedded in much larger simplicial complexes. The first move takes a part of (the boundary of) a tetrahedron, in this case a triangle. This triangle is replaced by the other part of the tetrahedron, which is a configuration of three triangles. This is called a $(1, 3)$ move, because it takes one simplex into three. The other move replaces two triangles with two other triangles and is similarly called a $(2, 2)$ move.

The possible moves in three dimensions are shown in figure 8.2 on the facing page. In four dimensions it becomes somewhat hard to draw. Figure 8.3 on page 110 shows the possible moves, but in the dual picture. This means that the simplices have been replaced by vertices and if two simplices were adjacent by sharing a face, the vertices of the dual graph have been connected. All information about the relative orientation of the simplices is lost.

We see that the moves always consist of replacing all the simplices containing a given subsimplex. In the $(1, 3)$ move, this subsimplex is a triangle itself, in the $(2, 2)$ move the two triangles next to an edge are replaced and the $(3, 1)$ move replaces the three triangles around a vertex. We can therefore choose a particular place in the configuration to apply a move by choosing a subsimplex with the correct number of simplices containing it. For a subsimplex of dimension $k$, this number is $d + 1 - k$. The move which might then be performed is a $(d + 1 - k, k + 1)$ move.

Figure 8.2. The possible $(k, l)$ moves in three dimensions

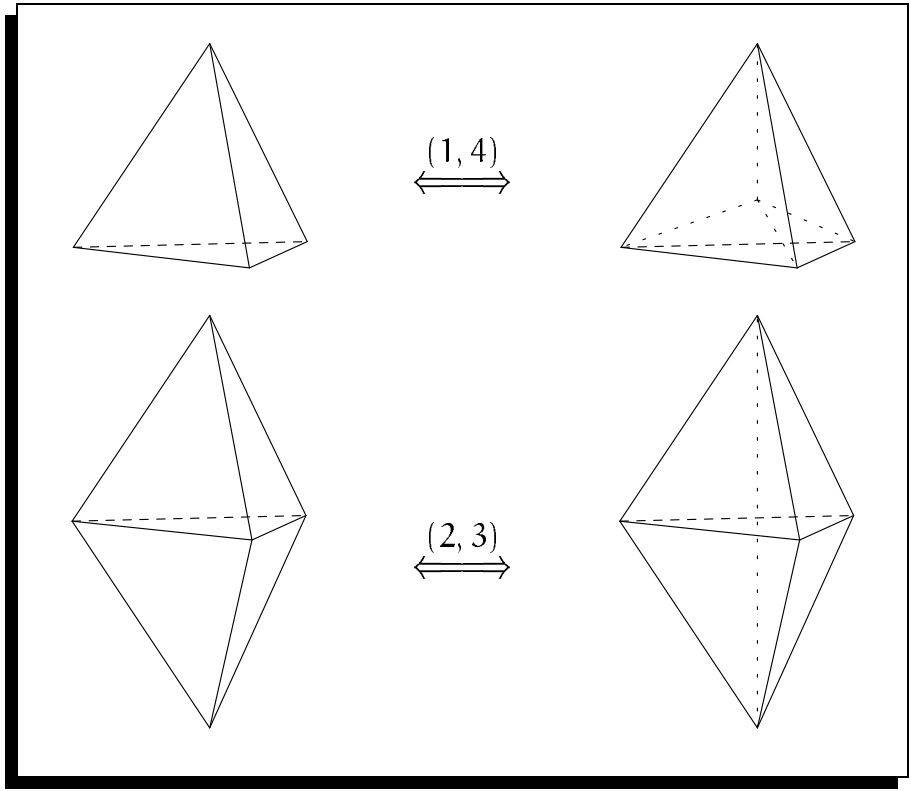

As has been explained on page 25 in section 1.9, these local moves can cause critical slowing down. One would like to use moves which are not local, i.e. moves which change large parts of the system at a time. One such set of moves has been implemented by others [Ambjørn & Jurkiewicz 1995b] and is known as baby universe surgery. These moves are not ergodic by themselves, so they have to be supplemented by the normal $(k, l)$ moves. A big move, as it is called, consists of cutting the space of simplices in two parts at a neck and reconnecting them. A neck is a place where part of the space is connected to the rest by the smallest possible boundary. In four dimensions this boundary is a three dimensional object consisting of five tetrahedrons which are all connected to each other. It is the same boundary as we would get when removing a single simplex from the triangulation.

Figure 8.3. The possible $(k, l)$ moves in four dimensions on the dual graph

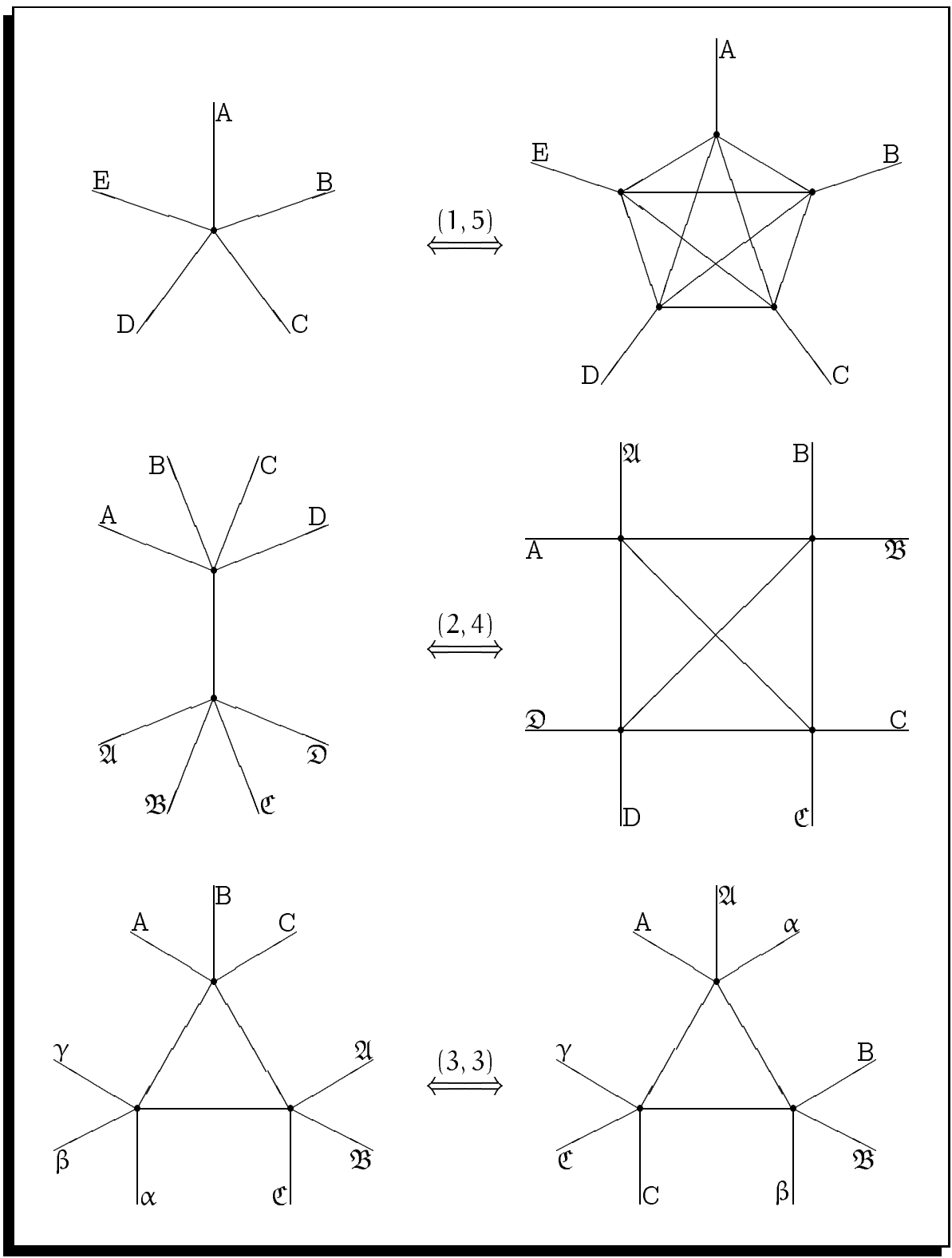

Adding these big moves to the normal $(k, l)$ moves is reported to drastically reduce the autocorrelation time near the phase transition, both in numbers of sweeps and (more importantly) in CPU time [Ambjørn & Jurkiewicz 1995b].

## 8.2 Detailed balance

To approximate the sum over triangulations correctly, we need to generate the configurations with the correct relative probabilities. A sufficient (but not necessary) condition for this is detailed balance (see e.g. [Creutz 1983]). This means that for two configurations A and B which should occur with relative probabilities $p(A)$ and $p(B)$ there should hold the equation

$$p(A)p(A \to B) = p(B)p(B \to A), \tag{8.1}$$

where $p(A \to B)$ is the probability that if we have configuration A that the next move will take it to B. The implementation of this condition is complicated by the fact that the number of possible moves changes with the configuration.

In our case, this probability $p(A \to B)$ is a product of two probabilities: the probability $p_{kl}$ that we will try the correct type of move to take A to B times the probability that we will choose the correct subsimplex whose move will take A to B.

$$p(A \to B) = p_{kl}p(\text{subsimplex}). \tag{8.2}$$

We will see below that the routine that performs the moves chooses a random subsimplex of a particular dimension by first choosing a random simplex and then a random subsimplex of this simplex. This means that the probability for a particular subsimplex to be chosen is proportional to the number of simplices containing it. If the geometry is the correct one to perform a $(k, l)$ move, this number is $k$. The number of $i$-dimensional subsimplices in a $d$-simplex is $\binom{d+1}{i+1}$. Thus, the probability to choose the particular subsimplex whose move will take A to B is

$$p(\text{subsimplex}) = \frac{k}{N_d(A)\binom{d+1}{k-1}}. \tag{8.3}$$

The desired probabilities $p(A)$ are proportional to $\exp(-S[A])$. The detailed balance equation (8.1) then becomes

$$\exp(-S[A])p_{kl}\frac{k}{\binom{d+1}{k-1}}\frac{1}{N_d(A)} = \exp(-S[B])p_{lk}\frac{l}{\binom{d+1}{l-1}}\frac{1}{N_d(B)}. \tag{8.4}$$

Because $k + l = d + 2$ it follows that

$$\frac{k}{\binom{d+1}{k-1}} = \frac{k!\, l!}{(d+1)!} = \frac{l}{\binom{d+1}{l-1}}, \qquad (8.5)$$

and equation (8.4) can be reduced to

$$\frac{p_{kl}}{p_{lk}} = \exp(S[A] - S[B]) \frac{N_d(A)}{N_d(B)}. \qquad (8.6)$$

If, as was the case in our simulations, the action $S[A]$ only depends on the $N_i$, the right hand side only depends on the current values of $N_i$ and how they change. The latter depends only on the type of move, not on where it is done. The conclusion is that we just need to try the different moves with suitable relative probabilities, which depend on the current configuration. No Metropolis acceptance step is needed. Note that we must not choose a type of move and then try to perform this type of move until it is accepted. Instead, if the move routine rejects a move (because the geometry does not fit), we must again randomly choose a type of move to try.

We also see that only the relative probability of opposite moves $(k, l)$ and $(l, k)$ is determined. We are free to choose the relative probability of e.g. a $(1, 3)$ and $(2, 2)$ move. One could try to tune this to minimize autocorrelation times. I have not done this, but simply used

$$p_{15} + p_{51} = p_{24} + p_{42} = p_{33} = \frac{1}{3}. \qquad (8.7)$$

If the best ratio is unknown, this one is certain not to be more than a factor of three slower. This would be the case if the best ratio would only use one type of move and others would not contribute at all to the decorrelation, a very unlikely event. Any other ratio could be worse, as it could try that particular (unknown) type less often.

## 8.3 Program

In this section I will describe the routine that performs the moves in the simplicial complex. This part of the program can be used for any dimension by a simple recompilation. It was based on ideas put forward in [Brügmann 1993], but developed independent of the very similar program described by Catterall in [Catterall 1994]. He did however in a private discussion provide me with the idea of writing code for arbitrary dimensions. The version included here has been shortened somewhat. In particular it lacks all the consistency checks on the configuration that could be compiled in conditionally for testing purposes.

There are two basic philosophies in programming dynamical triangulation. One can try to store the minimum amount of information. This is called minimal coding. Minimal coding makes finding a correct configuration to perform a move harder, but facilitates updating and saves memory. It is the approach that has been taken in [Brügmann 1993, Catterall 1994] and it is also what we do. On the other hand one can keep track of subsimplices of different dimensions for fast lookup of places where moves can be performed at the expense of larger updating complexity and memory usage. We call this maximal coding and it has been described in [Bilke *et al.* 1995]. The lattice sizes which are currently used in practice are bounded by the CPU time needed for large lattices, not by the available memory size. At the moment it is not clear which approach is more CPU efficient.

We start the program by defining the most important constants, the dimension and the maximum number of subsimplices of some particular dimension that a simplex can have, which is needed for some static arrays. The constant DIMEN is actually the number of dimensions d plus one, because this number is used much more often than d as it equals the number of vertices or faces of a simplex. The number of subsimplices of dimension k is simply the number of ways we can choose a subset of $k+1$ points of the $d+1$ points in the simplex. The constant MSUB is the maximum over k of this quantity, i.e.

$$\text{MSUB} = \left( \begin{array}{c} \text{DIMEN} \\ \lfloor \text{DIMEN}/2 \rfloor \end{array} \right). \tag{8.8}$$

So for four dimensions we have

```
#define DIMEN 5
#define MSUB 10
```

Now we define a type to contain numbers of vertices. RAM and disk usage significantly increase if we use a four byte type here. This will be necessary when $N_0$ becomes larger than $2^{16} = 65536$. In four dimensions we have $N_4/N_0 \gtrsim 5$ (the ratio strongly depends on $\kappa_2$), so the volumes we used (up to 64000) are still far away from this limit.

```
typedef unsigned short vertex;
```

The most important definition concerns the representation of a simplex in the computer. We use the following structure to represent a simplex:

```
struct ssimplex {
  struct ssimplex *next[DIMEN];
  vertex point[DIMEN];
  vertex nxtpnt[DIMEN];
  unsigned int flag;
};
typedef struct ssimplex simplex;
```

For each simplex s we keep track of its neighbours in s.next[DIMEN] and its vertices in s.point[DIMEN]. The vertex s.point[i] is always that vertex that is not contained in the neighbour s.next[i]. The array s.nxtpnt[i] contains the number of that vertex of neighbour s.next[i] which is not contained in the simplex s. It can easily be calculated using

$$
\texttt{nxtpnt[i]} = \sum_{j=0}^{\texttt{DIMEN}-1} \texttt{simplex.next[i]->point[j]}
$$
$$
- \sum_{j=0}^{\texttt{DIMEN}-1} \texttt{simplex.point[j]} + \texttt{simplex.point[i]}, \quad (8.9)
$$

but because it is so often needed we store it separately. To save disk space, however, it is not saved with the configurations, but calculated when reading them in. Finally, flag will be explained later where it is used. It is not needed for the representation of the simplicial complex, but only for the routine that performs the moves. As such, it does not have to be stored with the configurations on disk.

Because the number of simplices fluctuates, the simplex structures often become unused. This is marked by setting next[0] equal to 0. The free simplices are kept in a linked list and next[1] is used as a pointer to the next free simplex. Incidentally, the routine that saves the configuration also renumbers the simplices and vertices such that no free simplices are stored. This not only saves a little disk space, but also makes all the measurement routines slightly simpler because they do not have to take into account the possibility of unused simplices in the array.

We see here an important advantage of the programming language used. If this same program had been written in FORTRAN instead of C, the array of structures would typically have been coded as a set of different arrays. But we often need all the information in a particular simplex, and hardly ever the information of contiguously numbered simplices. This seems to make no difference, until we consider the data cache of a typical CPU. A simplex structure nicely fits into a few cache lines, while taking elements of several different arrays reads entire cache lines for a few bytes, while leaving large parts of those cache lines unused.

The next thing we use are several arrays that describe the simplex. These are initialized in `initmove()` and don't change thereafter. A move of type $m$ will denote a $(m + 1, d + 1 - m)$ move. The number of subsimplices of dimension $d - i$ (which is the dimension of the subsimplex that must be found for a move of type $i$) that a simplex has is kept in

```
static int numsubsim[DIMEN];
```

We will need lists of all the subsimplices in a simplex so we can choose one to perform a move around. The first $i$ members of `subsim[i][j][]` contain the neighbours that define (by their intersection with the base simplex) the $j$th subsimplex of dimension $d - i$, where $0 \leqslant j < \texttt{numsubsim[i]}$. The other $d + 1 - i$ members of `subsim[i][j][]` contain the other neighbours in no particular order. Of course one can also define subsimplices by the vertices they contain, but the scheme used happens to be what we need in the `ndmove()` routine below.

```
static int subsim[DIMEN][MSUB][DIMEN];
```

If we perform a move of type $i$, then $N_j$ will change by $\Delta N_j$. This quantity is kept in `deltan[i][j]`.

```
static int deltan[DIMEN][DIMEN];
```

Now we define some external variables that contain information about the configuration that other routines will need. The first is simply $N_i$.

```
int nsims[DIMEN];
```

The next thing is the number of allocated simplex structures. Some of these may be unused, so this is not $N_d$.

```
int simalloc;
```

The array that contains these simplices. Its size is `simalloc`.

```
simplex *lattice;
```

And the number of vertices allocated. Again, some of these may be unused.

```
vertex pntalloc;
```

## 8.3.1   Initialization

The first routine is a small function that calculates $\binom{n}{k}$ or returns 0 if $k < 0$ or $k > n$. It has not been reproduced here.

```
static int combi(int n, int k);
```

The next routine initializes the static arrays declared above. It should always be called before `ndmove()` can be used.

```
void initmove(void)
{
  int i;
```

We start by calculating the number of subsimplices of dimension $d-i$ in a simplex of dimension $d$. It is just $\binom{d+1}{i}$:

```
  numsubsim[0] = 1;
  for (i=1; i<DIMEN; i++)
    numsubsim[i] = numsubsim[i-1] * (DIMEN+1-i) / i;
```

The next part of this routine calculates those $\binom{d+1}{i}$ different sets and stores them in the `subsim[][][]` array.

```
  memset(subsim, -1, sizeof(subsim));
  for (i=0; i<DIMEN; i++) {
    int j;
    int nowsubs[DIMEN];
    for (j=0; j<i; j++)
      nowsubs[j] = j;
    for (j=i; j<DIMEN; j++)
      nowsubs[j] = -1;
    for (j=0; j<numsubsim[i]; j++) {
      int k;
      int neighbor = i-1;
      int ssind = 0, nsind = i;
      memcpy(subsim[i][j], nowsubs, sizeof(nowsubs));
      for (k=0; k<DIMEN; k++)
        if (subsim[i][j][ssind] == k)
          ssind++;
        else
```

```
      subsim[i][j][nsind++] = k;
    while (++nowsubs[neighbor] == DIMEN + 1 + neighbor - i)
      neighbor--;
    for (k=neighbor+1; k<i; k++)
      nowsubs[k] = nowsubs[k-1] + 1;
  }
}
```

The last thing to calculate is $\Delta N_j$ for a move of type i. This is of course equal to the number of j dimensional subsimplices that are only in the new $d + 1 - i$ simplices minus the number of j dimensional subsimplices that are only in the old $i + 1$ simplices.

```
for (i=0; i<DIMEN; i++) {
  int j;
  for (j=0; j<DIMEN; j++)
    deltan[i][j] = combi(DIMEN-i, DIMEN-j) -
      combi(i+1, DIMEN-j);
}
}
```

The next routine makes the smallest possible configuration, which is the boundary of a $(d + 1)$-dimensional simplex.

```
void makesmall(void) {
  int i;
```

This configuration uses $d + 2$ vertices and $d + 2$ simplices. Note that the code assumes that the `lattice` array is at the end of the data segment. One should give the stdio library its buffers using `setbuf()` before calling this routine or very strange errors will occur later on.

```
  simalloc = DIMEN + 1;
  lattice = (simplex *)sbrk(simalloc * sizeof(simplex));
  pntalloc = DIMEN + 1;
```

Just connect all the simplices to all the others. We set it up such that simplex i does contain all vertices except vertex i. This is not necessary, vertex numbers themselves have no meaning and can be permuted at will. Only the assignment to the elements of `point[]` is significant, because the move routine assumes that `lattice[i].next[k]` is the neighbour that does not contain the vertex `lattice[i].point[k]`.

```
for (i=0; i<DIMEN+1; i++) {
    int j;
    int k = 0;
    for (j=0; j<DIMEN+1; j++) {
        if (j == i)
            continue;
        lattice[i].next[k] = lattice + j;
        lattice[i].point[k] = j;
        k++;
    }
    for (j=0; j<DIMEN; j++)
        lattice[i].nxtpnt[j] = i;
    lattice[i].flag = 0;
}
```

In this configuration, any set of points is a subsimplex. Therefore $N_i$, the number of subsimplices of dimension i, is just the number of ways to choose $i + 1$ vertices out of the $d + 2$.

```
for (i=0; i<DIMEN; i++)
    nsims[i] = combi(DIMEN+1, i+1);
}
```

## 8.3.2 Geometry tests

We finally get to the routine that actually performs the moves. When called with the argument movetype, which we will denote by m, it tries to perform one $(m + 1, d + 1 - m)$ move. If it returns 0, the move succeeded, otherwise it did not. In the latter case the return value will indicate the test which was failed, so we can keep some statistics.

```
int ndmove(int movetype)
{
    int subs;
    simplex *base;
    simplex *moving[DIMEN+1];
    vertex mpoint[DIMEN+1];
    int i;
    int *heresub;
```

```
vertex newpoint;
static unsigned int recount = 0;
static simplex *first = 0;
static int frstpnt = 0;
#define MXFRPNT 8192
static vertex freepnts[MXFRPNT];
```

First of all we have to choose a random subsimplex of dimension $d - m$ to perform the move on. We do this by choosing a random simplex base (a member of the lattice array that has not been marked as free) and then a random subsimplex subs of this simplex. This means that the probability that a subsimplex will be chosen is proportional to the number of simplices in which it is contained. As was explained on page 111 in section 8.2 this has been accounted for in the calculation of the probabilities to get the various values of $m$. The routine randbb() is a random number generator that provides an unsigned int. Because most of the allocated simplices are actually used, the vast majority of cases will see the following loop only iterated once.

```
do {
    base = lattice + (randbb() % simalloc);
} while (!base->next[0]);
subs = randbb() % numsubsim[movetype];
heresub = subsim[movetype][subs];
```

The array heresub[m] will contain the numbers i that mark the $m$ simplices base->next[i] that, together with the base simplex, will be replaced by new simplices if the move succeeds. The other $d + 1 - m$ elements (those with indices $m \leqslant i < DIMEN$) of the array heresub[DIMEN] mark the other neighbours.

The $(1, d + 1)$ move (which has $m = 0$) is always possible, so none of the tests are necessary in that case. Otherwise, we need to check whether the move can be performed. This is done by three geometry tests.

```
if (movetype) {
    int frst, last;
    #define MXRNDSM 65536
    simplex *list[MXRNDSM];
    #define MAXPNTS 65536
    static char repnt[MAXPNTS];
```

Figure 8.4. Example of trying a $(3, 1)$ move around vertex A. If the first geometry test is passed, the picture on the left turns out to be the picture on the right.

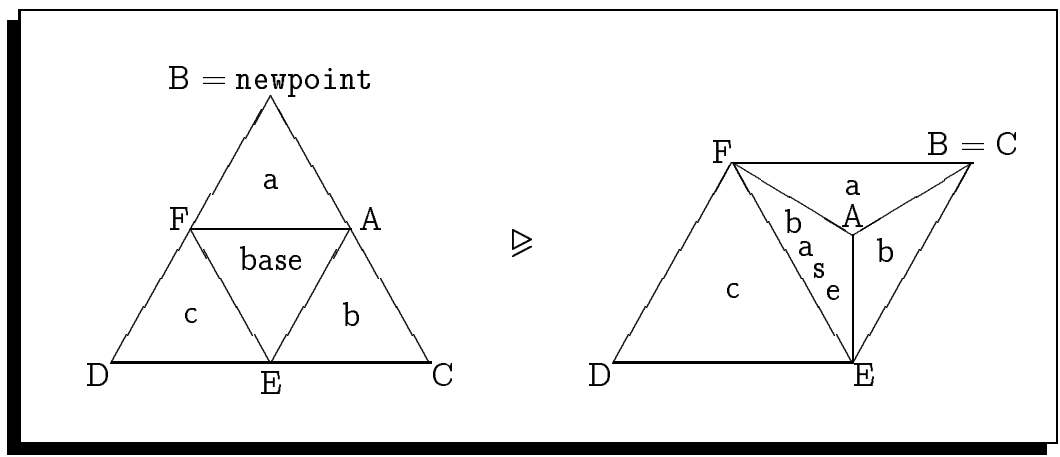

As has been explained on page 108 in section 8.1 a move can be seen as replacing a part of the boundary of a $d + 1$ dimensional simplex. This means that there are always $d + 2$ vertices involved in a move. One of these is not a vertex of the base simplex. This one is stored in `newpoint`.

```
newpoint = base->nxtpnt[heresub[0]];
```

The first test we will do is to see if the geometry around the subsimplex is the correct geometry to perform the move. To understand the code for this test, look at the example in figure 8.4. We are trying to perform a $(3, 1)$ move, and have chosen the neighbours a and b as potential simplices to replace. The move can only be performed if there are three triangles around the vertex A (this vertex is uniquely determined by the two neighbours). This is the case if the vertices B and C are actually the same vertex, making the edges AB and AC the same, because we do not allow two different edges to have the same vertices. In other words, B and C should both be the `newpoint`. We have now made `newpoint` equal to B and need to check if C is the same point.

```
for (i=1; i<movetype; i++)
  if (base->nxtpnt[heresub[i]] != newpoint)
    return 1;
```

The above loop does nothing if $m = 1$. This corresponds to a $(2, d)$ move. Such a move must always pass this test, because there are always exactly two simplices that share a face (which is a $d - 1$ dimensional subsimplex).

Figure 8.5. Example of a $(2, 2)$ move.

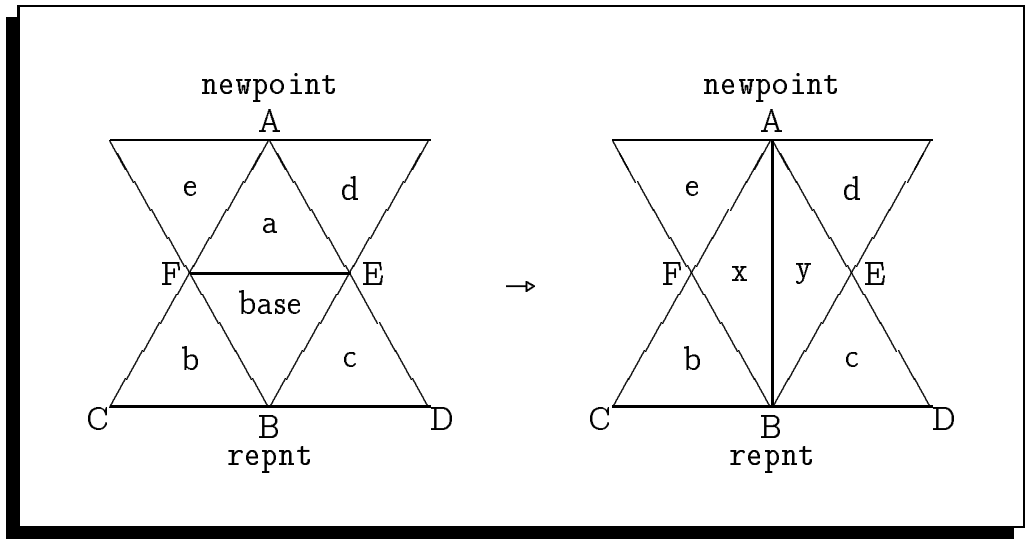

A move might create more than one subsimplex consisting of the same vertices. We do not allow these configurations, and we have to check whether any of the subsimplices that will be created would have the same vertices as any existing subsimplex.

We start by performing a quick test to see if any of the other vertices opposite the base simplex is the same as the newpoint. In the example of figure 8.4 on the preceding page, we check whether vertex D is equal to vertex B. This would mean that after the move we would get two triangles with the vertices B, E and F. One might think that D and B cannot be equal because this would imply the existence of two edges BE, but this is not true, because these can actually be the same edge. This test is implied by the next test. We could therefore just leave it out, but because it is faster and finds a large part of the cases, including it speeds up the program.

```
for (; i<DIMEN; i++)
  if (base->nxtpnt[heresub[i]] == newpoint)
    return 2;
```

Because most of the calls of this routine don't even get to this point, we see that it is essential to perform the first two tests as fast as possible. This is the reason we keep track of the nxtpnt[] array, instead of calculating this information when we need it.

As has already been suggested, the above test is not complete. This cannot be seen in the first example, but it can happen in the second example, which we see in figure 8.5 on the preceding page. Here we are trying to do a $(2, 2)$ move on the base triangle and triangle a, creating a new edge from vertex A to B. If, however, any of the triangles around vertex B also contains A, this would create two edges between the same pair of vertices. The test above only checks whether C or D is equal to A. See subsection 8.3.4 for a more general explanation.

The method is now to go through all the simplices which contain the vertex B and see whether they contain the newpoint. To do this we keep a queue of simplices to visit, adding for each visited simplex all its neighbours which also contain B (and have not yet been visited) to the queue. I have also tried to implement this with a recursive routine instead of a queue. This turned out to be significantly slower.

To mark a simplex as visited, we make its flag field equal to recount, a number that is incremented each time the test is run. If the number becomes 0, we have to reset all the flags, to make sure none of them is accidentally still equal to recount before the test is run. This happens only once every $2^{32} - 1$ times the test is run.

```
if (++recount == 0) {
  for (i=0; i<simalloc; i++)
    lattice[i].flag = 0;
  recount = 1;
}
base->flag = recount;
```

The variables frst and last are used to mark the start and end of the queue, which is kept in list[MAXRNDSIM]. Therefore, the constant MAXRNDSIM has to be larger than the maximum number of simplices around any particular subsimplex. This is not checked at runtime, because this would be tested so often as to result in significant slowdown. It does not matter if it is much too large, though.

```
frst = 0;
last = 0;
```

The array repnt[] is only 1 for those points that form the subsimplex around which we have to look. In the example, it will only be 1 for vertex B.

```
for (i = 0; i < movetype; i++)
  repnt[base->point[heresub[i]]] = 1;
```

We start by putting the simplices b and c in the queue. Any simplex which is put in the queue has its flag immediately set equal to recount to mark it such that it will not be put in the queue another time. We see that the vertices of b and c will be looked at another time, even though they have already been checked in the previous test. I tried to explicitly add their neighbours to the queue, while not looking at their vertices, but this did not result in any speedup. Presumably the extra code and resulting instruction cache loading outweigh the advantage of executing fewer instructions.

```
for (; i < DIMEN; i++)
    (list[last++] = base->next[heresub[i]])->flag = recount;
```

Go through the queue and see if any of the vertices of the simplices looked at is equal to newpoint. If that is the case, we cannot perform the move so clear the repnt[] array and return to the calling routine.

```
do {
    simplex *thissim = list[frst++];
    for (i = 0; i < DIMEN; i++)
      if (thissim->point[i] == newpoint) {
        for (i = 0; i < movetype; i++)
          repnt[base->point[heresub[i]]] = 0;
        return 3;
      }
```

Remember that vertex point[i] is never contained in simplex next[i]. Therefore, all neighbours that do not correspond to vertex B (which was marked in repnt[]) contain B and should be added to the queue, unless they have already been flagged as being in the queue.

```
    for (i = 0; i < DIMEN; i++) {
      if (repnt[thissim->point[i]] ||
          thissim->next[i]->flag == recount)
        continue;
      (list[last++] = thissim->next[i])->flag = recount;
    }
  } while (frst < last);
  for (i = 0; i < movetype; i++)
    repnt[base->point[heresub[i]]] = 0;
} else {
```

Else $m = 0$ and we are dealing with a $(1, d + 1)$ move. This is a barycentric subdivision of a simplex and creates a new vertex, which is now the newpoint. We keep a stack of free vertex numbers in freepnts[MXFRPNT], if the stack is empty, we use a new one.

```
if (frstpnt)
  newpoint = freepnts[--frstpnt];
else {
  newpoint = pntalloc++;
  if (!pntalloc) {
    fprintf(stderr,
            "Too many points for vertex data type!\n");
    exit(1);
  }
}
}
```

### 8.3.3  A move

All tests have been passed. We can now actually perform the move. The array moving[DIMEN+1] holds the $d + 2$ simplices in the boundary of the $(d+1)$-simplex that defines the move. The first $m + 1$ simplices already exist and are replaced by the others. We use free elements of the lattice array for the new simplices. These are kept in a linked list, the head of which is pointed to by the static variable first. It is admittedly somewhat inefficient to make a system call for each new allocated simplex, but this is not important because the number of times this happens is only of the order $N_d$, not of the order of the number of moves performed. Note also that this was run under AIX 3, where brk() never returns an error due to lack of memory.

```
moving[0] = base;
for (i=1; i<movetype+1; i++)
  moving[i] = base->next[heresub[i-1]];
for (; i<DIMEN+1; i++) {
  if (!first) {
    simalloc++;
    brk((char *)(lattice + simalloc));
    first = lattice + simalloc - 1;
    first->next[0] = 0;
```

```
      first->next[1] = 0;
      first->flag = 0;
    }
    moving[i] = first;
    first = first->next[1];
  }
```

Similarly, the array `mpoint[DIMEN+1]` holds the vertices of this $d + 1$ dimensional simplex boundary. As in a simplex structure, `mpoint[i]` is that vertex which is not a vertex of `moving[i]`.

```
  mpoint[0] = newpoint;
  for (i=1; i<DIMEN+1; i++)
    mpoint[i] = base->point[heresub[i-1]];
```

The new simplices i have to be given connections at all faces k. To do this we go through all the moving simplices j, connecting the simplex i to either another new simplex (if j is new) or to one of the simplices which are not themselves moving (if j will be replaced).

In the words of our second example in figure 8.5 on page 121 we have to connect the new simplex x to three other simplices. To do this we go through all four simplices involved in this move: base, a, x and y.

```
  for (i=movetype+1; i<DIMEN+1; i++) {
    int j;
    int k = 0;
    for (j=0; j<DIMEN+1; j++) {
```

The simplex x must not be connected to x itself, so in this case, do nothing.

```
      if (j == i)
        continue;
```

If j is another new simplex (y in the example) simply connect to it.

```
      if (j > movetype) {
        moving[i]->next[k] = moving[j];
        moving[i]->nxtpnt[k] = mpoint[i];
      } else {
```

If j is an old simplex, instead of connecting to it, we connect to the simplex that j is connected to. In the example this means that instead of connecting x to a, we connect it to e. And instead of connecting x to base, we connect it to b. The variable oldsim points to simplex a, oldtonon is the face of a connecting it to e, nonsim points to simplex e and nontoold is the face of e that connects it to a and which will have to become connected to x.

```
simplex *oldsim = moving[j];
int oldtonon;
simplex *nonsim;
int nontoold;
```

We recognize the simplex e (in lieu of d) as the correct neighbour of a by the fact that it does not contain the vertex E, the vertex which is also not in simplex x and therefore has its number contained in mpoint[i].

```
      for (oldtonon=0; oldtonon<DIMEN; oldtonon++)
        if (oldsim->point[oldtonon] == mpoint[i])
          break;
      nonsim = oldsim->next[oldtonon];
      for (nontoold=0; nontoold<DIMEN; nontoold++)
        if (nonsim->next[nontoold] == oldsim)
          break;
      moving[i]->next[k] = nonsim;
      moving[i]->nxtpnt[k] = nonsim->point[nontoold];
      nonsim->next[nontoold] = moving[i];
      nonsim->nxtpnt[nontoold] = mpoint[j];
    }
    moving[i]->point[k] = mpoint[j];
    k++;
  }
}
```

The move has been done and the old simplices that have been replaced should be marked as unused and put at the head of the linked list of free simplices.

```
  for (i=0; i<movetype+1; i++) {
    moving[i]->next[0] = 0;
    moving[i]->next[1] = first;
    first = moving[i];
  }
```

If this was a $(d+1,1)$ move, a vertex has been deleted and should be put on top of the stack of unused vertex numbers.

```
if (movetype == DIMEN-1) {
  if (frstpnt == MXFRPNT) {
    fprintf(stderr,"Size of freepnt too small!\n");
    exit(1);
  }
  freepnts[frstpnt++] = mpoint[DIMEN];
}
```

Finally, adjust $N_i$ and return 0, indicating success.

```
for (i=0; i<DIMEN; i++)
  nsims[i] += deltan[movetype][i];
return 0;
}
```

### 8.3.4  Double subsimplices

All that is left is to explain why the third geometry test does what we need. We will drop the convention used above to call any simplex of fewer dimensions than d a subsimplex. The aim of the test is to avoid creating a new simplex that consists of the same vertices that already define an existing simplex. To do this, let us divide the $d+2$ vertices that are involved in the move into two sets. The first set we shall call $a_i$ and consists of the vertices that make the simplex around which the move will be done. In the example of figure 8.5 on page 121 these are the vertices E and F. The second set $b_i$ are the other vertices. In the program these are the vertices newpoint and those that will be marked in the repnt[] array. In the example, these are the vertices A and B. The two sets are complementary in the sense that their roles are exactly reversed when the inverse move is done.

We will show in the next paragraph that the move can be performed if and only if there does not yet exist a simplex formed by the vertices $b_i$. Because this simplex will be a subsimplex of a d-simplex, all that is necessary is to look if any d-simplex contains all the vertices $b_i$. The program does this by going through all the d- simplices that contain the vertices marked in repnt[] (which are all the $b_i$ except for newpoint) and checking whether one of these contains newpoint.

The implication to the right is trivial. The move always creates the simplex formed by the $b_i$, so if this simplex does already exist, the move cannot be done. To prove the implication to the left, it is sufficient to show that any newly created simplex will contain all the vertices $b_i$. Because if such a simplex already exists, the simplex formed by $b_i$ must exist, as a simplicial complex contains all subsimplices of its simplices. Now suppose there is a newly created simplex $S$ that does not contain a particular vertex $b_k$. Because the $a_i$ are the intersection of all the old $d$-simplices involved in this move, there must be such a $d$-simplex $T$ that does not contain $b_k$, and therefore contains all the other $d+1$ vertices involved. But this is impossible, because $S$ already exists as a subsimplex of $T$ and can therefore not be newly created by this move. This leads to a contradiction, which completes our proof of the claim that the third test indeed exactly tests what we want it to test.

# Chapter Nine

# Outlook

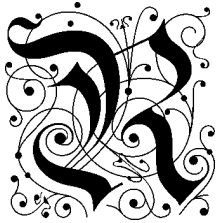 EADERS of this chapter are forewarned that much of it is speculation and not proven by the data, but only hinted at. I will first discuss our current view of the simplicial spacetimes that are generated by the model and then try to say something about the effective action of the resulting theory.

## 9.1  Simplicial spaces

It seems that the bare gravitational coupling constant $\kappa_2$ determines an effective curvature, that I will call $R_V$. A definition of $R_V$ has been explored in chapter three, but it would be nice if a better definition could be made than that one. To this curvature $R_V$ corresponds a curvature radius $r_V$, the magnitude of which is determined by the equation

$$|r_V| = \sqrt{\frac{12}{R_V}}. \tag{9.1}$$

The phase transition is the place where this radius is just sufficient to turn the space into a four-sphere. At larger values of $\kappa_2$ the radius $r_V$ decreases. Because the system has to accommodate the whole volume, which is larger than that of a four-sphere with radius $r_V$, it turns into a set of branching tubes with radius $r_V$. At large scales (compared to $r_V$), the tubes behave like a branched polymer with fractal dimension two.

So it appears that $\kappa_2$ determines the effective curvature radius $r_V$. On the other hand $\kappa_2$ is the bare gravitational constant, and one would think it determines the renormalized gravitational constant $G_R$ and therefore the Planck length $\ell_P = \sqrt{G_R}$.

If two scales are set by one parameter, we expect them to be related and therefore of the same order of magnitude. But in our universe we see that $r_V \ggg \ell_P$ by many orders of magnitude. The question is then how to separate the two scales with only one parameter.

A possible way out of this problem is the triviality discussed on page 65 in chapter three. This could allow us to set both the Planck length and the curvature radius by tuning not only $\kappa_2$, but also the number of simplices $N_4$. When the curvature radius $r_V$ is fixed, the shape of the space is fixed. Using more simplices then means that the simplices are smaller in physical units, that is in units of $r_V$. So the number of simplices $N_4$ determines $\ell/r_V$, i.e. the lattice spacing in physical units. In other words, by tuning the number of simplices $N_4$ we are tuning the lattice spacing $\ell$.

This idea seems to be in conflict with the usual view that the physical behaviour of the system is independent of the lattice spacing. This behaviour would indeed become independent in the continuum limit $\ell \to 0$. But at small finite lattice spacing we could still get continuum behaviour, where all the parameters that are irrelevant in the renormalization group sense (which go like $(\ell/r_V)^2$) have vanished but the marginal parameter $\ell_P/r_V$ (which goes only like $1/\ln(\ell/r_V)$) can be varied at fixed $r_V$.

To check whether something like this happens, it is necessary to measure the Planck length. Two possible ways have been described in this thesis. First, the Planck scale might be the scale where the running curvature defined in chapter three becomes large. Figure 3.12 shows a very slight hint that this scale becomes smaller compared to the system size, but this needs to be confirmed at other lattice sizes. A problem is that in the above scenario this scale decreases only logarithmically with the lattice spacing, making it very hard to see in the computer simulations.

A second way to measure the Planck length is to measure the renormalized gravitational constant $G_R = \ell_P^2$ directly. A possible way has been indicated in chapter five, where we measured binding energies. An important problem there is that the behaviour of the binding energy (as a function of the particle mass and the gravitational constant) does not seem to confirm to our guess (5.17). Perhaps this is due to the fact that the ratio $\ell_P/r_V$ is not (yet) very small.

Another measurement that deserves extension to larger lattices is the scaling analysis in chapter three. Compared to the paper [de Bakker & Smit 1995], where these findings where originally published, figure 3.10 has already been extended with a larger volume and the matching looks even better than it did. But it would also be interesting to extend the measurements for values of $\kappa_2$ far away from the phase transition and especially to confirm that the scaling dimension $d_s$, which is the dimension at small scales, is indeed four.

## 9.2   Effective action

As has been mentioned in chapter one the Euclidean action is not bounded from below. Although the path integral could still converge, due to the behaviour of the measure, this does raise a different problem. Assume for a moment that we can take a good continuum limit, as indicated by the evidence for scaling of chapter three. If the effective action describing semiclassical fluctuations around the resulting average space is the Euclidean gravity action, there will be a mode with negative eigenvalue, making that space unstable. Several scenarios seem possible.

First, the effective action that comes out is not Euclidean gravity as described by the action (1.5). There is no guarantee that the effective action that comes out of a lattice model is the one we started out with to discretize. For instance, in the continuum limit, the non-linear $O(4)$ sigma model mentioned in section 1.6 becomes the linear sigma model. The effective action could still be something describing gravity, perhaps Euclidean gravity with the conformal mode rotated [Gibbons *et al.* 1978, Schleich 1987].

As a second scenario, the instability might be physical. The universe is unstable at large time scales, slowly collapsing to black holes. Perhaps the instability is a manifestation of this phenomenon. The main problem with this line of thought is that naive arguments would predict the instability to be of the scale of the parameters of the theory, which is the Planck scale. The Planck length we would measure in the $S^4$-like space we found in chapter three would then be of the order of the size of the universe, making that universe very small. Although the model would in this case not be a good description of quantum cosmology, it could still describe quantum gravity at small scales.

It is interesting to compare this scenario to the minisuperspace model developed in [Hartle 1989, Birmingham 1995]. The model describes a universe with a boundary of only one simplex with varying complex edge length. It turns out that the model has Euclidean and Lorentzian stationary points, but the Euclidean stationary points all have universe sizes less than a critical value, while the Lorentzian ones have sizes greater than this value.

Another interesting phenomenon was observed in the 6j-symbol approach to three-dimensional simplicial quantum gravity in [Barrett & Foxon 1994]. The authors of this paper show that the model has both Lorentzian and Euclidean stationary points, in the sense that the geometry of the manifold is Euclidean or Lorentzian. The Lorentzian stationary points, however, are minima in the action, like one normally sees in a Euclidean model. Conversely, the Euclidean stationary points are oscillatory.

# Appendix One

# Summary

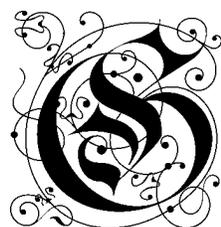RAVITY is the one fundamental force that eludes a quantum description. Creating a good theory of quantum gravity is one of the greatest problems in current high energy physics. A problem of quantizing gravity in the usual way is that general relativity is not renormalizable in perturbation theory. This by itself however implies little about its non-perturbative behaviour.

Simplicial quantum gravity is a candidate for a non-perturbative definition of quantum gravity. Starting point is the creating of a euclidean spacetime by glueing simplices together at the faces. In dynamical triangulation, the method used in this thesis, the path integral over metrics is defined by a sum over all possible ways to glue the simplices together to form a piecewise linear manifold, usually only taking those configuration with a fixed topology.

The model has two coupling constants, the gravitational constant and the cosmological constant. To take a continuum limit, the number of simplices has to go to infinity, fixing the bare (unrenormalized) cosmological constant. As a function of the gravitational constant, the model turns out to have two phases, a crumpled and an elongated phase.

In the crumpled phase distances between simplices are very small and the volume within a fixed distance increases exponentially with that distance. In the elongated phase the system makes narrow tubes that behave like a branched polymer. There is some evidence that the system scales, which means that the behaviour of the system becomes independent of the size of the simplices when that size becomes very small. At the phase transition the system looks in many respects like a four-dimensional sphere. At that phase transition correlations of the scalar curvature turn out to have a long range and to fall of with the fourth power of the distance. This could be taken as evidence for the existence of massless particles (gravitons?) in the model.

An interesting question is whether the model can reproduce gravitational binding. This has been investigated by studying the behaviour of a scalar field on the configurations. Although the data are still somewhat preliminary, there indeed seems to be a positive binding energy.

When doing computer simulations it is important that the moves used can reach all the possible configurations. If this is true we call the moves ergodic. In our simulations this is indeed the case. The number of moves needed, however, turns out not to be bounded by a computable function of the size of the system. It is not clear what the effects of this are in practice. I have looked numerically for such effects, but found none. This could mean that they are unimportant.

In most cases one only looks at configurations with fixed topology. We can also define a model where the sum over all topologies is taken. This model still has many problems, both in defining and in simulating it.

In the study of simplicial quantum gravity Monte Carlo simulations are very important. These are done by taking a small starting configuration and changing it by performing a sequence of moves. The probabilities with which these moves have to be done have to fulfill some conditions to make sure that the probability to reach a configuration is proportional to its Boltzmann weight. All this has been implemented in a C program.

# Bijlage Twee

# Samenvatting

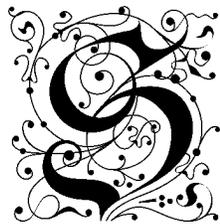IMPLICIALE QUANTUMGRAVITATIE. Zo luidt de Nederlandse titel van dit proefschrift. De quantumgravitatie, oftewel het maken van een quantummechanische beschrijving van de zwaartekracht, is een van de grootste problemen in de tegenwoordige hoge energie fysica. Een probleem bij de gewone quantisatie van zwaartekracht is dat de algemene relativiteitstheorie in storingsrekening niet renormaliseerbaar is. Dit zegt echter nog weinig over het niet-perturbatieve gedrag.

De simpliciale quantumgravitatie is een kandidaat voor een niet-perturbatieve definitie van de quantumgravitatie. Het uitgangspunt is het opbouwen van een Euclidische ruimtetijd door het aan elkaar plakken van simplices. In de dynamische triangulatie methode, waar dit proefschrift over handelt, wordt de padintegraal over metrieken dan gedefiniëerd door middel van een som over alle mogelijke manieren om de simplices tot een stuksgewijs lineaire variëteit aan elkaar te plakken, waarbij in het algemeen alleen de configuraties met een vaste topologie in aanmerking worden genomen.

Het model bevat twee koppelingsconstanten, de gravitatieconstante en de kosmologische constante. Om een continuumlimiet te nemen moet het aantal simplices naar oneindig gaan, hetgeen de kale (ongerenormaliseerde) kosmologische constante vastlegt. Als functie van de gravitatieconstante blijkt het model twee fases te hebben, een gekreukelde fase en een uitgerekte fase.

In de gekreukelde fase zijn de afstanden tussen de simplices erg klein en neemt het volume binnen een bol exponentieel toe met de straal ervan. In de uitgerekte fase vormt het systeem nauwe buisjes die zich gedragen als een polymeer met vertakkingen. Het lijkt er op dat het systeem schaalt, hetgeen wil zeggen dat het gedrag onafhankelijk wordt van de grootte van de simplices wanneer die erg klein worden. Op de fase-overgang lijkt het systeem in een aantal aspecten op een vierdimensionale bol. Bij die fase-overgang blijken de krommingscorrelaties een lange dracht te hebben en af te vallen als de vierde macht van de afstand. Dit is een aanwijzing voor het bestaan van massaloze deeltjes (gravitonen?) in het model.

Een interessante vraag is of het model gravitationele binding kan reproduceren. Dit is onderzocht door het gedrag van een scalair veld op de configuraties te bestuderen. Hoewel de data nog niet geheel sluitend zijn lijkt het erop dat er inderdaad een positieve bindingsenergie is.

Bij het uitvoeren van computersimulaties is het belangrijk of de gebruikte stappen in de configuratieruimte alle configuraties kunnen bereiken. We noemen deze stappen dan ergodisch. In onze simulaties is dit inderdaad het geval. Het aantal benodigde stappen blijkt echter niet begrensd te worden door een berekenbare functie van de grootte van de configuratie. Het is onduidelijk wat dit in de praktijk voor effect heeft. Ik heb numeriek naar deze effecten gezocht, maar ze niet gezien. Dat zou kunnen betekenen dat ze in de praktijk niet belangrijk zijn.

In het grootste deel van de gevallen wordt alleen gekeken naar een som over configuraties met dezelfde topologie. We kunnen echter ook een model definiëren waarin over topologieen gesommeerd wordt. Dit model heeft echter nog veel problemen, zowel in de definitie als in de simulatie ervan.

Bij de bestudering van de simpliciale quantumgravitatie spelen Monte Carlo simulaties een grote rol. Deze worden uitgevoerd door een kleine beginconfiguratie te nemen en deze steeds te wijzigen door er zetten op uit te voeren. De waarschijnlijkheden waarmee deze zetten worden uitgevoerd moeten voldoen aan bepaalde vergelijkingen die zorgen dat de kans om een configuratie te bereiken evenredig is aan het Boltzmann gewicht van die configuratie. Dit alles is geïmplementeerd in een C programma.

# Bijlage Drie

# Uitleg voor niet-fysici

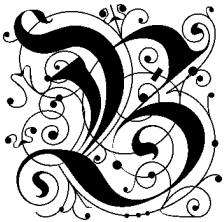IJLAGE drie heb ik geschreven in de hoop de geïnteresseerde leek een idee te geven waar dit proefschrift nu eigenlijk over gaat. Aan de hand van de woorden in de titel zal ik eerst iets vertellen over het onderwerp van onderzoek en uiteindelijk over mijn deel daarin. In de voetnoten staat wat extra informatie waarbij geen poging is gedaan deze zonder voorkennis begrijpelijk te maken.

## C.1   Zwaartekracht

Zwaartekracht (gravity in het Engels) kennen we allemaal. Het houdt ons op de grond, de maan bij de aarde en de aarde bij de zon.

Fysici beschrijven zwaartekracht als een kromming van de ruimte. Een gekromde ruimte is helaas niet iets wat we ons goed kunnen voorstellen. Daarom zullen we in veel voorbeelden doen alsof de ruimte twee-dimensionaal is, zoals een vel papier. Als we dit papier nu wat buigen of zelfs helemaal in elkaar frommelen hebben we een voorstelling van een gekromde ruimte.* Dat we die kromming van onze ruimte niet zien komt doordat deze erg klein is. In de ruimte kunnen we wel voorbeelden zien. Sommige sterren zien er (met behulp van grote telescopen) vervormd uit, omdat de ruimte ergens tussen ons en die ster sterk gekromd is. Ongeveer zoals bij een lachspiegel.

Hoe kan nu een kracht eigenlijk een kromming van de ruimte zijn? Ik zal dit illustreren met twee varende schepen. Kijk eens naar figuur C.1 op de pagina hierna. Op een dag beginnen twee schepen samen aan de lange reis naar het

---

*Strikt genomen hebben we pas een gekromde ruimte als we het papier bij het vervormen laten rekken.

Figuur C.1. Een gekromde ruimte veroorzaakt aantrekking.

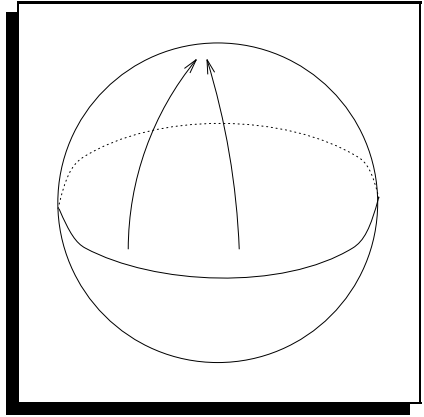

noorden. Om niet het gevaar te lopen tegen elkaar aan te botsen spreken ze (misschien wat onpraktisch) af om evenwijdig te beginnen en precies rechtdoor te blijven varen. Zo zouden ze dezelfde afstand moeten houden.

Zo gezegd, zo gedaan. Gelukkig is er op een van de schepen nog een oplettende matroos die niet veel van evenwijdige lijnen weet en de zaak niet zo vertrouwt. Na enige tijd merkt hij dat de schepen toch wel gevaarlijk dicht bij elkaar beginnen te komen. Zoals in de figuur te zien is, komt dit door de kromming van de aarde. Hoewel beide schepen dachten dezelfde afstand te zullen houden, gaan ze naar elkaar toe. Dit gaat steeds sneller. Op dezelfde manier werkt zwaartekracht. De ruimte is gekromd en zolang iets niet wordt tegengehouden (door bijvoorbeeld de vloer) zal het rechtdoor bewegen in die gekromde ruimte. Als ik van het dak af spring zullen de aarde en ik steeds sneller naar elkaar toe bewegen omdat we beide rechtdoor gaan in de gekromde ruimte.

In het voorbeeld met de varende schepen lag de kromming van te voren vast. Dit geldt niet voor de kromming van onze ruimte. Die ruimtekromming wordt veroorzaakt door alle voorwerpen. Hoe zwaarder iets is, hoe meer kromming het in zijn buurt veroorzaakt. Dit effect is maar heel klein, dus in het dagelijks leven merken we niets van de kromming die door gewone voorwerpen wordt veroorzaakt. Pas bij dingen zo groot als de aarde wordt dit belangrijk. In figuur C.2 op de rechter pagina is dit getekend. (Een gekromde ruimte is niet te tekenen. Daarom heb ik de tekening gemaakt alsof er maar twee dimensies zijn en de ruimte dus een gekromd oppervlak is.) In het midden bevindt zich een ster. De dikke lijn is de baan van een ruimteschip.* Hoewel het ruimteschip in de gekromde ruimte

---

*Natuurlijk veroorzaakt het ruimteschip zelf ook een kromming, maar omdat de ster zoveel zwaarder is kunnen we de kromming door het ruimteschip verwaarlozen.

Figuur C.2. Aantrekking tussen voorwerpen via ruimtekromming.

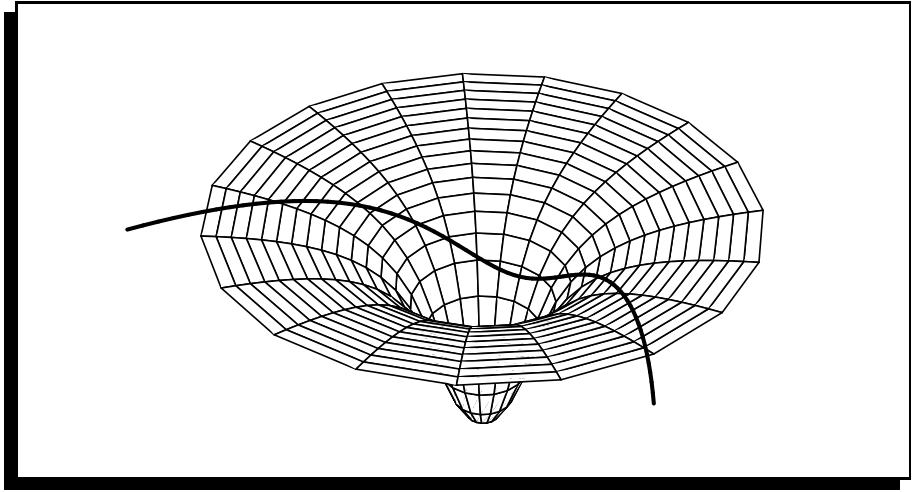

steeds rechtdoor gaat zal het uiteindelijk toch een andere richting hebben dan het mee begon en dus afgebogen zijn.

## C.2  Ruimtetijd

Als u goed over het bovenstaande nadenkt, dan zou u tot de conclusie moeten komen dat ik onzin sta te verkopen. Ik beweer in de vorige paragraaf dat als u een bal gooit dat deze rechtdoor gaat in de gekromde ruimte. Maar dat zou betekenen dat die bal altijd dezelfde baan in de gekromde ruimte volgt, of u hem nu hard of zacht gooit. Snel rechtdoor en langzaam rechtdoor maken tenslotte geen verschil in de baan die uiteindelijk gevolgd wordt. U weet wel beter: een bal die u harder gooit zal ergens anders terecht komen. In deze paragraaf zal ik proberen uit te leggen hoe dat komt.

In figuur C.3 op de pagina hierna heb ik een grafiek getekend van een fietstocht van Amsterdam naar Utrecht. Horizontaal staat de tijd, verticaal de afstand tot Amsterdam. Als de fietser pauzeert blijft deze enige tijd op dezelfde plek, de grafiek loopt dan horizontaal. Als hij fietst loopt de grafiek scheef, en hoe sneller hij fietst, hoe steiler de grafiek loopt.

Zo'n plaatje stelt de ruimtetijd voor. Een punt in de ruimtetijd komt overeen met een plaats en een tijdstip. Omdat we voor het tijdstip een getal extra hebben heeft de ruimtetijd een dimensie meer dan de ruimte. In het voorbeeld hadden we twee dimensies nodig voor het plaatje, terwijl er maar één ruimtedimensie was, namelijk de afstand tot Amsterdam. Onze wereld is drie-dimensionaal. We Figuur C.3. De ruimtetijd.

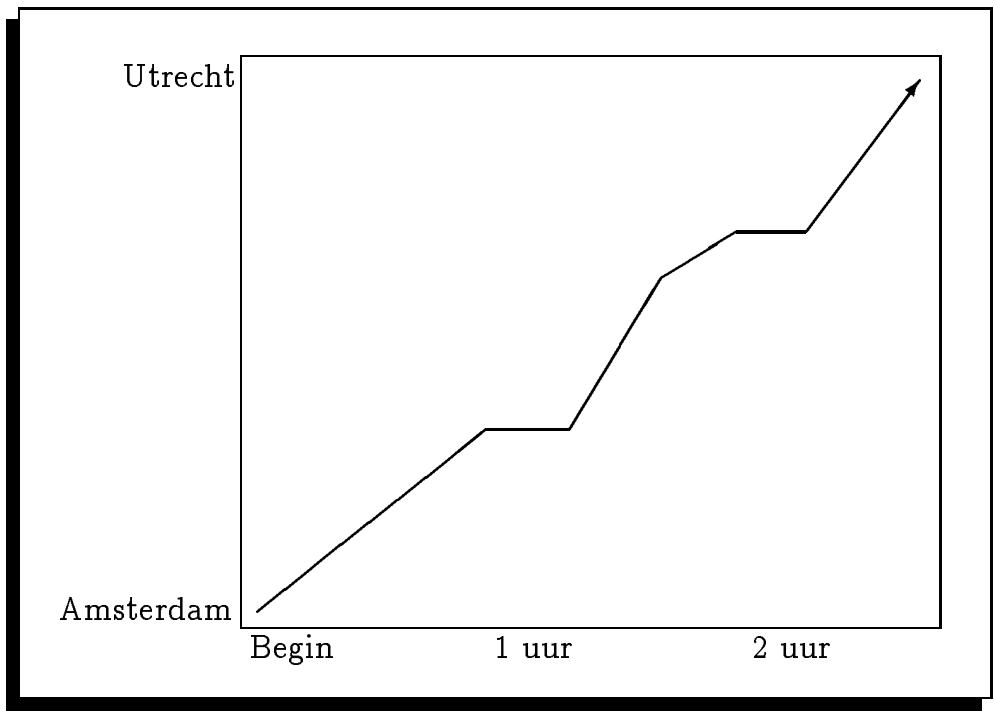

leven dus in drie ruimtedimensies en daarom heeft onze ruimtetijd vier dimensies. Eigenlijk is niet alleen de ruimte gekromd, maar de hele ruimtetijd.

We zien nog iets in het plaatje. Wanneer je sneller gaat dan loopt je grafiek steiler en ga je in die ruimtetijd dus een andere kant op. Een andere snelheid is hetzelfde als een andere richting in de ruimtetijd. Omdat de ruimtetijd gekromd is, hangt het dus ook van je snelheid af wat "rechtdoor in de gekromde ruimte" precies is en dus welke baan je volgt.

In het plaatje van de fietstocht heb ik de schaal eigenlijk niet goed getekend. De ruimtetijd zit zo in elkaar dat we met 45 graden schuin vooruit gaan in de ruimtetijd wanneer we bewegen met de lichtsnelheid (die is 300.000 kilometer per seconde). Met andere woorden: een seconde naar rechts (in de tijdrichting) is even ver in de ruimtetijd als 300.000 kilometer naar boven. Dat betekent dus dat we in de tijdrichting zeer grote "afstanden" aan het afleggen zijn. Zo'n zeer grote afstand door een ruimtetijd die overal maar een klein beetje krom is betekent dat het totale effect toch groot kan zijn. Daarom hoeft de ruimte maar een heel klein beetje gekromd te zijn om de zwaartekracht die we om ons heen zien te veroorzaken.

De conclusie is dat we voor een beschrijving van de zwaartekracht naar vierdi-

Figuur C.4. Een benadering van een bol door middel van platte stukjes.

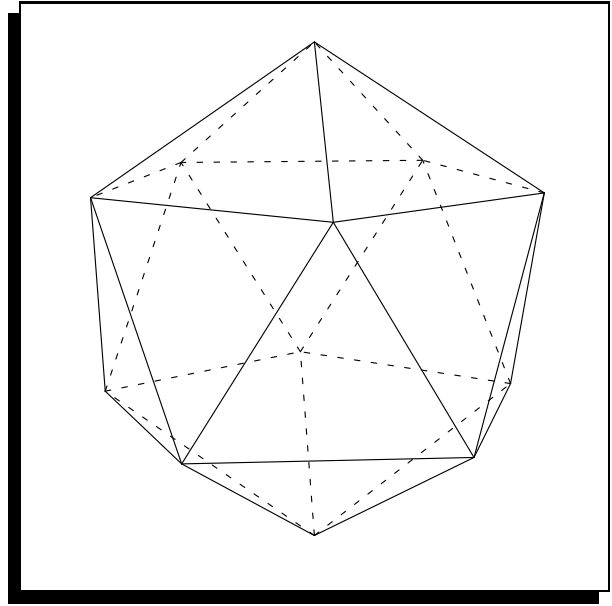

mensionale gekromde ruimtetijden moeten kijken. Ik zal deze voor het gemak weer ruimtes noemen, want de tijddimensie is niet anders dan een ruimtedimensie. Een gekromd oppervlak is een gekromd oppervlak, of we het nu een twee-dimensionale ruimte noemen of een ruimtetijd met één ruimtedimensie.*

## C.3   Simplices

Om te kijken wat de effecten van de zwaartekracht zijn willen we nu graag deze gekromde ruimte op de computer simuleren. Daartoe zouden we van elk punt van de ruimte moeten bijhouden hoe de kromming loopt. Maar zelfs een eindig stukje ruimte (bijvoorbeeld een kubieke centimeter) heeft oneindig veel punten. Om van al die punten de kromming bij te houden zouden we een computer nodig hebben met oneindig veel geheugen en die bestaat niet. Daarom doen we net alsof de ruimte uit een groot aantal (bijvoorbeeld tienduizend) platte stukjes bestaat. We hoeven dan alleen maar bij te houden welk stukje met welk ander stukje verbonden is. In figuur C.4 zien we hoe we op die manier met platte stukjes een bol kunnen benaderen.

Als platte stukjes gebruiken we simplices, vandaar de titel. Hoe een simplex

---

*Helaas maakt dit niet alleen eigenlijk toch uit, het is zelfs een van de grootste problemen van de quantumgravitatie.

Figuur C.5. Simplices in verschillende dimensies.

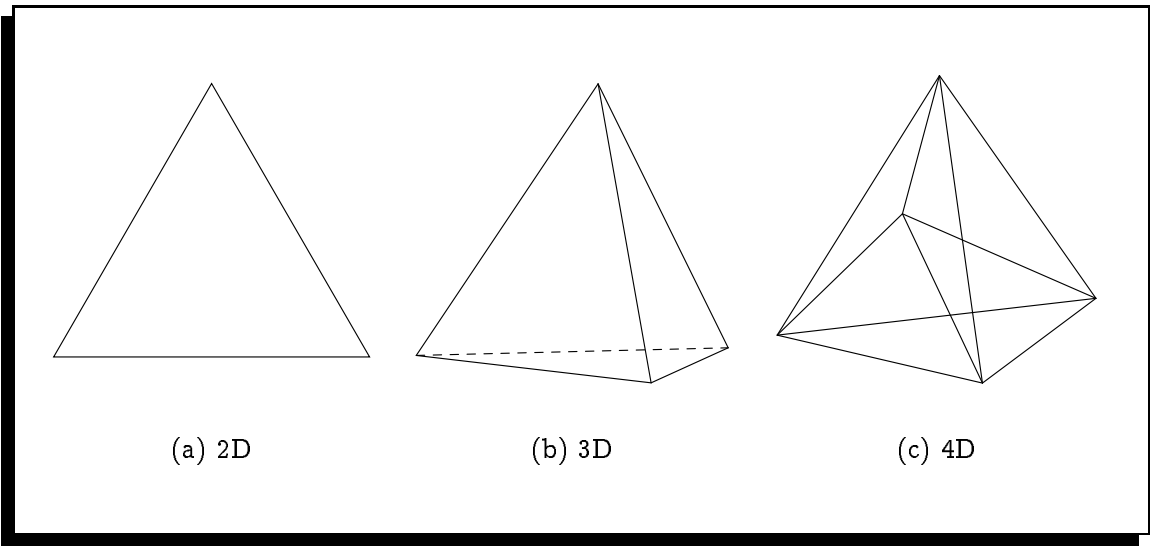

(a) 2D          (b) 3D          (c) 4D

eruit ziet hangt af van het aantal dimensies dat je bekijkt. Enkele voorbeelden zijn te zien in figuur C.5. In twee dimensies is een simplex gewoon een driehoek. In drie dimensies is het een viervlak, ook wel een driezijdige piramide genoemd. In vier dimensies noemen we het een 4-simplex. Dit is niet meer voor te stellen. Desgewenst kunt u in de rest van de tekst voor "simplices" gewoon "driehoekjes" lezen.

Als we nu maar steeds kleinere simplices gebruiken om de ruimte uit op te bouwen dan kunnen we steeds beter de werkelijkheid benaderen. Helaas hebben we er dan ook steeds meer nodig en past het al vlug niet meer in onze computers.

Terzijde: misschien heeft u eens gehoord dat quantummechanica te maken heeft met het opdelen in stappen van dingen als lading en energie, waardoor deze niet meer traploos verstelbaar zijn. Voor alle duidelijkheid: dat heeft hier niets mee te maken. Het opdelen van de ruimte in stukjes is hier alleen maar een manier om die ruimte in de computer te krijgen.*

## C.4   Quantumfysica

In de zogenaamde klassieke fysica (dat is niet-quantum fysica) is, gegeven de situatie op een zeker moment, alles precies bepaald. Als ik een bal opgooi, zegt de klassieke fysica mij precies hoe die bal zich zal bewegen. In de quantumfysica is dit niet zo. De bal kan alle mogelijke routes volgen, maar sommige routes zijn waarschijnlijker dan andere. De quantumfysica is voor zover we weten de eigenlijke werkelijkheid. De klassieke fysica werkt echter zo goed, omdat de kans dat de bal meer dan een heel klein beetje van de klassieke route afwijkt bijzonder klein is. Ik zal die bal dus nooit plotseling heen en weer zien zigzaggen in plaats van zijn gewone boogje te zien volgen.

---

*En ook om het model überhaupt te definiëren.

Als we echter naar heel kleine dingen kijken, zoals atomen, dan is een kleine afwijking toch erg belangrijk. Het is dan niet meer voldoende om alleen naar het gedrag volgens de klassieke fysica te kijken. We moeten alle mogelijke manieren waarop het atoom kan bewegen bekijken en daar een gemiddelde van nemen. Willen we bijvoorbeeld de snelheid weten, dan berekenen we de gemiddelde snelheid van al die routes. Zo'n gemiddelde is een gewogen gemiddelde. Sommige banen hebben meer kans dan andere.

Om dingen te berekenen in quantumgravitatie moeten we dus kijken naar alle mogelijke gekromde ruimtes en daar een gewogen gemiddelde van nemen. Het grote probleem is echter dat niemand weet hoe je een gemiddelde van "alle mogelijke gekromde ruimtes" neemt.

Probeer maar eens het gemiddelde van oneindig veel getallen te nemen. Eerst tellen we ze allemaal op, dan krijgen we oneindig, en dan delen we de som door het aantal, ook oneindig. Je kunt echter niet oneindig door oneindig delen. Gelukkig is het niet zo erg, er zijn in veel gevallen wel manieren om zo'n gemiddelde te berekenen. Maar een gemiddelde over alle gekromde ruimtes snapt niemand.

## C.5   Simplicial Quantum Gravity

Het idee van de zogenaamde *simplicial quantum gravity* waar dit proefschrift over gaat is nu om het gewogen gemiddelde te nemen van alle manieren om simplices aan elkaar te plakken. Een zo'n manier zag u in figuur C.4 op pagina 141, maar er zijn er uiteraard nog veel meer. Gelukkig is dit wel een eindig aantal, dus nu weten we hoe we zo'n gemiddelde moeten nemen en kunnen we de computer dit uit laten rekenen.

Dit concept was al bekend toen ik aan mijn onderzoek begon. Ik heb nader bekeken in hoeverre er redelijke dingen uit dit model tevoorschijn komen. In deze paragraaf zal ik iets zeggen over een deel van de dingen die ik heb gedaan.

Ik heb hierboven gezegd dat hoe kleiner we de simplices maken, hoe beter we de werkelijkheid benaderen. Wat we echter niet weten is of we ook het goede gemiddelde over alle gekromde ruimtes benaderen. Als alles goed is dan moeten

Figuur C.6. Steeds gekkere ruimtes?

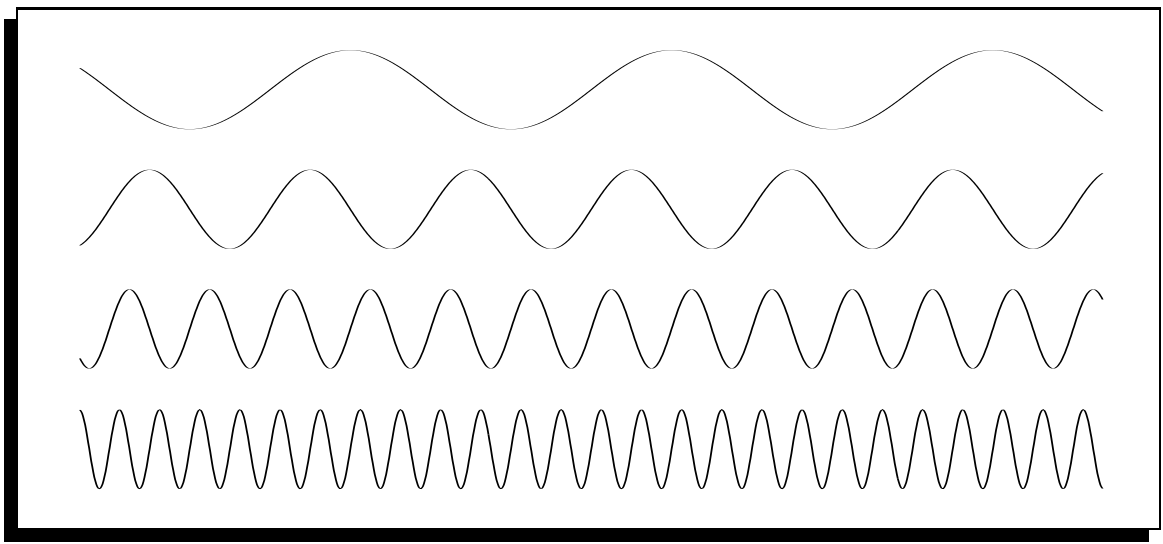

de dingen die we uit de simulatie krijgen als we steeds kleinere simplices gebruiken naar een bepaalde waarde gaan. Dan zeggen we dat die waarde de echte is. Het zou echter best kunnen gebeuren dat wat er uit de computer komt steeds maar groter of kleiner blijft worden. Dan is het model gewoon niet goed.

Andere mensen hebben beweerd dat als je de simplices steeds kleiner maakt dat dan de gemiddelde ruimtekromming steeds maar groter blijft worden. Ik beweer echter dat je daar anders naar moet kijken. In figuur C.6 heb ik enkele stukken golfplaat van de zijkant getekend. Eenvoudig gesteld is dit wat er gebeurt als je steeds kleinere simplices neemt. Inderdaad wordt de gemiddelde ruimte die we met de simplices maken steeds gekker en zou je kunnen concluderen dat hier nooit een mooie gladde ruimte (lees: oppervlak) uit gaat komen. Al die golfjes in de ruimte zijn echter veel kleiner dan een atoom* en die zien we dus nooit. Als we van een grote afstand kijken dan zijn al die stukken golfplaat gewoon plat. Ik meen dat het ook zo werkt met de ruimte die uit de simulaties tevoorschijn komt.

Iets anders waar ik naar gekeken heb is of dingen elkaar inderdaad aantrekken in dit model. Ook dat is van tevoren niet duidelijk. Een model van zwaartekracht waarin we niet kunnen zien dat dingen elkaar aantrekken kunnen we uiteraard net zo goed in de prullenmand gooien. Gelukkig hebben we inderdaad aantrekking gezien.

De conclusie is dat het model er op dit moment hoopvol uitziet. Of het klopt,

---

*Een atoom is ongeveer $10^{-10}$ m groot en een proton ongeveer $10^{-15}$ m, terwijl de ruimtetijd fluctuaties zich afspelen op de zogenaamde Planck schaal van ongeveer $10^{-34}$ m.

dat wil zeggen dat het een goede beschrijving van de werkelijkheid geeft, weten we nog niet.

## C.6   Waarom?

Ik heb nu geprobeerd uit te leggen waar dit proefschrift over gaat, maar de vraag doet zich voor wat er nu eigenlijk interessant is aan die quantumgravitatie. U zou kunnen denken: "Als die afwijkingen veel kleiner zijn dan een atoom, wat kan ons die dan schelen?" en u zou nog gelijk hebben ook. Toch zijn er enkele redenen om quantumgravitatie te bestuderen.

We zien dat alle melkwegstelsels in het heelal zich van elkaar af bewegen. Daaruit (en uit nog veel meer aanwijzingen) concluderen we dat ze dus vroeger dichter bij elkaar zaten. Zo dicht zelfs, dat bij het begin van het heelal alles in een punt bij elkaar zat. Vlak na het begin (de zogenaamde Big Bang) was het heelal dus zo klein, dat zelfs die hele kleine afwijkingen die de quantumgravitatie beschrijft erg belangrijk waren. Dus om te weten wat er in het begin gebeurde hebben we quantumgravitatie nodig.

De tweede reden is dat fysici gewoon grenzeloos nieuwsgierig zijn en graag willen weten hoe de wereld in elkaar zit, zelfs al merken we niets van dat aspect ervan.

Er zijn vele voorbeelden in de geschiedenis van de natuurkunde waarbij dingen waar eerst niemand iets nuttigs in zag later toch op onvoorziene wijze toegepast werden in alledaagse gebruiksvoorwerpen zoals televisies. Velen menen dat de quantumgravitatie zó obscuur is dat daar echt nooit iets mee gedaan zal kunnen worden. Daarom wil ik dit hoofdstuk eindigen met een uitermate speculatief idee. Aan de ene kant is het bekend dat men energie uit roterende zwarte gaten kan halen door er dingen (bijvoorbeeld afval) vlak langs te gooien. Aan de andere kant weten we dat vlak bij zwarte gaten quantumgravitatie een belangrijke rol moet spelen. Wie weet ligt hier dus ooit eens een toepassing.

# Dankwoord

ALHOEWEL het nogal als een cliché klinkt, wil ik graag mijn oprechte dank betuigen aan Jan Smit, zonder wiens begeleiding van dit proefschrift weinig terecht zou zijn gekomen. Men hoort wel eens verhalen over begeleiders die hun promovendi nauwelijks zien en niet weten waar die mee bezig zijn. Niets was minder het geval. Hoewel ik de vele malen dat hij met enthousiaste ideeen mijn kamer binnenliep of zelfs van elders opbelde op het moment in kwestie wel eens als storend ervoer, waren ze zeer stimulerend.

Mijn instituutsgenoten dank ik voor het scheppen van een fijne plek om te werken. Een paar mensen wil ik in het bijzonder noemen. Bernard Nienhuis, vanwege de enthousiaste manier waarop hij soms met alledaagse, niet direct aan het onderzoek gerelateerde, fysische onderwerpen kwam. Mijn kamergenote Nathalie Muller, vanwege het gezelschap, het gebak en het verdragen van mijn onaflatend computergebruik. I would like to thank Piotr Białas for the pleasurable cooperation.

De systeembeheerders van de vele computers die ik gebruikte wil ik bedanken voor hun medewerking aan mijn soms bovengemiddelde wensen, voor het terughalen van de files die ik per ongeluk verwijderde en voor het niet opmerken wanneer ik meer deed dan zij mogelijk achtten.

Willemien wil ik bedanken voor alles, zoals men dat noemt, en in het bijzonder voor de opbouwende kritiek op de uitleg voor niet-fysici.

# Publications

This thesis is partly based on the following articles:

- ★ B.V. de Bakker and J. Smit, *Euclidean gravity attracts*, Nucl. Phys. B (Proc. Suppl.) 34 (1994) 739.

- ★ B.V. de Bakker and J. Smit, *Volume dependence of the phase boundary in 4D dynamical triangulation*, Phys. Lett. B 334 (1994) 304.

- ★ B.V. de Bakker and J. Smit, *Curvature and scaling in 4D dynamical triangulation*, Nucl. Phys. B 439 (1995) 239.

- ★ B.V. de Bakker, *Dynamical triangulation with fluctuating topology*, Nucl. Phys. B (Proc. Suppl.) 42 (1995) 716.

- ★ B.V. de Bakker and J. Smit, *Exploring curvature and scaling in 4D dynamical triangulation*, Nucl. Phys. B (Proc. Suppl.) 42 (1995) 719.

- ★ B.V. de Bakker, *Absence of barriers in dynamical triangulation*, Phys. Lett. B 348 (1995) 35.

- ★ B.V. de Bakker and J. Smit, *Two-point functions in 4D dynamical triangulation*, preprint ITFA-95-1, submitted to Nucl. Phys. B.

# Index

# Production notes

This thesis was edited using GNU Emacs and typeset with the help of TeX and many macro packages, in particular LaTeX 2$_\varepsilon$ and $\mathcal{AMS}$-LaTeX. The typeface used in the main text is Concrete, while that in the formulas is AMS Euler. The initials, made by Yannis Haralambous, are simply called yinit. The **boldface** used in headings and on the front cover is my own adaptation to Concrete of **Computer Modern Bold Extended**. The cyrillic typeface is my own adaptation to Concrete of Computer Modern Cyrillic. All these were made with the help of METAFONT.

Drawings were made with xfig and converted to postscript by fig2dev. Plots were made using gnuplot, together with the enhpost driver. In both cases text and formulas were put into them using psfrag. Special thanks go to GNU make, which I finally coerced into running LaTeX exactly as many times as needed to get the cross references stable.

All computer programs mentioned above, as well as all fonts used in this thesis, are free software.

"There is a danger that authors —
who are now able to typeset their own books with TeX
— will attempt to do their own designs."

DONALD E. KNUTH, *The TeXbook*